\documentclass[final,3p,times]{elsarticle}

\usepackage{amssymb}

\usepackage{amsmath}
\usepackage{amssymb}
\usepackage{mathrsfs}
\usepackage{algpseudocode}
\usepackage{algorithm}

\usepackage{subcaption}
\usepackage{enumitem}
\usepackage[colorlinks=true,
            linkcolor=black,
            citecolor=black,
            urlcolor=black,
            bookmarksnumbered=true,
            pdfstartpage=1,
            pdfpagemode=UseOutlines]{hyperref}

\usepackage{xcolor} 

\newcommand{\tens}[1]{\boldsymbol{#1}}    
\newcommand{\dmat}[1]{\mathbf{#1}}

\begin{document}

\begin{frontmatter}


\title{Midplane based 3D single pass unbiased segment-to-segment contact interaction using penalty method}
\author[affiliation_label]{Indrajeet Sahu\corref{cor1}} 
\ead{indrajeet.sahu@eng.ox.ac.uk}
\cortext[cor1]{Corresponding author}
\author[affiliation_label]{Nik Petrinic}
\affiliation[affiliation_label]{
            addressline={Department of Engineering Science},
            organization={University of Oxford},
            city={Oxford},
            postcode={OX1 3PJ},
            state={Oxfordshire},
            country={United Kingdom}}

\title{}


\begin{abstract}
This work introduces a contact interaction methodology for an unbiased treatment of contacting surfaces without assigning surfaces as master and slave. The contact tractions between interacting discrete segments are evaluated with respect to a midplane in a single pass, inherently maintaining the equilibrium of tractions. These tractions are based on the penalisation of true interpenetration between opposite surfaces, and the procedure of their integral for discrete contacting segments is described in this paper. A meticulous examination of the different possible geometric configurations of interacting 3D segments is presented to develop visual understanding and better traction evaluation accuracy. The accuracy and robustness of the proposed method are validated against the analytical solutions of the contact patch test, two-beam bending, Hertzian contact, and flat punch test, thus proving the capability to reproduce contact between flat surfaces, curved surfaces, and sharp corners in contact, respectively. The method passes the contact patch test with the uniform transmission of contact pressure matching the accuracy levels of finite elements. It converges towards the analytical solution with mesh refinement and a suitably high penalty factor. The effectiveness of the proposed algorithm also extends to self-contact problems and has been tested for self-contact between flat and curved surfaces with inelastic material. Dynamic problems of elastic and inelastic collisions between bars, as well as oblique collisions of cylinders, are also presented. The ability of the algorithm to resolve contacts between flat and curved surfaces for nonconformal meshes with high accuracy demonstrates its versatility in general contact problems. 

\end{abstract}

\begin{highlights}
\item A midplane-based high-accuracy unbiased contact methodology is proposed for fair treatment of surfaces without labelling them as master and slave

\item The proposed methodology evaluates contact traction unbiasedly in a single pass instead of the dual-pass approach typically used to offset biasing in the master-slave treatment of surfaces

\item The work also explains the different possible geometrical configurations that contacting segments can have for the evaluation of contact forces using the penalty method for nonconformal meshes

\item The unbiased nature of the proposed methodology is also of direct benefit to the problems involving self-contact and large deformation
\end{highlights}


\end{frontmatter}



\section{Introduction}

The design and development of complex systems in engineering applications, like manufacturing, automotive, and aerospace industries, have accelerated in the last few decades due to the increasing capability to simulate physical problems at larger scales with higher accuracy. While in any system, there will be an effect of non-contact forces, e.g. gravity and electromagnetic forces, all components in assemblies are generally either supported in equilibrium or maintained in a desired motion by contact forces. The analysis of contact problems poses a critical challenge due to the nonlinearities arising from the lack of prior knowledge of contact area and the contact traction distribution \cite{johnsonContactMechanics1985}\cite{munjizaCombinedFiniteDiscrete1995}. The finite element method (FEM) has been a widely accepted methodology for numerical solutions to structural problems \cite{liuEightyYearsFinite2022}. Computational solutions using finite elements can generally proceed through a route of either implicit or explicit time integration, with the explicit method being generally preferred for problems with high dynamic nature or nonlinearities \cite{batheKJBatheFinite2014}\cite{laursenComputationalContactImpact2003}. The nonlinearities in the system can be induced by both material and geometric behaviour, with contact conditions often playing a critical role. 
Having a highly accurate contact interaction methodology will offer a competitive advantage over traditional methods in a broader range of applications, and is a core theme of this work.

In continuously evolving systems, the changing positions and configurations of all bodies imply that certain surfaces might have physical interactions. Numerical solutions for such physical interactions between discretised domains involve two aspects: contact discretisation and contact constraint enforcement. Contact discretisation describes the procedure of forming contact pairs between discrete entities on opposite surfaces. The motion of these contacting discrete entities is resisted against interpenetration using contact constraint enforcement techniques such as the widely used Lagrange multiplier and penalty methods. While the Lagrange multiplier method seeks an exact enforcement of constraints, the increased degrees of freedom add to the computational cost. On the other hand, the penalty method penalises interpenetration between contacting surfaces by using a penalty factor without increasing degrees of freedom in the system \cite{wriggersComputationalContactMechanics2002}. It is not only relatively easier to implement the penalty method in any solver, but it also becomes particularly advantageous to use in large-scale problems where the computational cost of a large number of degrees of freedom may be a concern.

For finite element solutions, multiple contact discretisation schemes have been used - node-to-node(NTN), node-to-segment (NTS), and segment-to-segment (STS). The NTN method, a very primitive approach, forms contact pairs between nodes on conforming meshes with matching nodes \cite{hughesFiniteElementMethod1976}, and it also passes the contact patch test for transferring uniform pressure along the contact interface. However, it is restricted to infinitesimally small sliding, as any large deformation or finite sliding leads to nonconformity between the meshes, requiring the contacting entities to be remeshed to form node-to-node pairs. Recent works in \cite{kimIntroductionNonlinearFinite2015}\cite{jinNodetonodeSchemeAid2015}\cite{jinNodetonodeSchemeThreedimensional2016} present 2-dimensional (2D) and 3-dimensional (3D) NTN schemes by employing variable node elements, where hexahedral elements are replaced by polyhedral elements, and it passes the contact patch test while also working with large deformations. Polyhedral elements for the NTN scheme based on the scaled boundary finite element method have also been used to convert nonconformal meshes (non-matching nodes) into conformal ones and pass the contact patch test \cite{xingScaledBoundaryFinite2018}\cite{xingNodetonodeSchemeThreedimensional2019}.

The NTS scheme forms contact pairs between nodes of one surface labelled as slave and the segments on the other surface labelled as master \cite{hughesFiniteElementMethod1976,hughes1977finite}\cite{hallquistNIKE2DImplicitFinitedeformation1979}. It opposes the penetration of slave nodes into the master surface, thereby resulting in contact traction between the two surfaces. Unlike NTN, it is not limited to conformal meshes but can also be applied to nonconformal meshes experiencing sliding with large deformations \cite{hallquistSlidingInterfacesContactimpact1985}. It is well suited for linear solid elements with a single integration point \cite{flanaganUniformStrainHexahedron1981}. The NTS scheme with a single pass suffers from bias due to the master/slave approach, allowing the master nodes to penetrate the slave surface without generating a contact force. Regardless of using the Lagrange multiplier or penalty method, it fails the contact patch test, with the mesh refinement not alleviating this issue \cite{taylorPatchTestContact1991}. The one-pass approach can be improved by adding a virtual slave node with the penalty method to satisfy the contact patch test \cite{zavariseModifiedNodetosegmentAlgorithm2009}\cite{zavariseNodetosegmentAlgorithm2D2009}, by smoothed FEM \cite{sunNovelNodetosegmentAlgorithm2023}, by area regularisation techniques \cite{kangImprovedAreaRegularization2023}, or by a dual-pass strategy with Lagrange multipliers \cite{pusoDualPassMortar2020}\cite{netoSurfaceSmoothingProcedures2017}. In the explicit framework, a double pass-based approach using the penalty method can be referred to in \cite{danielsonCurvedNodetofaceContact2022}. The NTS scheme, however, suffers issues in the detection phase with non-smooth $C^0$ continuous surfaces for unique projection of slave nodes \cite{yastrebovNumericalMethodsContact2013}, leading to jumps in nodal forces at the contact interface. Additionally, it also suffers from locking and over-constraining problems \cite{zavariseNodetosegmentAlgorithm2D2009}.

In the STS approach initially proposed in 2D \cite{simoPerturbedLagrangianFormulation1985}, contact constraints are enforced between pairs of segments on opposite surfaces. With the work of \cite{zavariseSegmenttosegmentContactStrategy1998}, the STS method transitioned from pointwise imposition of contact constraints to continuous imposition over contacting segments. Initial approaches also involved the imposition of constraints over the integration points of contacting segments \cite{el-abbasiStabilityPatchTest2001}. Generally, in STS methods, one surface is projected over another, and contact constraints are applied on a common intersection area \cite{pusoMortarSegmenttosegmentContact2004}. While the NTS method suffers from a sudden jump in nodal force for the sliding of nodes from one segment to another, the STS method exhibits a gradual change in the volume of interpenetration, which predicts a physically smooth variation of the generated contact forces. This is particularly advantageous with large deformations, finite sliding, and curved surfaces, leading to improved accuracy in solutions. 
Consequently, even with a lower number of elements, the results on the contact interface are qualitatively well represented \cite{lorenzisIsogeometricContactReview2014}.
Mortar methods, initially developed for domain decomposition problems \cite{bernardiCouplingFiniteElement1990}, have also been extensively studied with STS approach \cite{puso3DMortarMethod2004}\cite{pusoMortarSegmenttosegmentContact2004}\cite{poppDualMortarApproach2010}\cite{carvalhoEfficientAlgorithmRigid2022} for different integration strategies \cite{farahSegmentbasedVsElementbased2015}, and with dual pass approach for self-contact \cite{pusoDualPassMortar2020}. The dual pass approach is generally used to eliminate the biasing of the master-slave labelling in NTS and STS methods by having an additional contact contribution coming from swapping of slave and master surfaces in each iteration or step of the solution \cite{laursenContinuumbasedFiniteElement1993}; however, this comes at the cost of additional computation due to the second pass. The enforcement of contact constraints has also been studied with intermediate surfaces in 2D for unbiased contact traction evaluation in mortar methods \cite{simoPerturbedLagrangianFormulation1985}\cite{mcdevittMortarfiniteElementFormulation2000}.

This paper proposes an unbiased approach for contact traction evaluation in a single pass. By defining a midplane, true interpenetration between the contacting segments is penalised through nodal forces obtained by integrating traction over the midplane, which is in contrast to the interpolation of nodal constraints \cite{pusoMortarSegmenttosegmentContact2004}\cite{pusoSegmenttosegmentMortarContact2008}. 
As the methodology is inherently unbiased, it can also maintain consistency in contact constraint enforcement in the case of self-contact. The presented contact formulation is implemented with an explicit finite element scheme using first-order elements based on central difference time integration and HRZ lumping of nodal masses \cite{zienkiewiczFiniteElementMethod2013}\cite{danielsonCurvedNodetofaceContact2022}.

The work in this paper is organized as follows - section 2 describes the analytical problem and development of its weak form for finite element solution. Section 3 presents the midplane-based contact methodology applied with the penalty method for contact traction. The solution procedure with the algorithm for the proposed method is described in section 4. Finally, numerous benchmark tests in section 5 demonstrate the robustness of the proposed algorithm.

\section{Problem description }  
The problem of solids in motion in continuum mechanics often involves physical contact between them, resulting in the transfer of loads and subsequent deformation of the solids. These contacts impose external constraints on the solids, influencing the evolution of state changes in the system. This section provides a mathematical description of the laws governing the motion and constraint-induced deformation of solids. Initially, the problem is described in its strong form,  outlining the fundamental equations that govern the system's behaviour. Subsequently, a variational principle-based weak solution is introduced to formulate its finite element-based solution. This mathematical description forms the basis for the proposed solution methodology, which is presented in the subsequent sections.

\subsection{Strong formulation of the continuum problem of contact}
\begin{figure}[t!]
    \centering
    \includegraphics[width=0.80\linewidth]{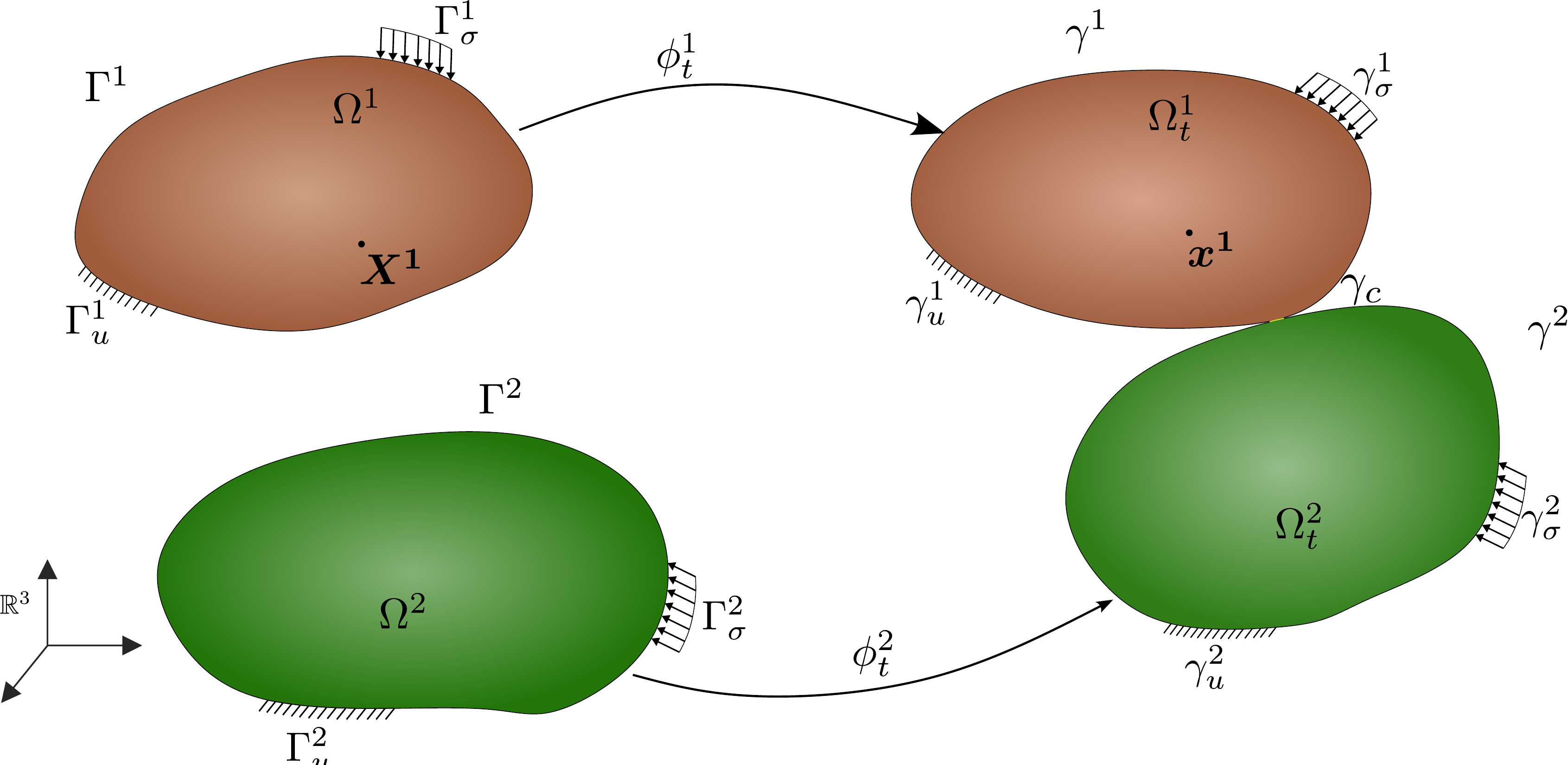}
    \caption{A general finite deformation frictional contact problem}
    \label{fig:Strong_Form_Contact_Problem_label}
\end{figure}

Consider a general case of a system of two deformable solids $\Omega^i$ $(i=1,2)$  in $\mathbb{R}^{n_{sd}}$ (where $n_{sd}=3$ for a 3D problem) undergoing motion and deformation leading to a mutual 
imposition of contact constraint as shown in Fig.~\ref{fig:Strong_Form_Contact_Problem_label}. The
position of any point $\tens{x}^i$ in either of the two bodies $\Omega^i$ at any 
time $t$ is related to its undeformed position $\tens{X}_i$ by displacement $\tens{u}$ 
as $ \tens{x} = \tens{X} + \tens{u}(\tens{X},t) $. In the 3D Euclidean space, the boundary value problem at any point $\tens{x}^i$ for each solid $\Omega^i $ will be represented through the following equations:

\begin{equation}
\begin{aligned}
    \nabla \cdot \boldsymbol{\sigma}^i + \tens{b}^i &=  \rho^i\partial^2_t \boldsymbol{x} &\quad   &\forall\ \tens{x}^i \in {\Omega}^i \times (0,T) \\
    \tens{u}^i(\tens{x}^i) &= \overline{\tens{u}}^i &\quad    &\forall\  \tens{x}^i \in \gamma_{u}^i \times (0,T) \\    
    \boldsymbol{\sigma}^i\tens{n}^i &= \overline{\tens{t}}_{\sigma}^i &\quad       &\forall\  \tens{x}^i \in \gamma_{\sigma}^i \times (0,T)
\end{aligned}
\end{equation}

The first equation follows from the conservation of momentum, with the symbols $\boldsymbol{\sigma}^i$ and $\tens{b}^i$ representing Cauchy stress and body force per unit volume. The second and third equations represent the imposed Dirichlet and Neumann boundary conditions on $\gamma_{u}^i$ and $\gamma_{\sigma}^i$, respectively, where the overbars in the notation $\overline{(\cdot)}$ are used to denote prescribed boundary conditions of imposed displacements and surface tractions over the two domains ($\gamma^i = \partial\Omega^i$). Here, $\gamma^i_{\sigma}\cap\gamma_u^i=\gamma^i_{\sigma}\cap\gamma_c^i=\gamma^i_u\cap\gamma_c^i=\emptyset$. The above set of equations governs the change of positions of the points in both domains. 
In the event the two bodies come into contact, the traction $\tens{t}_c^i $ generated at any point in their contact interface $\gamma_c$ ($=\gamma_1\cap\gamma_2 $) can be decomposed into a normal component $\tens{t}^i_N $ and a tangential component $\tens{t}^i_T $. The normal contact traction can be written as
\begin{align}
    \tens{t}_N^i = (\tens{t}_c^i \cdot \tens{n}^i) \tens{n}^i 
\end{align}
where $\tens{n}^i$ denotes the outward normal at the contact point on $\Omega^i$. By projection on the tangential plane, the frictional traction vector, if the problem also involves frictional effect, can be written as
\begin{equation}   
    \tens{t}_T^i =  (\tens{I}-\tens{n}^i\otimes\tens{n}^i)\tens{t}_c^i   
\end{equation}
The action-reaction principle implies that the tractions acting on the two bodies at any infinitesimally small area $d\gamma_c$ is equal in magnitude and opposite in direction, i.e., $\tens{t}_c^1 =-\tens{t}_c^2 $. The forthcoming proposition of the contact formulation satisfies this fundamental requirement in a single-pass approach. In this work, only frictionless surfaces in contact are considered, so determining $\tens{t}_N$ will be the sole focus of the work presented in this paper.

\subsection{Variational form of the solution}

By applying the variational principle to the strong formulation of the problem, a finite element solution to the nonlinear boundary value problem is developed that maintains the dynamic equilibrium at any instant for the given boundary conditions on the bodies. To develop the variational form, consider the space $\mathscr{U}^i$ containing all kinematically admissible deformations in the solid $\Omega^i$ to define the solution space 
\begin{equation}
    \mathscr{U}^i = \{  \tens{u}^i : \Omega^i \rightarrow \mathbb{R}^3 | \tens{u}^i \in [H^1(\Omega^i)]^{n_{sd}} , \tens{u}^i =  \overline{\tens{u}}^i \ \  \forall \tens{x}\in \gamma_{u}^i     \}     \\    
\end{equation}
and the space of kinematically admissible variations to define the weighting space as
\begin{equation}
    \mathscr{V}^i = \{  \delta \tens{u}^i : \Omega^i \rightarrow \mathbb{R}^3 | \delta\tens{u}^i\in[H^1(\Omega^i)]^{n_{sd}} ,  \delta\tens{u}^i=\tens{0} \ \ \forall  \tens{x} \in \gamma_{u}^i    \}
\end{equation}
where $H^1(\Omega^i)$ denotes the Sobolev space of functions which are square integrable functions over its values and their first derivatives. The weak form representing the virtual work for dynamic equilibrium at any instant to determine the deformation $\tens{u}^i$ can be written as
\begin{equation}
\int_{\Omega^{i}}\rho^i\partial_{t}^{2}\tens{u^{i}}\cdot\delta\tens{u^{i}}dV+\int_{\Omega^{i}}\boldsymbol{\sigma^{i}}(\tens{u}^i):\nabla(\delta\tens{u})^{i}dV=\int_{\Omega^{i}}\rho^i\tens{b}^{i}\cdot\delta\tens{u}^{i}dV+\int_{\gamma^i_{\sigma}}\tens{t}_{\sigma}^{i}\cdot\delta\tens{u}^{i}d\gamma^i + \int_{\gamma_{c}^i}\tens{t}_c^{i}\cdot\delta\tens{u}^{i}d\gamma^i \ , \ \ \ \forall \delta\tens{u}^i \in \mathscr{V}^i
\end{equation}
Here, the terms on the left side represent the internal virtual work ($\delta\Pi_{int}$) due to the inertia and the strain energy of the body, respectively. The terms on the right side denote the external virtual work ($\delta\Pi_{ext}$) due to the body forces($\tens{b}$), external surface tractions($\tens{t_{\sigma}}$), and the contact traction ($\tens{t}_c$) on this body $\Omega^i$, respectively. The Cauchy stress tensor $\tens{\sigma}^i$ in $\Omega^i$ is a function of deformation $\tens{u}^i$ through the constitutive model for the material of the body. The contact traction $\tens{t}_c^i$ is generated with the changing domain states that bring the two solids into contact and constrain their motion due to the so-established contact interface. In the traditional description, the potential contact surface $\gamma_c^i$ is decomposed into the so-called active contact surface $\gamma_A^i$ and inactive contact surface $\gamma_I^i=\gamma_C^i\setminus\gamma_A^i$ that are possibly changing continuously with time. Determining $\gamma_A^i$ and $\gamma_I^i$, along with $\tens{t}_c^i$, is the objective of the contact problem. Notice that displacement and its first derivatives both appear in the weak form and are contained in the Sobolev spaces defined above, which put a condition of square integrability.

Summing up the virtual work due to contact constraint over the contact interface $\gamma_c$( $=\gamma^1_c=\gamma^2_c$) we get
\begin{align}
    \delta\Pi_c(\tens{u},\delta\tens{u}) &= \int_{\gamma_{c}^1}\tens{t}_{c}^{1}\cdot\delta\tens{u}^{1}d\gamma+\int_{\gamma_{c}^2}\tens{t}_{c}^{2}\cdot\delta\tens{u}^{2}d\gamma      \label{eq:contact_virtual_work}       \\
     &= \int_{\gamma_{c}}\tens{t}_{c}^{1}\cdot(\delta\tens{u}^{1}-\delta\tens{u}^{2})d\gamma=\int_{\gamma_{c}}(||\tens{t}_{N}||\delta g_{N}+\tens{t}_{T}\cdot\delta \tens{g}_{T})d\gamma
\end{align}
where $\delta g_N$ and $\delta g_T$ represent variation in normal and tangential gaps, respectively. The quantity $\delta\Pi_C$ represents the virtual work due to contact traction acting between the two surfaces and presents a challenge for its calculation in the finite element framework.

\subsection{Contact constraints}
    The boundaries $\gamma^1, \gamma^2$ of the two interacting bodies $ \Omega_i, (i=1,2)  $ will have a normal gap between them, which is mathematically expressed in the traditional master-slave form as 
    \begin{align}\label{eq:normal_gap_x1x2cpp}
        g_N = ( {\tilde{\tens{x}}}_2 - \tilde{\tens{x}}_1 ) \cdot {\tens{n}_1}   
    \end{align}
    Here, $\tilde{\tens{x}}_1 $, $\tilde{\tens{x}}_2$ are points on the opposite surfaces having the smallest gap along the normal $\tens{{n_1}}$ at $\tilde{\tens{x}}_1 $. For any physical contact between two solid bodies, the kinematical constraint on their boundaries will not allow for any interpenetration ($g_N<0$) between the two surfaces, as compressive traction ($t_N=\tens{t}_N\cdot\tens{n}$) acting between the two bodies restricts them with $g_N=0$.  This condition, known as the Karush-Kuhn-Tucker (KKT) condition, is expressed as
     \begin{align}
         g_N \geq 0, \   {t}_N \leq 0, \ t_N g_N = 0  
     \end{align}
    The contact law is non-smooth and multivalued when $g_N=0$, as there are infinite possible values of $t_N$, which can be understood through the fact that the contact pressure is a reaction force generated to maintain the equilibrium between the two solids. So, techniques from optimisation theory, like the Lagrange multiplier and the penalty method, can be utilised to solve this problem involving inequality constraints. It should be noted that while the condition of impenetrability is, theoretically, exactly enforced by Lagrangian methods, it is inherently violated in the penalty methods, and the numerical solutions depend upon the choice of the penalty. Although an infinitely large penalty could theoretically enforce impenetrability, it is impractical in numerical approaches.

\section{Midplane based segment-to-segment contact discretisation}
    In this section, we propose the methodology for imposing contact constraints unbiasedly using the penalty method. Initially, the approach is presented in the continuum framework, followed by its description in the discretised setting. Here, the focus is on defining the contact interface between the two surfaces, where the contact traction and its integral are evaluated, as presented in the last subsection. The entire conceptual development follows the key idea of ensuring unbiased treatment of both surfaces.
    
    \subsection{Continuum based description}  
\begin{figure}[bhtp!]
        \centering
        \includegraphics[width=0.75\linewidth]{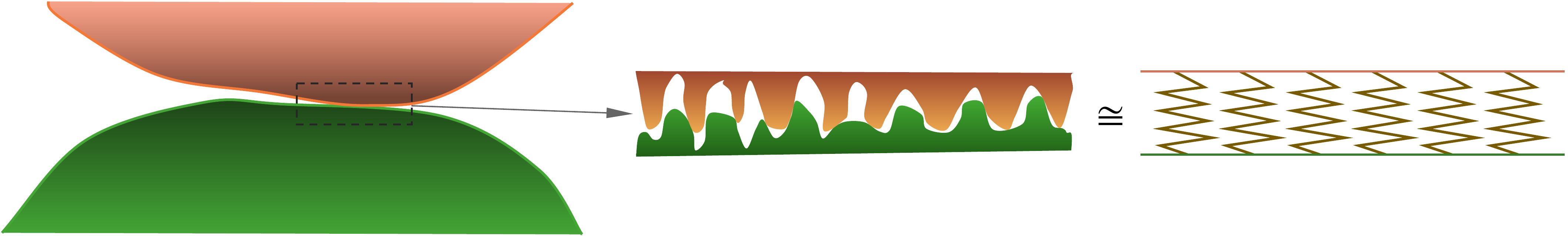}
        \caption{Physical analogy of penalty method with a bed of springs}
        \label{fig:bed_of_spring_analogysvg}
    \end{figure}

    For the two bodies $\Omega^1$ and $\Omega^2$ coming into contact over their boundaries $ \gamma^1$ and $\gamma^2 $, respectively, there will be a shared part of the two boundaries defining the contact interface $\gamma_c$($=\gamma^1_c=\gamma^2_c= \gamma^1 \bigcap \gamma^2$). With the ongoing motion and deformation of the two bodies, the sub-regions of the two boundaries coming into contact ($\gamma^1_{c}$ and $\gamma^2_{c}$) forming the contact interface ($\gamma_c$) will change continuously. To enforce the contact constraint between the interacting surfaces, this work employs penalty method-based restrictions to oppose interpenetration of the contacting surfaces. The interpenetration is penalised by opposing traction that acts to restrict or limit this overlap between the two domains. Physically, the penalty method can be seen as analogous to a bed of springs between the surfaces resembling the asperities \cite{zavariseRealContactMechanisms1992}, and the material stiffness of these asperities opposes their compression, Fig.~\ref{fig:bed_of_spring_analogysvg}, \cite{zavariseModifiedNodetosegmentAlgorithm2009}. 

\begin{figure}[thbp!]
        \centering
        \begin{subfigure}{0.79\linewidth}
            \centering
            \includegraphics[]{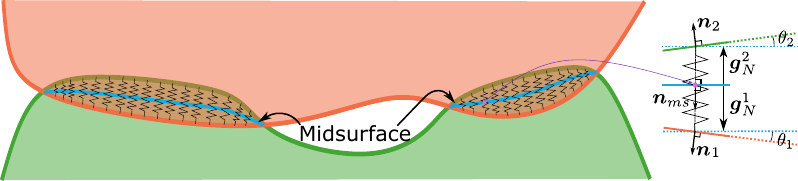}
        \end{subfigure}
        \begin{subfigure}{0.19\linewidth}
            \centering
            \includegraphics[]{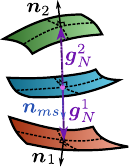}
        \end{subfigure}
        \caption{2D analogue of exaggerated geometrical interpenetration of two solids during contact and the gap between surfaces and corresponding 3D elements}
        \label{fig:penalty_interpenetration_spring}
    \end{figure}

    To evaluate contact tractions in an unbiased manner, the geometrical intersection of the two bodies can be interpreted to be occurring with respect to a common interface, referred to as midsurface, which follows the geometry of both surfaces $\gamma^1_c$ and $\gamma_c^2$, Fig.~\ref{fig:penalty_interpenetration_spring}. Here, the definition of the normal gap $g_N$ is slightly altered from the traditional form stated in eq.~\ref{eq:normal_gap_x1x2cpp} that utilises the closest point of projection, which often poses a challenge of its own. The normal gap $g_N$, considering the two end points of the analogous spring on both surfaces $\tilde{\tens{x}}_1$ and $\tilde{\tens{x}}_2$, is
    \begin{align}
        {g}_N = ( \tilde{\tens{x}}_2 - \tilde{\tens{x}}_1 ) \cdot \tens{n}_{\text{ms}}
    \end{align}    
    where $\tens{n}_{\text{ms}}$ is the normal to the midsurface and the gap $g_N$ is negative only in the case of interpenetration. 

    Using the penalty method, the gap of interpenetration is utilised to evaluate the normal traction $t_N$. This normal traction or contact pressure is proportional to both the normal gap $g_N$ and the stiffness of the analogous springs, which is contained in the penalty factor $\epsilon_N$ and written as -
    \begin{align}\label{eq:traction_definition}
        ||\tens{t}_N|| = -\epsilon_N g_N  H(-g_N)
    \end{align}
    where $H(\cdot)$ is the Heaviside function. The analogous springs, generated following the interpenetration, are normal to this midsurface which allows splitting the normal gap vector $\tens{g}_N$ in two parts $\tens{g_N}^1$ and $\tens{g_N}^2$ with each representing the interpenetration with respect to the midsurface point $\tens{x}_m$ as
    \begin{align}
    \tens{g}_N = \tens{g}_N^2 - \tens{g}_N^1 = (\tilde{\tens{x}}_2 - \tens{x}_m) - (\tilde{\tens{x}}_1 - \tens{x}_m) = \tilde{\tens{x}}_2 - \tilde{\tens{x}}_1
    \end{align}

    Using this description of the normal traction, the net normal force acting between the two bodies at the contact interface $\gamma_c$ can be evaluated using 
    \begin{equation}
        \tens{F}_N = \int_{\gamma_c} \tens{t}_N d\gamma = \int_{\gamma_c} -\epsilon_N \tens{g}_N H(-g_N) d\gamma
    \end{equation}

\subsection{Description in 3D discretised domains}
    Proceeding towards the discretised setting of the contact problem, consider the domains $\Omega_h^1$ ($\simeq \Omega^1$)and $\Omega_h^2$ ($\simeq\Omega^2$) with their discrete surfaces $\gamma^i_h$ described using parameterisation of the form 
    \begin{equation}
        \tilde{\tens{x}}^i = \sum_{j=1}^{m^i} \psi_j^i(\xi_1,\xi_2) \tilde{\dmat{x}}_j^i     \ \  \text{where}, \ \tilde{\tens{x}}^i, \tilde{\dmat{x}}_j^i \in \gamma^i_h
    \end{equation}
    where $\psi_j^i$ represents the shape function associated with the $j^{th}$ node (having position $\tilde{\tens{x}}_j^i $) in the $i^{th}$ discretised boundary with the parameterisation of $(\xi_1,\xi_2)$. The symbol $(\tilde{\cdot})$ has been used to denote that these points are over the boundaries of the respective domains. Any physical interpenetration between $\Omega_h^1$ and $\Omega_h^2$ results in the geometrical overlap between surfaces $\tilde{\tens{x}}^1$ and $\tilde{\tens{x}}^2$. This geometrical overlap for contact interaction is considered in a piecewise-pairwise manner by taking discrete facets (or segments) of finite elements on opposite boundaries. So, only the support of the basis functions over the facets is considered here for their pairwise interaction.

    For each potentially contacting facet pair, a local midplane is defined in this discretised setting for the system in its current solution step, and the interpenetration of both facets with respect to this midplane is utilised in evaluating the contact traction. The construction of the midplane and the application of this contact traction is described in the following sections.
    
    \subsubsection{Construction of midplane per facet pair}    
    Consider two interpenetrating facets $A$ and $B$ on the opposite boundaries with the region $\gamma^1_{h_A}$ and  $\gamma^2_{h_B}$, respectively, the contact interface in a current solution step is chosen as a region over the midplane between the two facets and is written as $\gamma_{h_{c_{A,B}}}$. With this, the overall contact interface for boundaries $\gamma_h^1$ and $\gamma_h^2$ becomes  $   \underset{\substack{A,B}}{\bigcup}  \gamma_{h_{c_{A,B}}} $.    

\begin{figure}[htbp!]
        \begin{subfigure}[b]{0.61\linewidth}
            \centering
            \includegraphics[width=1.0\linewidth]{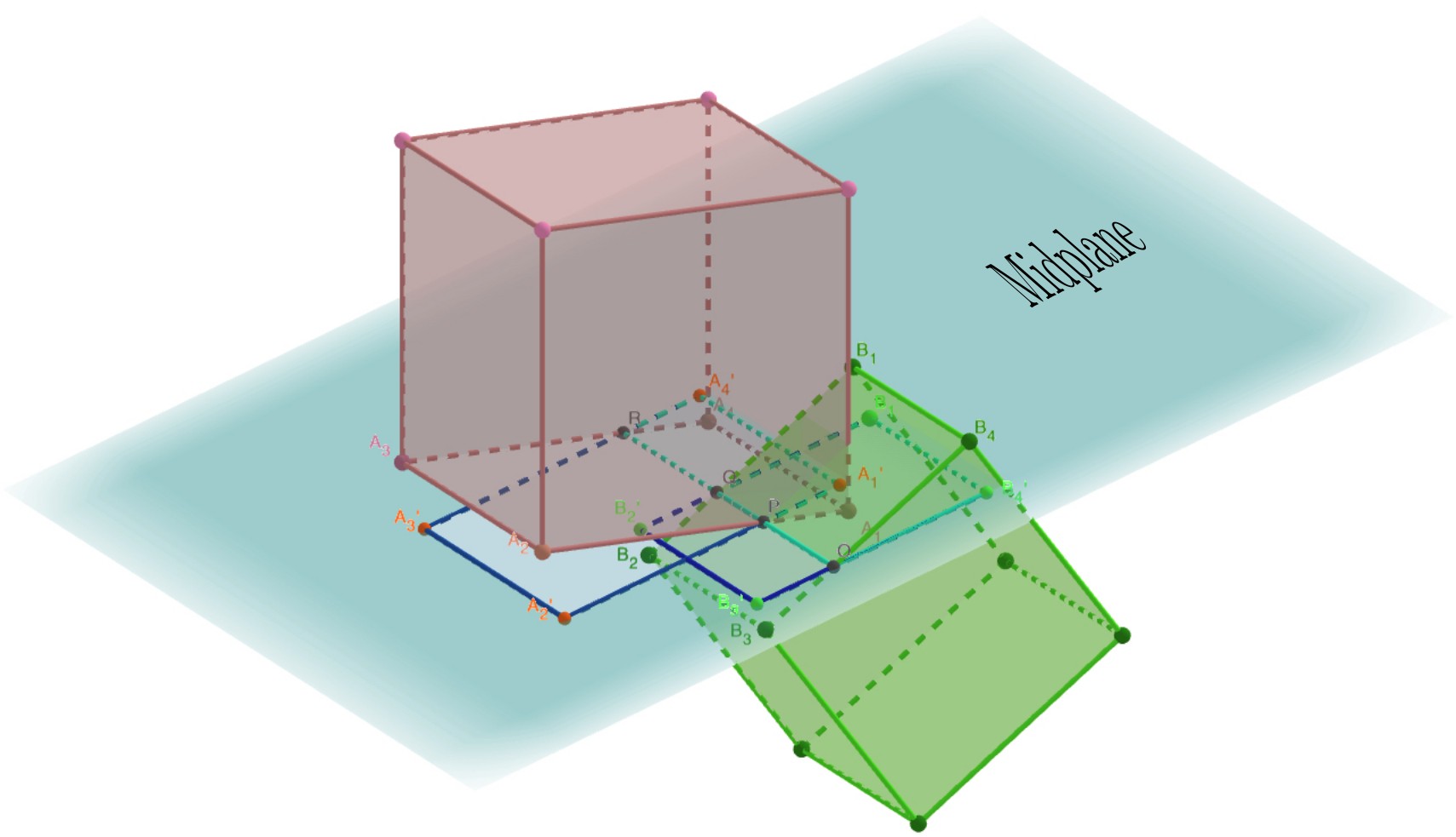}
            \caption{}

            \label{fig:twocubes_interpenetration}
        \end{subfigure}
        \hfil
        \begin{subfigure}[b]{0.38\linewidth}
            \centering
            \includegraphics[width=0.82\linewidth]{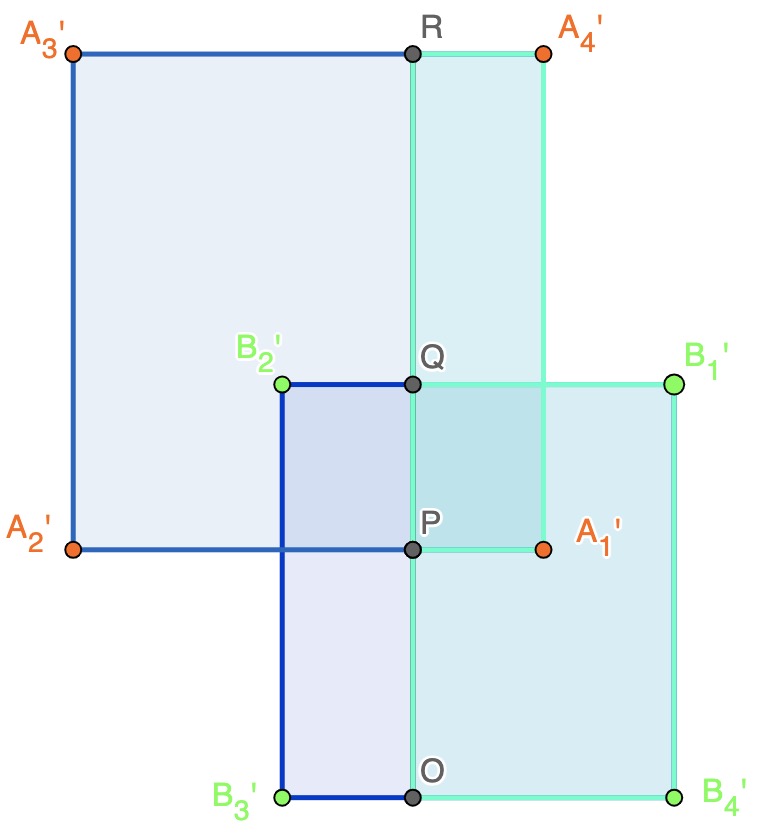}
            \caption{}

            \label{fig:interpenetrating_cubes_projections}
        \end{subfigure}
        \caption{Simplified illustration of interpenetration between contacting elements shown for a pair of facets (a), and the overlap of the projections (b). Notice that only a sub-region of the intersection polygon corresponds to interpenetration between facets.}
        \label{fig:two_cubes_midplane_two_figs}
    \end{figure}

    For the sake of initial explanation, a simplified illustration of this midplane concept is shown in Fig.~\ref{fig:two_cubes_midplane_two_figs} for two orthogonal hexahedral elements having interpenetration between their facets labelled as A and B. The midplane equi-inclined to both facets is constructed to define the contact interface and evaluate the contact traction for this pair of facets. The contact interface is defined by taking the intersection of the polygons of projections of both facets over the midplane. However, as shown in Fig.~\ref{fig:interpenetrating_cubes_projections}, only a sub-region of the intersection corresponds to the interpenetration. This sub-region is the intersection of the projections of penetrating portions of both facets.

    However, facet nodes are not necessarily co-planar in the general finite element representation of facets based on interpolation, so representative facet planes are constructed for two contacting facets to define their midplane. The orientation of the representative plane for each facet is calculated by averaging the cross-products of edge vectors at all corners of the facet. So, the unit normal perpendicular to this plane can be written as
    \begin{align}\label{eq:normal_rp}
        \tens{n}_{\text{rp}} = \frac{ \frac{1}{4} \sum_{i=1}^{4} \dmat{n}_i}{\left\Vert \left( \frac{1}{4} \sum_{i=1}^{4} \dmat{n}_i \right) \right\Vert } 
        \text{\quad where } \dmat{n}_i = \dmat{l}_i \times \dmat{l}_{i-1}
    \end{align}
    with $\dmat{l}_{i}$ and $\dmat{l}_{i-1}$ denoting the edge vectors at the corner $i$, both being directed away from the corner node. Both $\dmat{l}_{i}$ and $\dmat{l}_{i-1}$ are ordered in a such way that each $\dmat{n}_i$ gets directed outside the element. Each representative facet plane is considered to be passing through the mean position of the nodes ($x_0$), so its equation takes the form
    \begin{align}
        \tens{n}_{\text{rp}}\cdot\tens{x} - d = 0,  \ \ \text{where } d = \tens{n}_{\text{rp}}\cdot x_0
    \end{align}
    
\begin{figure}[htb!]
        \centering
        \includegraphics[width=0.65\linewidth]{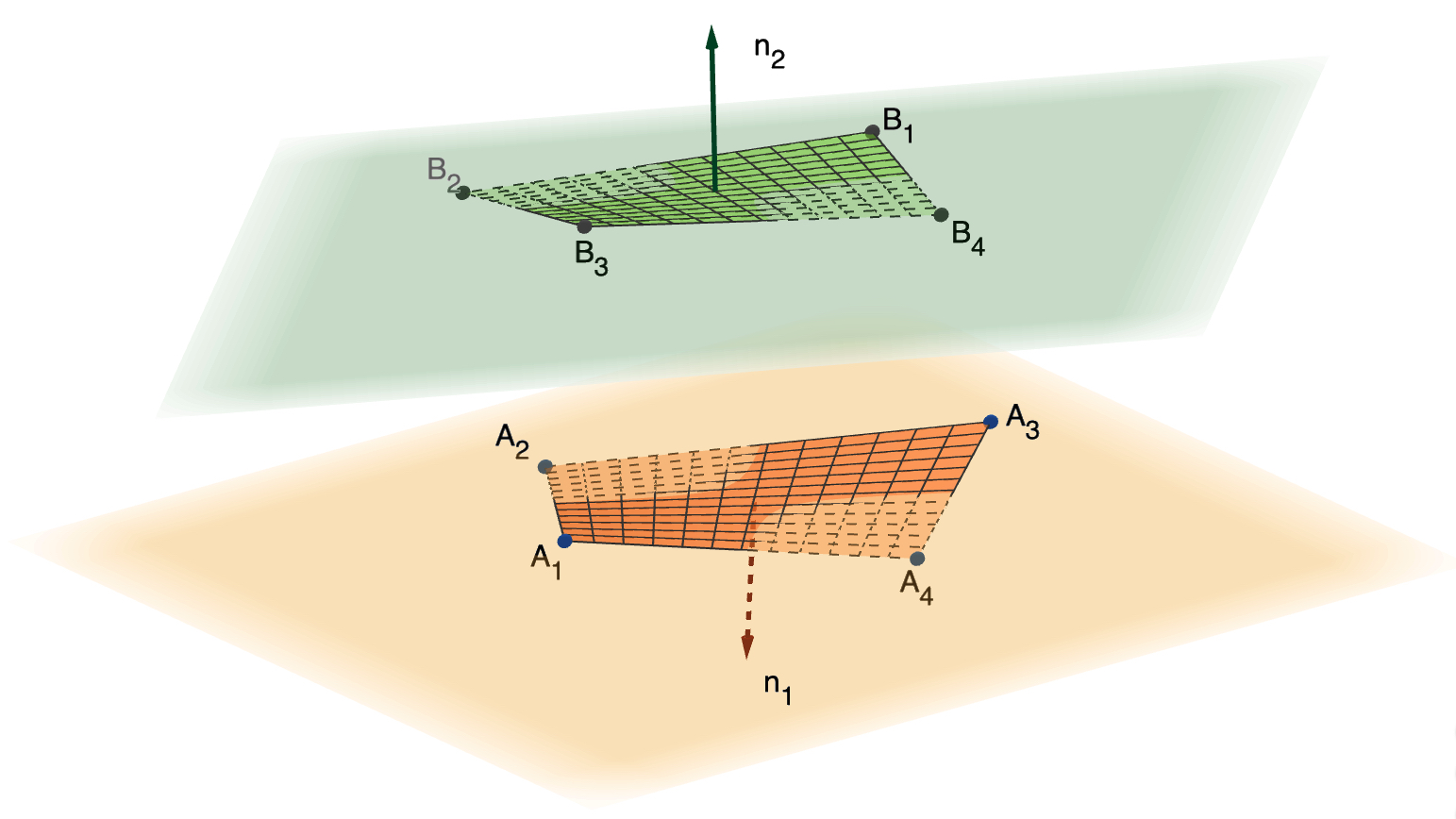}
        \caption{Two contacting physical facets with their representative planes. Also shown are the outward normal directions for both planes. \\ }
        \label{fig:two_facets_bilinear_interpenetration}
    \end{figure}

    In this definition of the representative plane, the distances of all four nodes from the representative plane are equal. This definition also turns out to be the same as the concept of using normal to the plane formed by parametric lines joining the mid-points on the opposite edges \cite{papadopoulosSimpleAlgorithmThreedimensional1993}. An example of the two interpenetrating physical facets with their respective representative planes is shown in Fig.~\ref{fig:two_facets_bilinear_interpenetration}.
    
    The midplane for the facet pair is defined to be equi-inclined to the two representative planes, so the unit normal to the midplane $\tens{n}_{\text{mp}}$ is written as
    \begin{align}
        \tens{n}_{\text{mp}} = \frac{ \left( {\tens{n}_1} - {\tens{n}_2} \right) }  { \left\Vert \left( {\tens{n}_1} - {\tens{n}_2} \right) \right\Vert }        
    \end{align}
    where $\tens{n_1},{\tens{n_2}}$ are unit normals perpendicular to the representative facet planes for the first and second facet, respectively, as described in eq.~\ref{eq:normal_rp}. The midplane is taken to pass through the intersection of the two facet planes if they are non-parallel or through a point equidistant to both representative planes in case they are parallel. However, as will be shown later, only the direction of the midplane affects the solution and not its position. It is evident that for well-formed surface facets undergoing interpenetration during solution, their continuously changing spatial configurations will result in smooth changes in the midplane direction.

\subsubsection{Interpenetration zone between facets}
    The interpenetration between the per-facet pairs is evaluated using the midplane defined in the previous section and its orientation. This interpenetration is studied through two choices, and only the second one is selected as it yields true interpenetration.
    
\begin{figure}[htb]
        \centering
        
        \begin{subfigure}[b]{0.47\linewidth}
            \centering
            \includegraphics[width=1.0\linewidth]{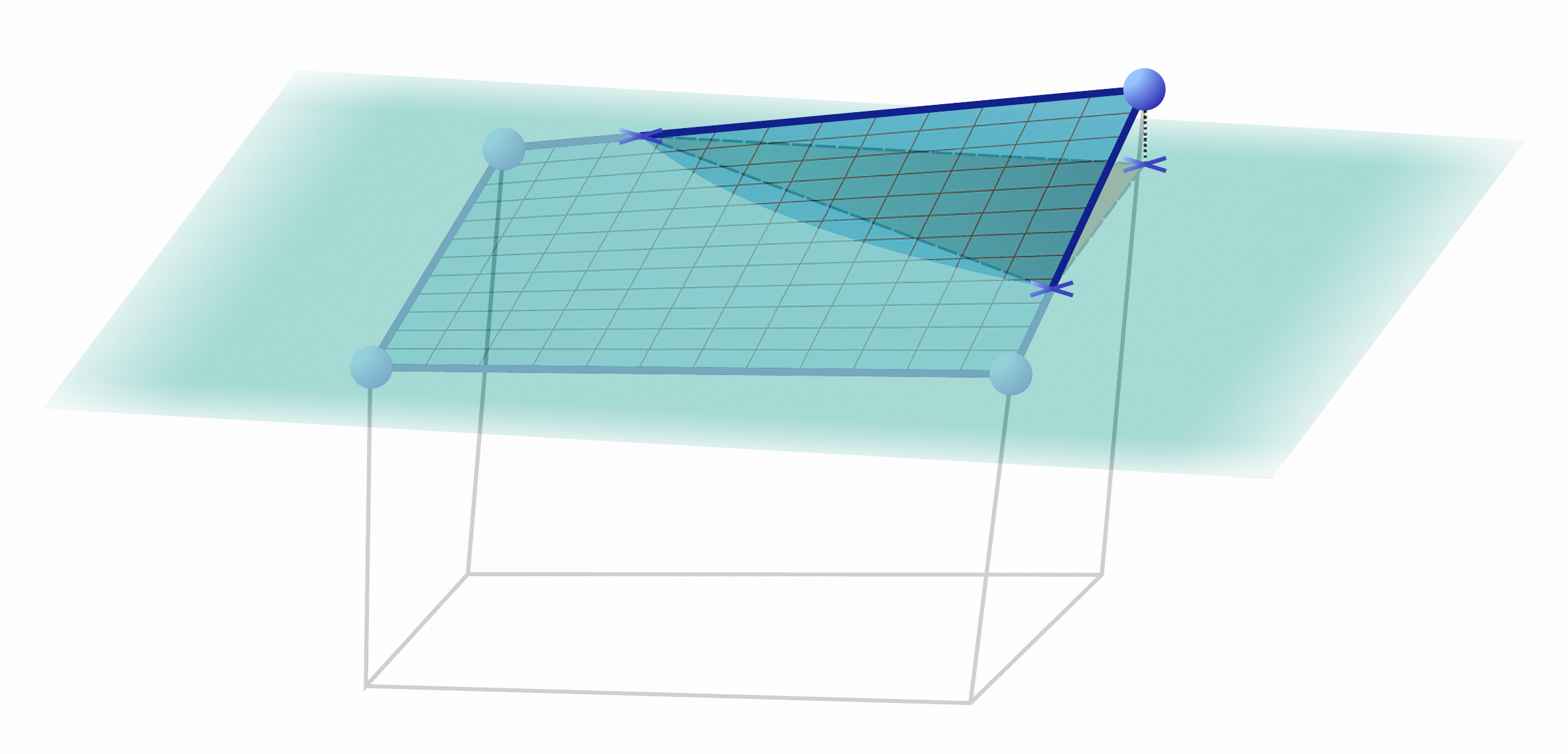}
            \caption{One node penetrating the midplane}
        \end{subfigure}
        \hfil
        \begin{subfigure}[b]{0.47\linewidth}
            \centering
            \includegraphics[width=1.0\linewidth]{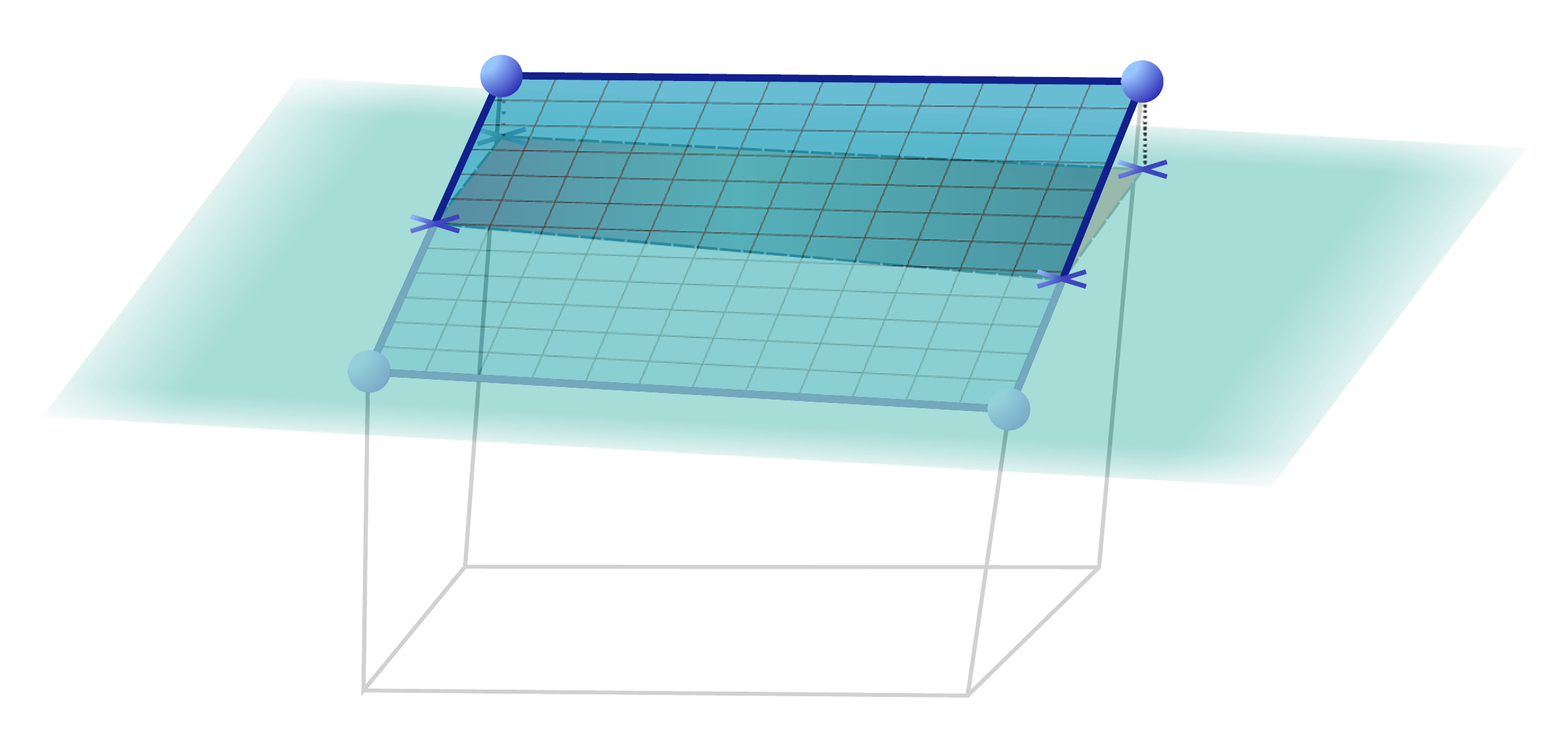}
            \caption{Two adjacent nodes penetrating the midplane}
        \end{subfigure}
    
        \begin{subfigure}[b]{0.47\linewidth}
            \centering
            \includegraphics[width=1.0\linewidth]{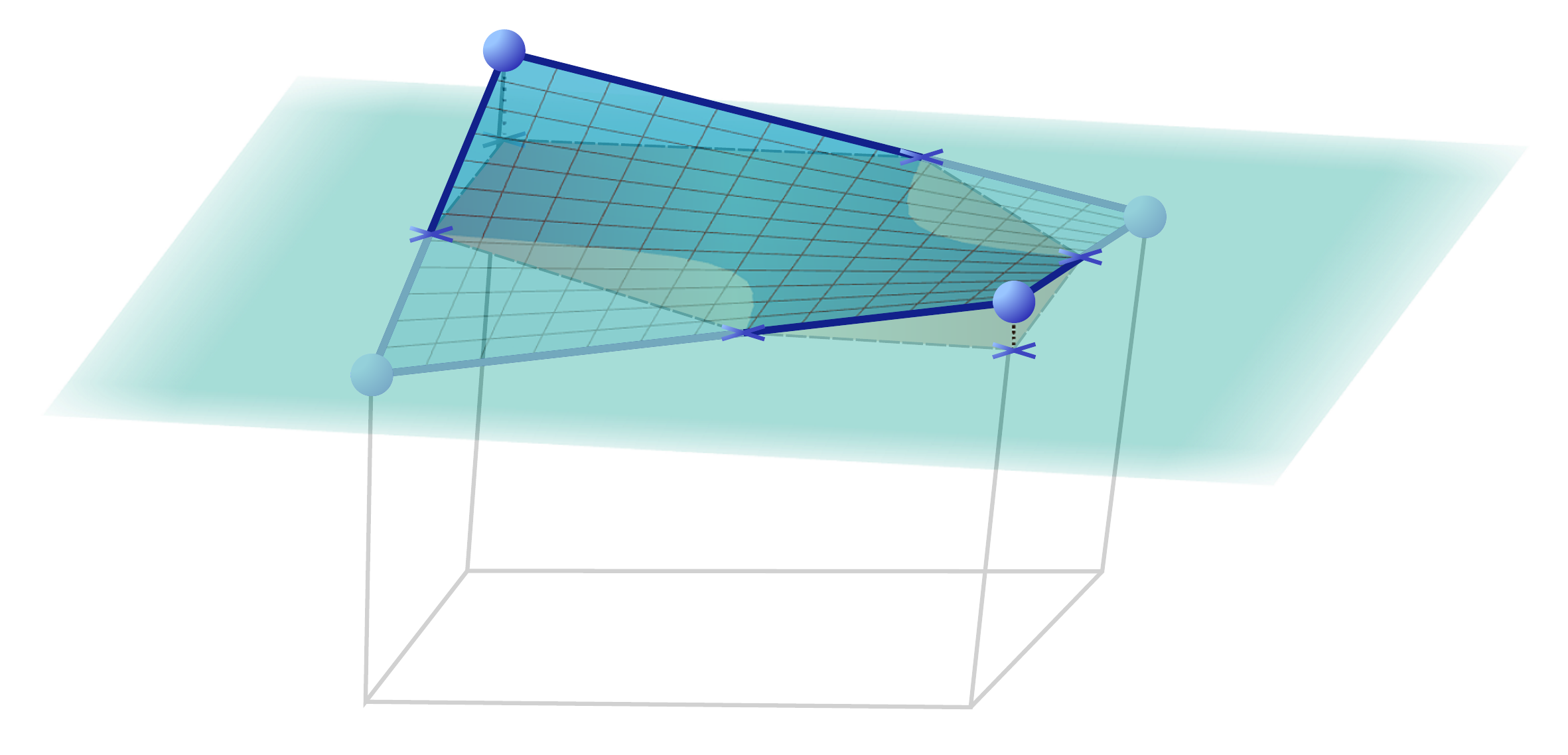}
            \caption{Opposite nodes penetrating the midplane}
        \end{subfigure}
        \hfil
        \begin{subfigure}[b]{0.47\linewidth}
            \centering
            \includegraphics[width=1.0\linewidth]{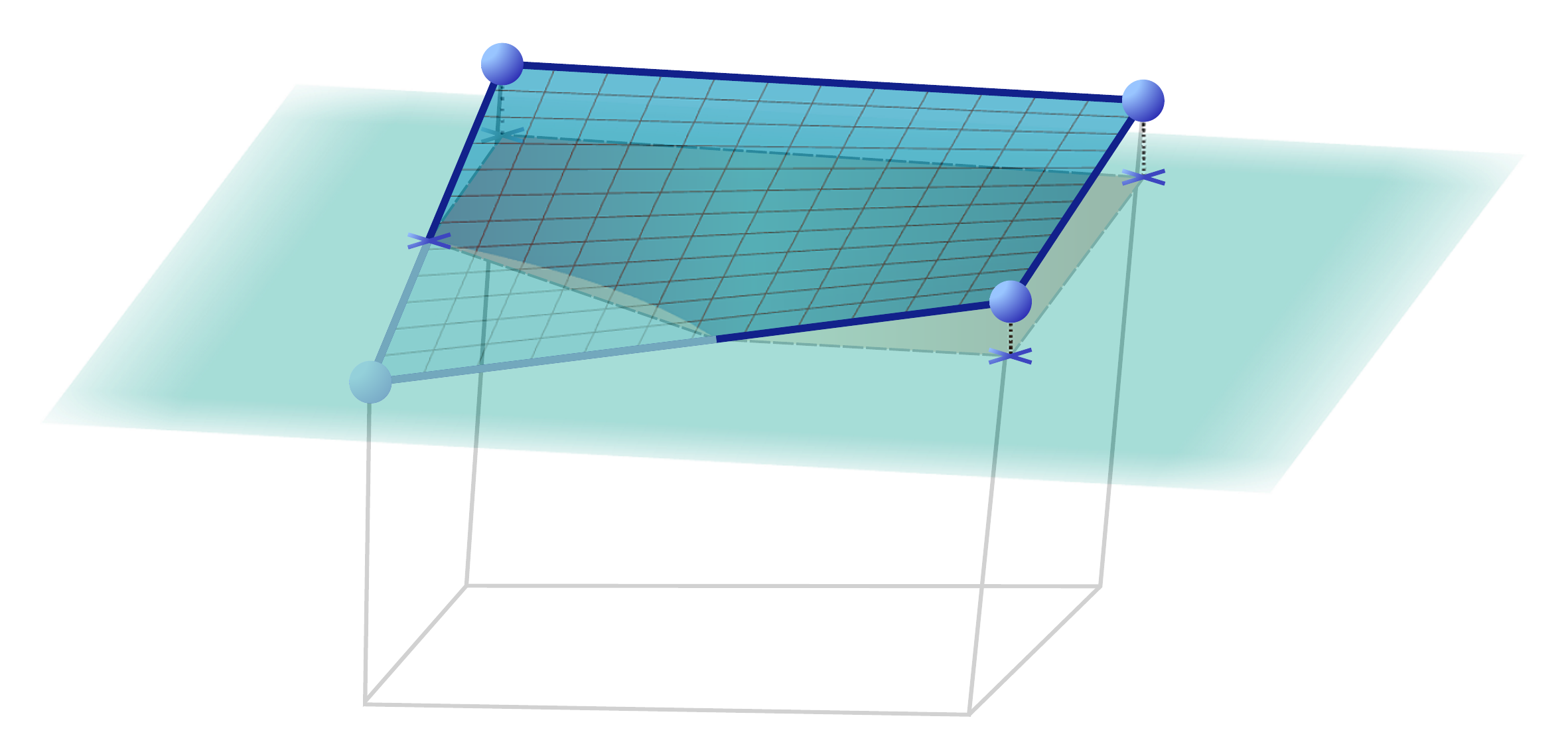}
            \caption{Three nodes penetrating the midplane}
        \end{subfigure}
    
        \begin{subfigure}[b]{0.47\linewidth}
            \centering
            \includegraphics[width=1.0\linewidth]{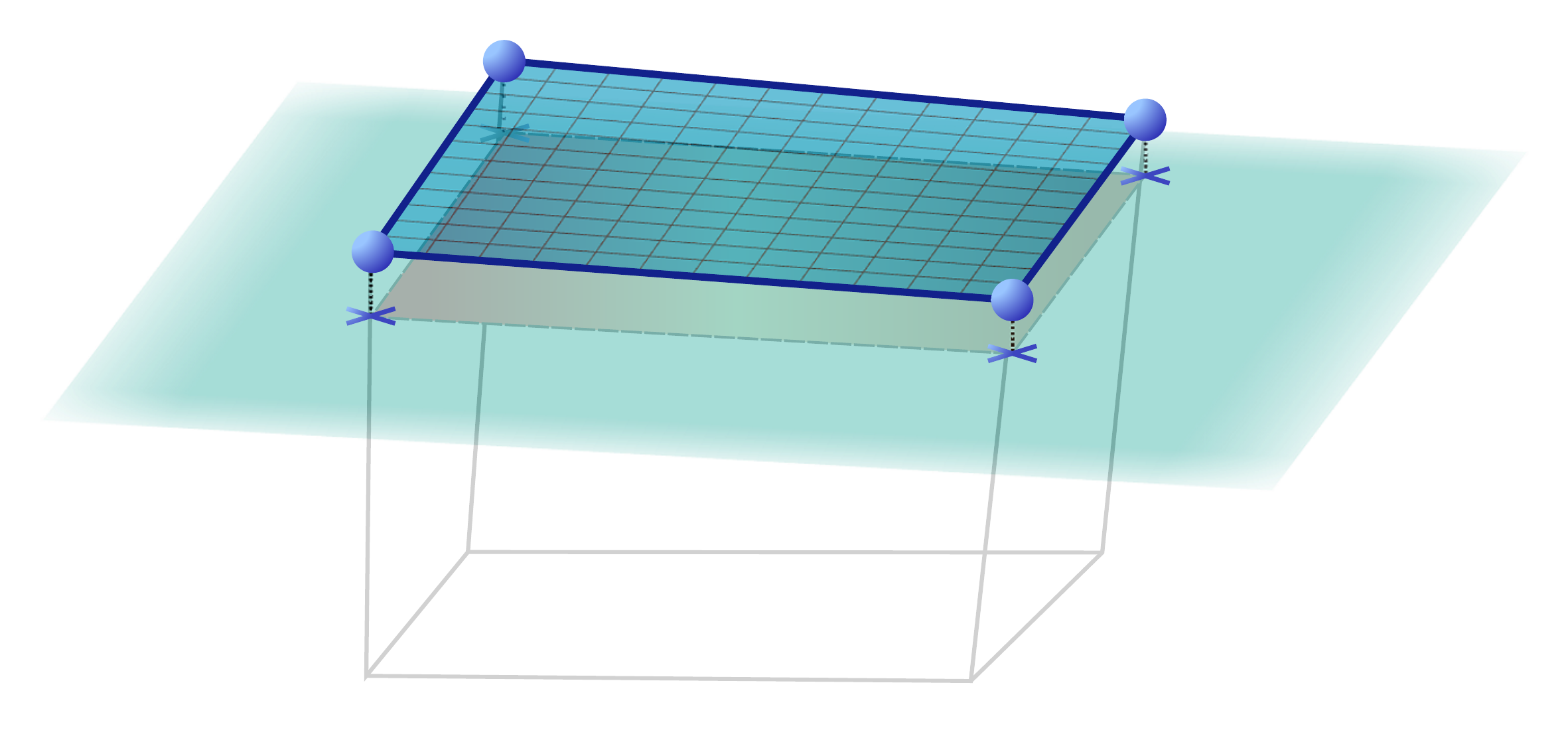}
            \caption{All four nodes penetrating the midplane}
        \end{subfigure}
    
        \caption{Different possible configurations of a facet penetrating the midplane. Also shown are the projection polygons of penetrating portions}
        \label{fig:facet_penetration_possible_configurations}
    \end{figure}

   In the first possibility, if only the penetration through the midplane is considered for defining the interpenetration between facets, then facets of hexahedral elements having bilinear interpolation can have five different possible configurations of penetration through the midplane as shown in Fig.~\ref{fig:facet_penetration_possible_configurations}. The approximated projections of the penetrating portion for defining the contact region are also shown. The construction of these projection polygons progresses as follows: first, the relative positions of all nodes of the facets are evaluated with respect to the midplane. Any node penetrating the midplane will be normally projected onto it. Next, if the nodes at both ends of any edge are on opposite sides of the midplane, then the intersection of this edge with the midplane is evaluated. Finally, a convex hull constructed over all these points defines the approximated projection polygon.

    Now, to define the region on the midplane over which contact tractions will be evaluated, two possible regions of interest exist. One option is to consider the intersection between the projection polygons of midplane penetrating portions of both facets, as described in Fig.~\ref{fig:facet_penetration_possible_configurations}. The alternative is to consider the intersection between the quad polygons, formed by full projections of both facets over the midplane.
    
\begin{figure}[bthp]
        \centering
        \includegraphics[width=0.40\linewidth]{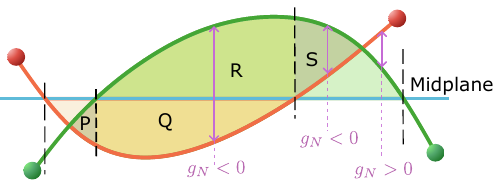}
        \caption{Cross-sectional view of a possible interpenetration between curved facets. Here, P+Q+R+S corresponds to actual interpenetration between facets, whereas only Q+R is accounted for if the projections of only the midplane penetrating portions of both facets are considered.}
        \label{fig:simplified_2D_interpenetration_volume}
    \end{figure}

    While projections of penetrating portions will capture the interpenetration between facets for many cases, there is a possibility that the curvature of the opposite facets results in some interpenetration outside the midplane. One such differentiating scenario between the two choices is shown through a cross-sectional view of the interpenetration between 3-dimensional facets in Fig.~\ref{fig:simplified_2D_interpenetration_volume}. If two different projections are considered - full projection of the facets and projection of only the penetrating portion of the facets, then both will lead to different contact interface definitions. While the full projection will capture the entire interpenetration between the two facets (P+Q+R+S), the projection of only the portions penetrating the midplane will only capture the region Q+R. Although the interpenetration in regions P and S does not involve the midplane, the full projection of both facets has been chosen in this work for the evaluation of contact tractions. This will also allow us to capture the overlapping region outside the midplane, thereby corresponding to the \textit{true }interpenetration. The projection of the interpenetrating volume over the midplane defines the contact region $\gamma_{h_{c_{A,B}}}$ for this facet pair. Note that irrespective of the choice of projections of full facets or only their midplane penetrating portions, the overall contact interface $\underset{\substack{A,B}}{\bigcup}  \gamma_{h_{c_{A,B}}}$ between full surfaces $\gamma_h^1$ and $\gamma_h^2$ will not be necessarily continuous.

    So, to establish local contact interface for the facets $\gamma^1_{h_A}$ and $\gamma^2_{h_B}$ having a midplane $\gamma_{\text{mp}}^{A,B}$, the projections of both facets over the midplane can be written as
    \begin{align}
        \gamma^{1,\text{mp}}_{h_A} = \text{Proj}_{\gamma_\text{mp}} \left( \gamma^1_{h_A} \right) \ , \ \ \text{where } \gamma^1_{h_A}= \sum_{j=1}^{w^A} \psi_j^A\dmat{x}_j^A \\
        \gamma^{2,\text{mp}}_{h_A} = \text{Proj}_{\gamma_\text{mp}} \left( \gamma^2_{h_A} \right) \ , \ \ \text{where } \gamma^2_{h_A}= \sum_{j=1}^{w^B} \psi_j^B\dmat{x}_j^B
    \end{align}
    where $w^A$ and $w^B$ are the number of nodes on both facets. These projections can be directly evaluated by projecting facet nodes over the midplane and using the same bilinear interpolation functions. Thus, the contact interface, local to this pair of facets $A$ and $B$, can be written as 
    \begin{align}
        \gamma_{h_{c_{A,B}}} = \gamma^{1,\text{mp}}_{h_A} \bigcap \gamma^{1,\text{mp}}_{h_B}
    \end{align}
    and the area corresponding to the true interpenetration will be its subset (i.e. $ \subseteq\gamma_{h_{c_{A,B}}}$).
    
\subsubsection{Contact traction}
The interpenetration-based traction evaluation per facet pair in the discretised setting, following the original definition presented in eq.~\ref{eq:traction_definition}, utilises the gap between facets measured normal to any point on the midplane. The split gap vector of either facet $i$ $\left(i\in\{1,2\}\right)$ with respect to any point $\tens{x}_m$ on the midplane $\gamma_{\text{mp}}$ are written as

\begin{align}\label{eq:split_gap_vector_ith_facet}
    \tens{g_N}^i(\tens{x}_m) =  \sum_{j=1}^{w^i} \psi_j^i\bigg|_{\left(\xi^i_1,\xi^i_2\right)} \dmat{x}_j^i  - \tens{x}_m \,\text{, where}, \left(\xi_1^i,\xi_2^i\right): \text{Proj}_{\gamma_{\text{mp}}} \left( \sum_{j=1}^{w^i} \psi_j^i\bigg|_{\left(\xi^i_1,\xi^i_2\right)}  \dmat{x}_j^i \right) = \tens{x}_m ,  \tens{x}_m\in\gamma_{\text{mp}}
\end{align}
where $w^i$ is the number of nodes. Here, the parametric coordinates $(\xi_1^i,\xi_2^i)$ on the $i^{th}$ physical surface correspond to the physical point lying on the normal to the midplane. With this knowledge, the gap vector between facets with respect to $\tens{x}_m$ becomes 
\begin{align}
        \tens{g}_N = \tens{g}_N^2 - \tens{g}_N ^1 = \sum_{j=1}^{w^2} \psi_j^2(\xi_1^2,\xi_2^2)  \dmat{x}_j^2   - \sum_{j=1}^{w^1} \psi_j^1(\xi_1^1,\xi_2^1) \dmat{x}_j^1  
    \end{align}
The gap vector decides whether the facets are interpenetrating at this point (i.e. $g_N=\tens{g}_N\cdot\tens{n}_{\text{mp}}<0$). Also, note that it satisfies $\tens{g}_N^i \times \tens{n}_{\text{mp}} = 0,  \forall \ i\in\{1,2\}$. Only the normal direction of the midplane is relevant in calculating the gap, and all parallel planes will yield the same gap. This definition of normal gap captures the true continuous interpenetration between the surfaces, unlike the definition based on interpolation of the discrete normal gaps at nodes \cite{pusoSegmenttosegmentMortarContact2008}.

With this definition of the gap vector, the virtual work of contact defined in eq.~\ref{eq:contact_virtual_work}, knowing that the variation in surface displacements and positions are the same (i.e., $\delta\tens{u}=\delta\tens{x}$), can now be written as
\begin{align}    
        \delta \Pi_c \approxeq \delta \Pi_{c_h} = \sum_{\gamma_h} \int_{\gamma_{h_c}}\tens{t}_{c}^{1}\cdot(\delta\tens{u}^{1}-\delta\tens{u}^{2})d\gamma   
        = \sum_{\gamma_h} \int_{\gamma_{h_c}}\tens{t}_{N}^{1}\cdot(\delta\tens{x}^{1}-\delta\tens{x}^{2})d\gamma \\
        = \sum_{\gamma_h} \int_{\gamma_{h_c}}\tens{t}_{N}^{1}\cdot \left(\sum_{j=1}^{w^1} \psi_j^1 \delta\tens{x}_j^1 - \sum_{j=1}^{w^2} \psi_j^2 \delta\tens{x}_j^2 \right)d\gamma = \sum_{\gamma_h} \int_{\gamma_{h_c}} \tens{t}_N^1 \cdot \delta \tens{g}_N d\gamma     
\end{align}
  So, the normal normal traction at any point on the midplane $\gamma_{\text{mp}}$ can be written as 
  \begin{align}
      \tens{t}_N = -\epsilon_N \tens{g}_N H(-(\tens{g}_N\cdot\tens{n}_{\text{mp}}))d\gamma
  \end{align}
and it is non-zero only at the points corresponding to true interpenetration between surfaces, i.e. at $\gamma_{h_{{c}_{A,B}}}$. This traction is applied equal and opposite on the two interacting facets along the direction of the midplane normal, thereby being compressive in nature, i.e.,
    \begin{align}
        \tens{t}_N^1 = -\tens{t}_N, \quad \tens{t}_N^2 = \tens{t}_N
    \end{align}

\subsection{Contact integral}    
    Contact tractions, evaluated using per facet pair interpenetration described in the previous section, provide the necessary contact constraint on the entire surfaces $\gamma_{h}^1$ and $\gamma_h^2$. These tractions are, however, defined pairwise between facets and act normal to their midplanes. For the finite element approach, these tractions are converted into effective nodal forces and the net nodal forces due to contributions by all pairs constraints the contact between bodies $\Omega_h^1$ and $\Omega_h^2$. 

    Mathematically, the distribution of nodal forces due to any traction $\tens{t}$ on a facet $\gamma_{h_e}$ is given by 
    \begin{align}
        \dmat{R} = \int_{\gamma_{h_e}} \dmat{\Psi} \tens{t} d\gamma 
    \end{align}
    where $\dmat{\Psi}$ contains the shape functions corresponding to all nodes on the facet $\gamma_{h_e}$, and $\dmat{R}$ contains forces that will act upon these nodes \cite{batheKJBatheFinite2014}. Individually, the $i^{th}$ nodal force $\left(i\in \{1,2,..,w^e\}, w^e= \text{total number of facet nodes} \right)$ on facet $A$ can be written using its associated shape function $\psi_{i_A}$ as 
    \begin{align}\label{eq:nodal_forces_integral}
    \dmat{R}_{i_A} = \int_{\gamma_{h_e}} \psi_{i_A} \tens{t}  d\gamma
    \end{align}
    Using the above eq.~\ref{eq:nodal_forces_integral}, the contribution of this interaction's normal traction $\tens{t}_N\left(= \epsilon_N g_NH(-g_N) \tens{n}_{\text{mp}}\right)$ towards the nodal forces on facet $\gamma_{h_A}$ is given by
    \begin{align}
        \dmat{R}_{i_A}^N = \int_{\gamma_{h_A}} \psi_{i_A} \tens{t}_N  d\gamma = \int_{\gamma_{h_{\text{mp}}}} \psi_{i_A} \epsilon_N g_N H(-g_N) \tens{n}_{\text{mp}}  d\gamma
    \end{align}
    where $g_N$ is the interpenetration gap between the facets and $\gamma_{h_{c_{A,B}}}$ is the polygon of the intersection of projections of both facets over the midplane. Similarly, nodal forces $\dmat{R}_{i_B}^N$ on the opposite facet $\gamma_{h_B}$ can be written using its shape functions $\psi_{i_B}$ and will act in the opposite direction, so  $\dmat{R}^N_{i_A}$ and $\dmat{R}^N_{i_B}$ will be together compressive in nature for the contact interface. In this work, the minimum of the bulk moduli of the two contacting surfaces is chosen as the penalty factor, which is further amplified by a scaling factor $f_s$. The choice of a single penalty parameter for all contact pairs provides a consistency in traction calculation throughout the interacting discretised surfaces \footnote{The traditional definition of the penalty parameter involves the contact area between the STS pair, the volumes of the contacting elements, and the bulk modulus, defined as $\epsilon = \frac{f_s A^2 K}{V}$ \cite{belytschkoContactimpactPinballAlgorithm1991, batisticFiniteVolumePenalty2022}, which exhibits dependency on element sizes. In this work, the bulk modulus is used directly as a reference for penalty stiffness. Although this choice does not preserve unit consistency, it serves as a material-dependent, heuristic baseline and is treated as a tunable parameter chosen for numerical convenience. While only compressible materials have been studied in this work, elastic modulus as penalty can be studied for incompressible materials.}.

\begin{figure}[htbp!]
        \centering
        \begin{subfigure}[b]{0.41\linewidth}
            \centering
            \includegraphics[width=1.0\linewidth]{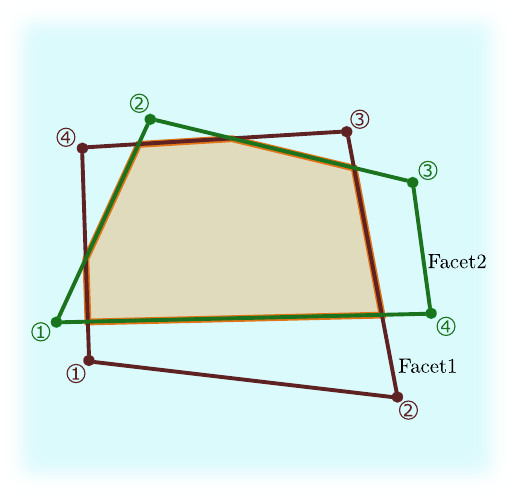}
            \caption{Projections of the two quad facets onto the midplane}
            \label{fig:qdFct1qdFct2FuullProjIntersection}
        \end{subfigure}
        \hfill
        \begin{subfigure}[b]{0.41\linewidth}
            \centering
            \includegraphics[width=1.0\linewidth]{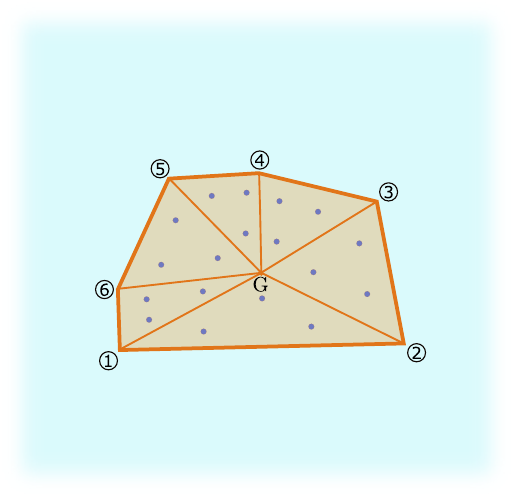}
            \caption{Triangulation of the overlapping region and the quadrature points inside it}
            \label{fig:intersec_polygn_triangulation_integration_points}
        \end{subfigure}    
        \hfill
        \vfill
        
        \begin{subfigure}[b]{0.41\linewidth}
            \centering
            \includegraphics[width=1.0\linewidth]{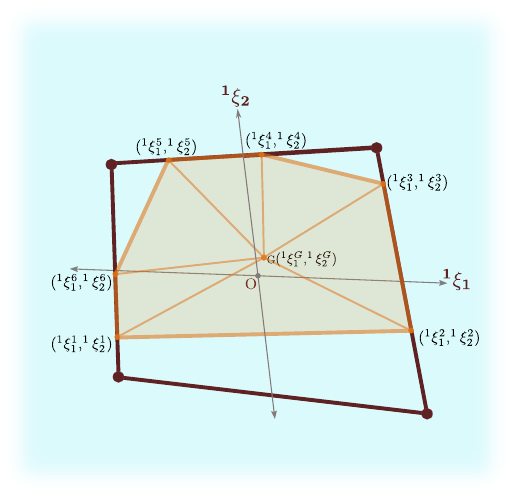}
            \caption{Parametric coordinates of points in the overlapping region with respect to facet1}
            \label{fig:facet1_intersecpolyg_parametric_verts}
        \end{subfigure}
        \hfill
        \begin{subfigure}[b]{0.41\linewidth}
            \centering
            \includegraphics[width=1.0\linewidth]{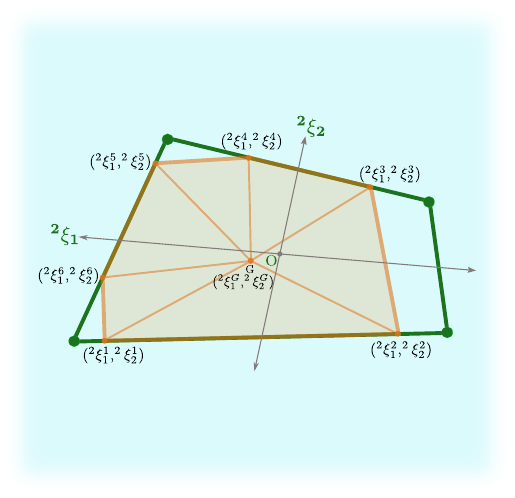}
            \caption{Parametric coordinates of points in the overlapping region with respect to facet2}
            \label{fig:facet2_intersecpolyg_parametric_verts}\textbf{}
        \end{subfigure}
        
        \vfill
        \centering
        \begin{subfigure}[b]{0.48\linewidth}
            \centering

            \includegraphics[width=1.0\linewidth]{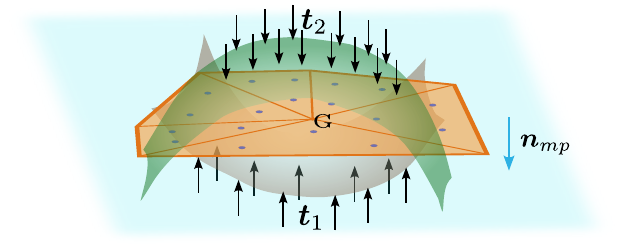}
            \caption{Tilted view of the intersection polygon on the midplane in 3D with the interpenetration of opposite surfaces}            
            \label{fig:tilted_intersec_polygn_triangulation_integration_points}
        \end{subfigure}    
        
    \caption{Illustration of projections of both facets on the midplane, triangulation of the overlapping region, and inverse bilinear mapping in both projected quadrilaterals for parametric coordinates of quadrature points.}
    \label{fig:qdFct1qdFct2Projection_triangulation}
    \end{figure}

     The intersection of the polygons of full facet projections over the midplane $\gamma_{h_{\text{mp}}}^I\left(\in\gamma_{h_{\text{mp}}}\right)$ represents the common span of both facets (with possible interpenetration) where contact traction is evaluated. The contact traction is zero elsewhere on the midplane as this region is outside the possible interpenetration volume between the bounded surfaces of facets. To evaluate the integrals numerically, the intersection polygon is subdivided into triangles ($\gamma_{h_{\text{mp}}}^I= \Delta_1 + \Delta_2 + ... + \Delta_k$) formed by its edges and centroid, as shown in Fig.~\ref{fig:qdFct1qdFct2Projection_triangulation}, allowing to write the nodal forces as 
     \begin{align}\label{eq:integral_triangulation_subdivision}
        \dmat{R}_{i_A}^N = f_s \epsilon_N \tens{n}_{\text{mp}}\left[ \int_{\Delta_1} \psi_i g_N H(-g_N) d\gamma  + \int_{\Delta_2} \psi_i g_N  H(-g_N) d\gamma  + ... + \int_{\Delta_k} \psi_i g_N  H(-g_N) d\gamma    \right]
     \end{align}
     
     The intersection polygon is decomposed into triangles using the centre-based triangulation over the Delaunay triangulation as the Delaunay triangulation is not a unique procedure \cite{farahSegmentbasedVsElementbased2015}. With this subdivision, the integrals are evaluated using the quadrature points over each triangle using the numerical integration technique described in \cite{cowperGaussianQuadratureFormulas1973},\cite{dunavantHighDegreeEfficient1985}. One of the possible integration schemes from different numbers of quadrature points over triangles can be chosen depending on the desired level of accuracy.
     
    The interpenetration gap $g_N$ at any quadrature point on the midplane is determined by employing both split gap vectors $\tens{g}_N^1$ and $\tens{g}_N^2$ as described in the eq.~\ref{eq:split_gap_vector_ith_facet}. This requires finding the pair of parametric coordinates for each facet where a line normal to the quadrature point intersects the facet. For a first-order interpolation, the parametric coordinates of any point on the physical facet with respect to the physical facet and the parametric coordinates of this point's projection on the midplane with respect to the projected facet are the same. So, while the positions of the quadrature points on each triangle are evaluated using the physical coordinates of the vertices of the intersection polygon, their relative locations with respect to each projected quadrilateral are written using the parametric coordinates of bilinear mapping over the respective projected quadrilateral. These parametric coordinates ($\xi_1,\xi_2$) for each quadrature point are calculated using the inverse mapping of the bilinear interpolation as described in \cite{sahuBilinearInverseMapperAnalyticalSolution2024}. After determining $\left(\xi_1,\xi_2\right)$, physical points on both facets are determined using shape functions.
    
    At each quadrature point $q$, the interpenetration gap $g_N$ is evaluated using both split gap vectors $\tens{g}_N^1\left(^1\xi_1,^1\xi_2\right)$ and $\tens{g}_N^2\left(^2\xi_1,^2\xi_2\right)$. It is only negative in the case of interpenetration, as explained previously and illustrated in Fig.~\ref{fig:simplified_2D_interpenetration_volume}.   
    
    Using $m$ quadrature points per triangle, these integrals inside $[\cdot]$ in the eq.~\ref{eq:integral_triangulation_subdivision} can be written as 
    \begin{align}
        \sum_{j=1}^{k} \int_{\Delta_j} \psi_i g_N H(-g_N) d\gamma = \sum_{j=1}^{k} A_{\Delta_j} \left[ \Sigma_{q=1}^m \psi_{i_q} w_q g_{N_q}  H(-g_{N_q}) \right] = \left[ \Psi_q V_{encl} \right]_i
    \end{align}
    with the $w_q$'s denoting the weights of the associated quadrature points, $\psi_{i_q}$, the value of the shape function corresponding to the $i^{th}$ node at this quadrature point, and $g_{N_q}$,  the interpenetration gap between the facets at the $q^{th}$ quadrature point. The expression $\left[\Psi_q V_{encl}\right]_i$ is being used as a shorthand notation for the quantity on the left, which will correspond to the normal force on the $i^{th}$ node of the selected facet.
    
    The presented formulation inherently maintains the force-moment equivalency in the forces applied on the nodes with the acting contact traction. It can also be observed that the nodal forces calculated using this midplane-based method depend upon the interpenetration between facets, so changes in nodal forces due to the time-based configurational changes in the system follow the smoothly changing interpenetration volume between facets. Thus, the penalisation of interpenetration provides a smooth variation of nodal forces during deformation and avoids the jumping of contact forces, as is typical in NTS schemes.

\section{Solution procedure}
    Computational solution of contact problems for given discretised geometries undergoing motion and deformation is a two-stage methodology. In the first stage, a contact search is performed to establish contact pairs of facets having interpenetration, whereas in the second stage, contact constraint enforcement is carried out for physically interacting boundaries. 

   \begin{figure}
        \centering
        \includegraphics[width=0.40\linewidth]{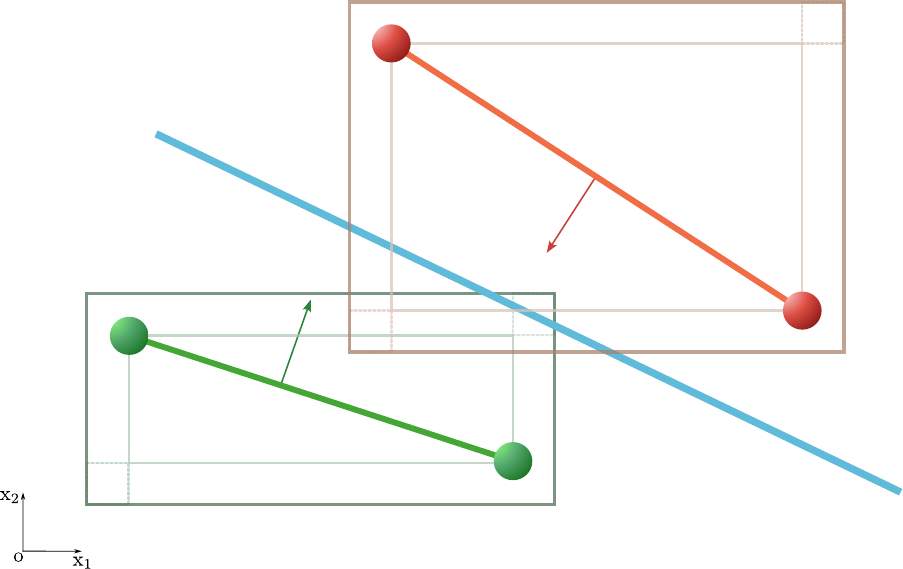}
        \caption{A 2D analogue of geometrical intersection check for a facet pair along with its midplane}
        \label{fig:geom_intersec_2D_analogue}
    \end{figure}

    Contact search involves finding out which parts of the surfaces of solids are in contact. It involves two phases: first, spatial search to establish potential contact pairs and second, detection to confirm the contact pairs with a geometrical intersection. This work uses the bounding box algorithm for spatial search to form a list of facet pairs \cite{bonetAlternatingDigitalTree1991} whose Axis Aligned Bounding Boxes (AABB) have an intersection. These bounding boxes are expanded by a pre-decided width in each dimension to avoid possible degeneracy before being tested for their intersection to form the list of potential contact pairs. This list of potential contact pairs of facets also considers facets on the same body, thus accounting for self-contact problems. Once this list is generated, a geometrical intersection algorithm is used to confirm the facet pairs having interpenetration. 
    
    The geometrical intersection algorithm primarily involves three sequential contact checks to confirm the potential contact facet pairs with interpenetration. In the first check, facets with their representative planes having angles greater than a pre-decided value (e.g., $80^{\circ}$ in this work) will be discarded. For facet pairs passing the first check, the positions of nodes on each facet will be checked with respect to the midplane. If all nodes of both facets lie on the opposite sides of the midplane (i.e., no interpenetration), then this second contact check will discard such facet pairs. One such facet pair, with the bounding boxes of both facets having their nodes on the opposite sides of the midplane, are shown for a 2D analogue in Fig.~\ref{fig:geom_intersec_2D_analogue}. Lastly, the remaining facet pairs will be tested for the intersection of the polygons of projections of both facets over the midplane using a clipping algorithm \cite{sutherlandReentrantPolygonClipping1974}. Only the facet pairs with non-zero areas of intersection of projection polygons are considered for the contact traction evaluation.

    Contact traction evaluation for any interpenetrating facet pair is carried out over the intersection polygon on the midplane, and it will act normal to the midplane. The intersection polygon is triangulated, and quantities of interest are calculated at quadrature points of each triangle for numerical integration. This algorithm for evaluating contact forces acting on nodes of both facets is described in Algorithm\ref{alg:trianguation_algorithm}. One important property of bilinear interpolation of quadrilaterals that has been realised in the presented methodology is that when a bilinear quadrilateral surface in 3D is projected over a plane to form a 2D quadrilateral, the bilinear parametric lines of quadrilateral surface in 3D also get projected on to the bilinear parametric lines of the 2D quadrilateral formed by projection. Thus, using parametric coordinates of quadrature points with respect to both 2D quadrilaterals to find the shape functions of both facets is a valid approach.

    \begin{algorithm}[htbp!]
        \caption{Multipoint integration of contact traction for interpenetrating facet pairs} 
        \begin{algorithmic}[1]
            \State get the polygon of the intersection of full projections of opposite facets
            \State triangulate the intersection polygon using edges and its centroid
            \State find the area of all triangles
            \State Set $\left[\Psi_q V_{encl}\right]_i = 0$ 
            \For{each triangle}
                \For{each quadrature point}
                    \State locate this quadrature point on this triangle using the physical coordinates of vertices
                    \State {find the parametric coordinates $(^{i}\xi_1,^{i}\xi_2)$ for this quadrature point w.r.t both projected quadrilaterals.}
                    \State find the physical points on both facets at this quadrature point using the parametric coordinates
                    \State find the split gap vectors  $\tens{g}_N^1$ and $\tens{g}_N^2$ at this quadrature point on the midplane
                    \State find the interpenetration gap $g_N (=\tens{g_N}\cdot{n_{\text{mp}}})$
                    \If{$g_N<0$}
                        \State continue to the next quadrature point as there is no interpenetration
                    \Else
                        \State update the quantity  $\left[\Psi V_{encl}\right]_i$ for both facets by taking contribution from this quadrature point.  
                    \EndIf
                \EndFor
            \EndFor
            \State Calculate the normal forces  $f_s\epsilon_N\left[\Psi_q V_{encl}\right]_i$ on each node of both facets in this pair, with the direction being decided by the midplane.                    
        \end{algorithmic}
        \label{alg:trianguation_algorithm}
    \end{algorithm}         

    The contact forces, calculated for each node of the interacting facets, are assembled and added to the global column matrix containing the external nodal forces acting on the system, treating the contact constraints as boundary conditions on the interacting bodies. These contact forces are computed at each solution step and added before calculating the new kinematic quantities in the current step. Adding contact forces to the external column matrix induces the effect of contact constraints on the system configuration changes during the current step. So, the contact constraints applied at each step are based on the interpenetration resulting from the previous step. This procedure is repeated in each solution step, focusing only on the interpenetrating STS pairs at that step and penalising their true interpenetration.

\section{Some Numerical Experiments}
The above-conceptualised segment-to-segment contact constraint enforcement strategy is analysed for its accuracy and validity through multiple benchmarks for contacts between solids composed of first-order elements by implementation in the in-house solver DEST \cite{bronikDESTDocumentation2022}. The presented tests consider contact between both flat and curved surfaces, and 13 quadrature points are used per triangle for integration in all cases. Different sets of strong discretisation schemes and material properties have been employed to demonstrate the versatility and robustness of the proposed scheme. 

To achieve the quasi-static nature of deformation in the initial five tests, the dynamic relaxation technique is used, which helps in attaining the static solution by considering the steady state part of the transient response \cite{belytschko1985computational}. All the plots showing contact pressure variation are based on directly extracting stress states from the subsurface quadrature points inside finite elements of the contacting surface. The stress measure reported in all examples is Cauchy stress. Visualisations in this paper are presented using Paraview \cite{ahrensParaViewEnduserTool2005}\cite{ayachitParaViewGuideParallel2015}.

\subsection{\textit{Contact Patch Test}}
\begin{figure}[!tbp]
    \centering
    \hfill
    \begin{subfigure}[b]{0.28\linewidth}
        \centering
        \includegraphics[width=\linewidth]{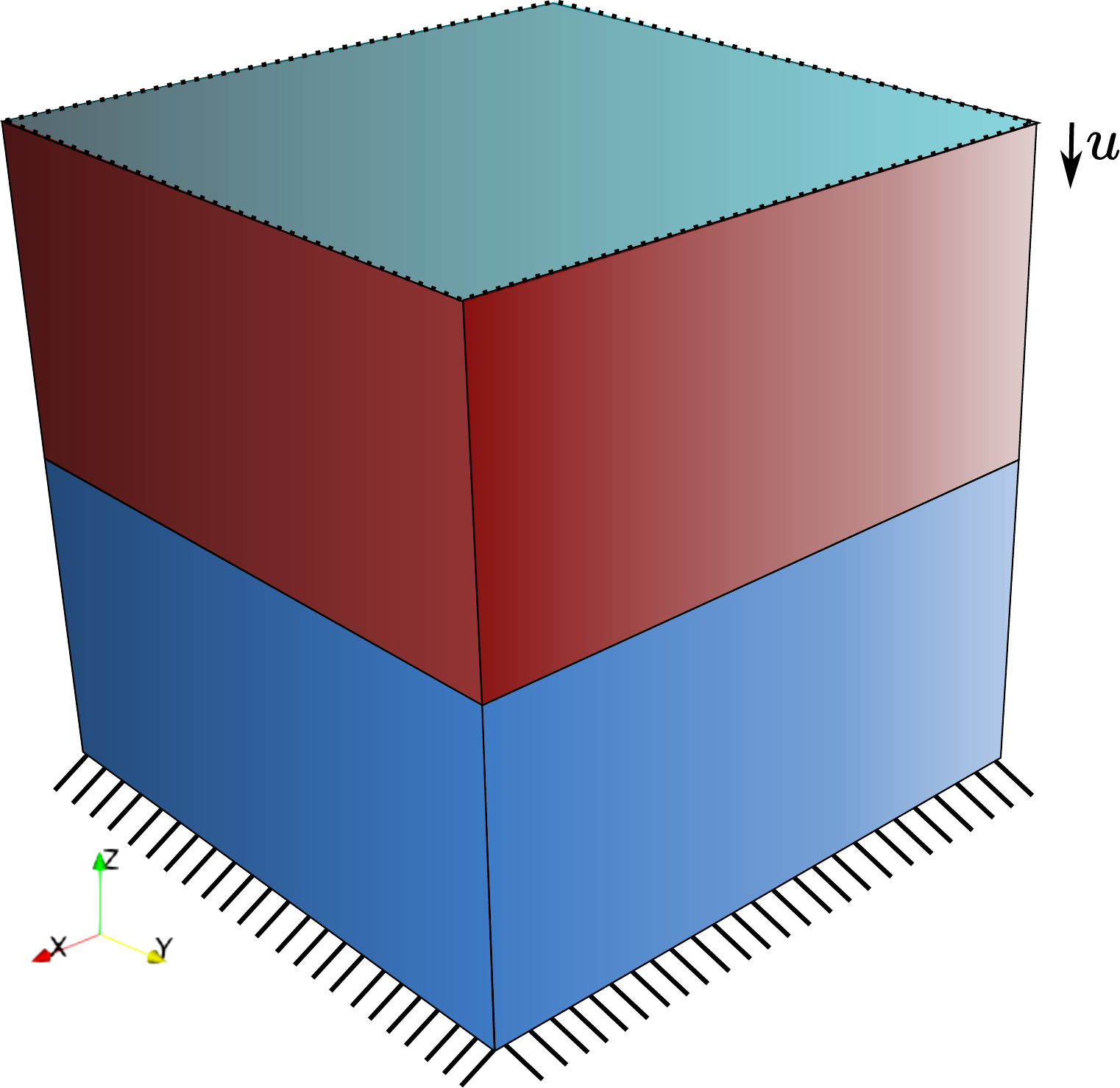}
        \caption{}
        \label{fig:contact_patch_test_geometry}
    \end{subfigure}
    \hfill
    \begin{subfigure}[b]{0.25\linewidth}
        \centering
        \includegraphics[width=\linewidth]{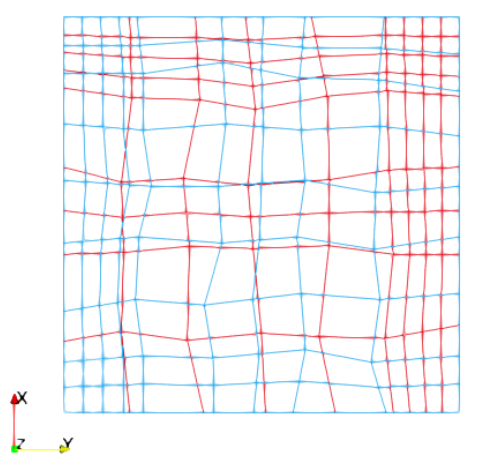}
        \caption{}
        \label{fig:contact_patch_test_strong_contrast_interface_wireframe}
    \end{subfigure}
    \hfill    
    \begin{subfigure}[b]{0.45\linewidth}
        \centering
        \includegraphics[width=1.0\linewidth]{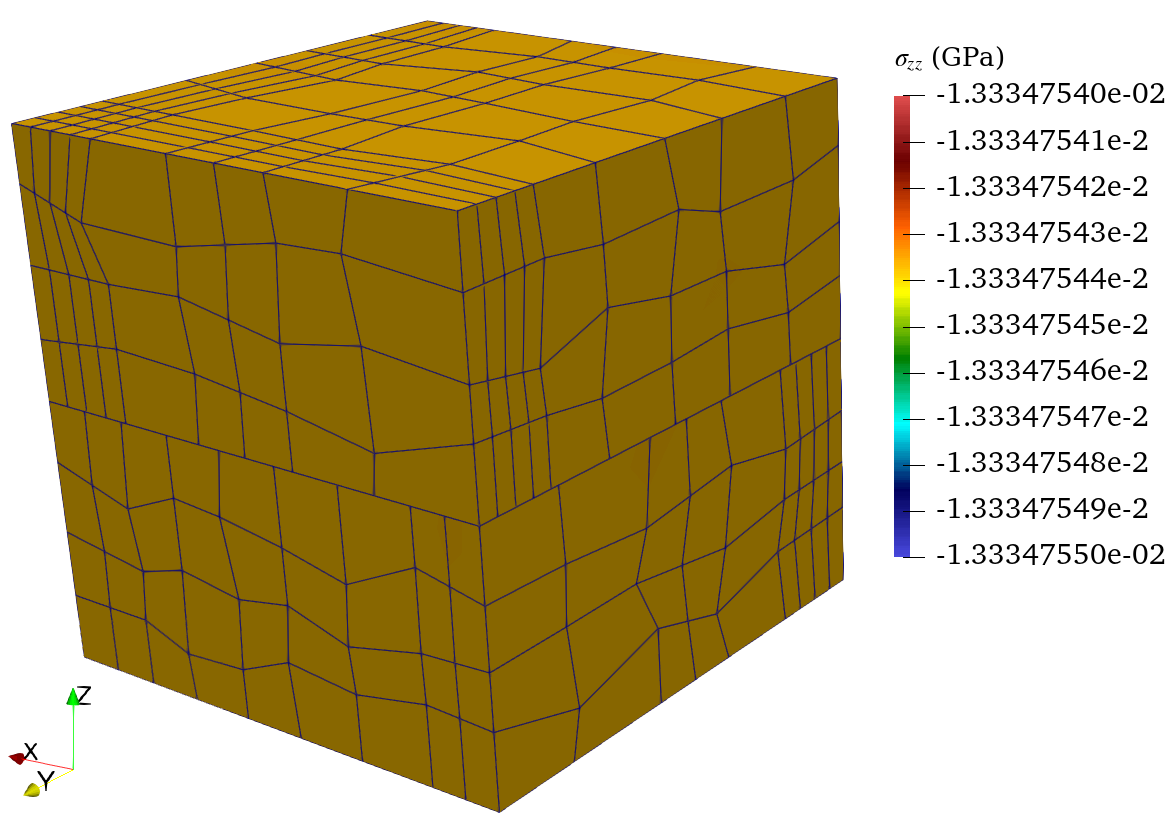}
        \caption{}
        \label{fig:contact_patch_test_341341_strongly_nonconformal}
    \end{subfigure}

    \caption{Contact Patch test: (a) schematic of the symmetric compression of the upper block while all lateral sides and bottom surface of the lower block are constrained, (b) wireframe of interface facets coming in contact, and (c) result from proposed methodology.}
    \label{fig:contact_patch_test}
\end{figure}

This test, originally proposed in  \cite{taylorPatchTestContact1991}, \cite{el-abbasiStabilityPatchTest2001} checks for the transmission of uniform pressure between two flat surfaces in contact. This test is also critical for slow applications like abrasion/wear studies. In this work, a version of the contact patch test used in \cite{netoSurfaceSmoothingProcedures2017} is employed for verifying the transmission of uniform pressure. Here, two blocks with identical geometries are compressed against each other with frictionless conditions; see Fig.~\ref{fig:contact_patch_test_geometry}. The bottom surface of the lower block is restricted against downward motion, while the lateral faces of both upper and lower blocks are constrained against motion in directions normal to those faces. The top surface of the upper block is uniformly pressed downwards, and theoretically, a uniform stress state is generated in both blocks.

For the discretised problem, both blocks have size  $20\times20\times10 \ \text{mm}^3 $ and elastic materials with the Young's modulus of elasticity, E$_1= 10$ GPa, E$_2\in\{10,100,1000\}$ GPa, and Poisson's ratio, $\nu_1=0.3$, $\nu_2=0.35$. The combinations of materials with a range of stiffness ratios and penalty scaling factors provide a diverse variety in testing the unbiased nature of the contact algorithm.
The top surface of the upper block is pressed downwards by 0.01 mm, with the penalty scaling factor in contact being $f_s\in\{1,10,100\}$. Here, both blocks are discretised with non-regular meshes having a strong contrast in discretisation at the interface, as shown in Fig.~\ref{fig:contact_patch_test_strong_contrast_interface_wireframe}. In all cases, the proposed scheme passed the contact patch test, with uniform stress states generated in both blocks. On the contrary, the use of NTS schemes results in hot spots or oscillations observed near the contact interface \cite{netoSurfaceSmoothingProcedures2017}\cite{kwonFullyNonlinearThreedimensional2023}. The stress values in all cases were the same up to 10 or 11 digits after the decimal point, which matches the accuracy of the element patch test that has the same 11 digits, implying that the contact algorithm is as accurate as the element patch test. 
The result of the proposed scheme using material stiffness ratio E$_2:$ E$_1=100$ and $f_s=100$ is shown in Fig.~\ref{fig:contact_patch_test_341341_strongly_nonconformal}.

\subsection{\textit{Two-beam bending test}}
\begin{figure}[!tbhp]
    \centering
    \hfill
    \begin{subfigure}[b]{0.4\linewidth}
        \centering
        \includegraphics[width=\linewidth]{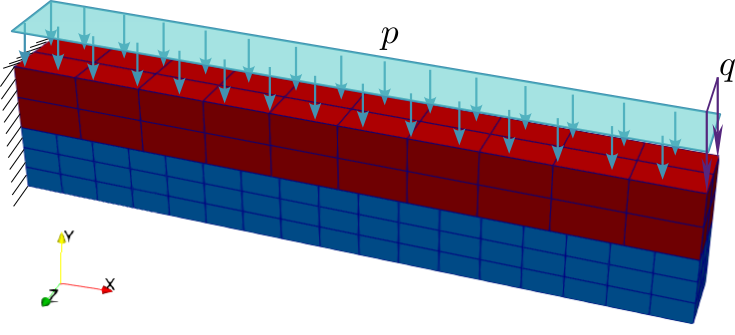}
        \caption{}
        \label{fig:two_beam_test_flat_interface_Underformed_mesh}
    \end{subfigure}
    \hfill    
    \begin{subfigure}[b]{0.4\linewidth}
    \centering
        \includegraphics[width=\linewidth]{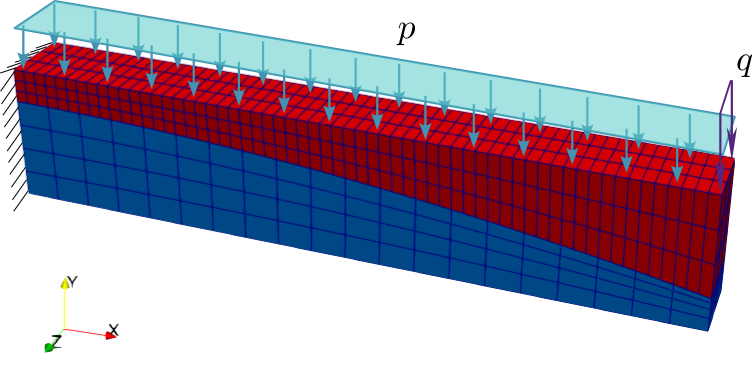}
        \caption{}
        \label{fig:two_beam_test_curved_interface_Underformed_mesh}
    \end{subfigure}
    \hfill
    \vfill
    \hfill    
    \caption{Two-beam bending test to validate the locking-free behaviour: (a),(b) show undeformed meshes with flat and curved contact interface.}
    \label{fig:two_beam_bending_figs_undeformed}
\end{figure}

\begin{figure}[bthp]
    \centering
    \hfill
    \begin{subfigure}[b]{0.4\linewidth}
        \centering
        \includegraphics[width=\linewidth]{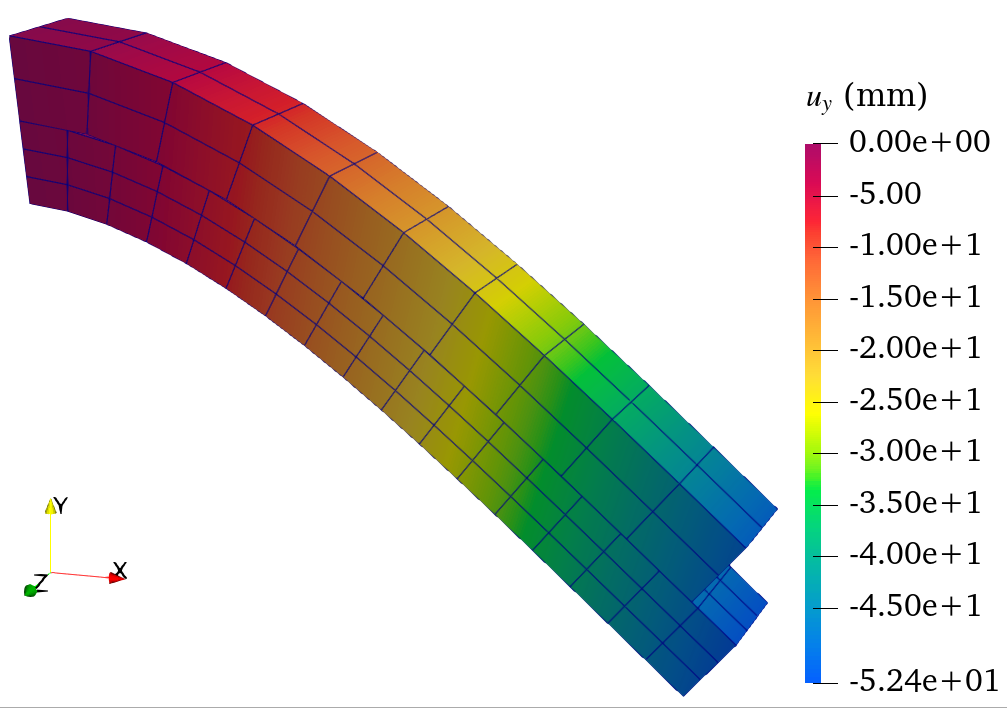}
        \caption{}
        \label{fig:two_beam_test_flat_interface_result_displacement_y}
    \end{subfigure}
    \hfill
    \begin{subfigure}[b]{0.4\linewidth}
        \centering
        \includegraphics[width=\linewidth]{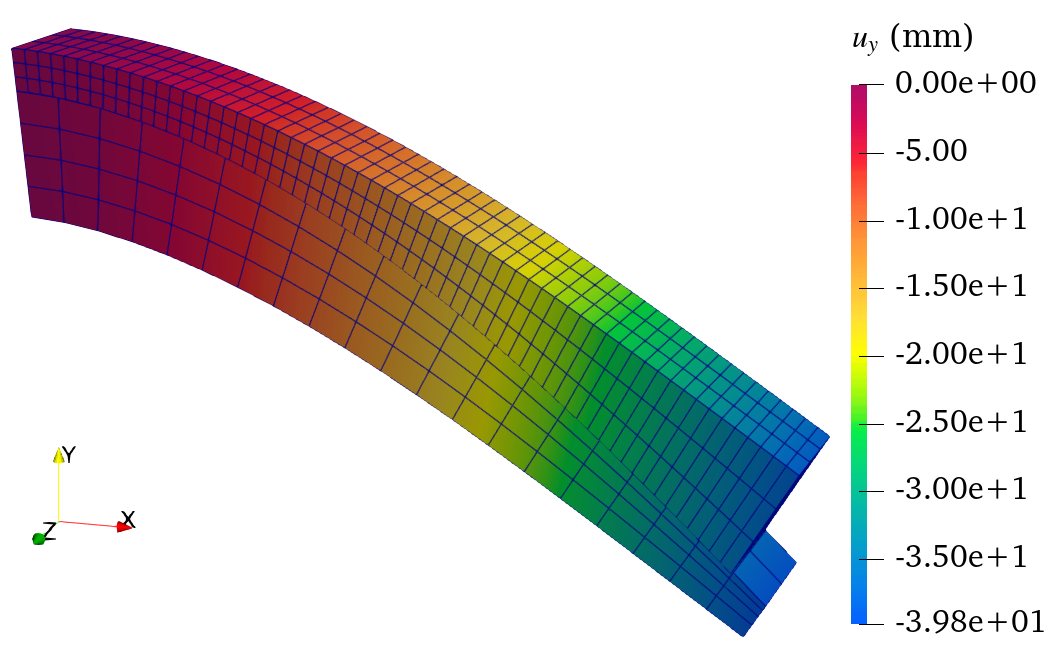}
        \caption{}
        \label{fig:two_beam_test_curved_interface_result_displacement_y}
    \end{subfigure}
    \hfill
    \vfill
    \hfill
    \begin{subfigure}[b]{0.4\linewidth}
        \centering
        \includegraphics[width=\linewidth]{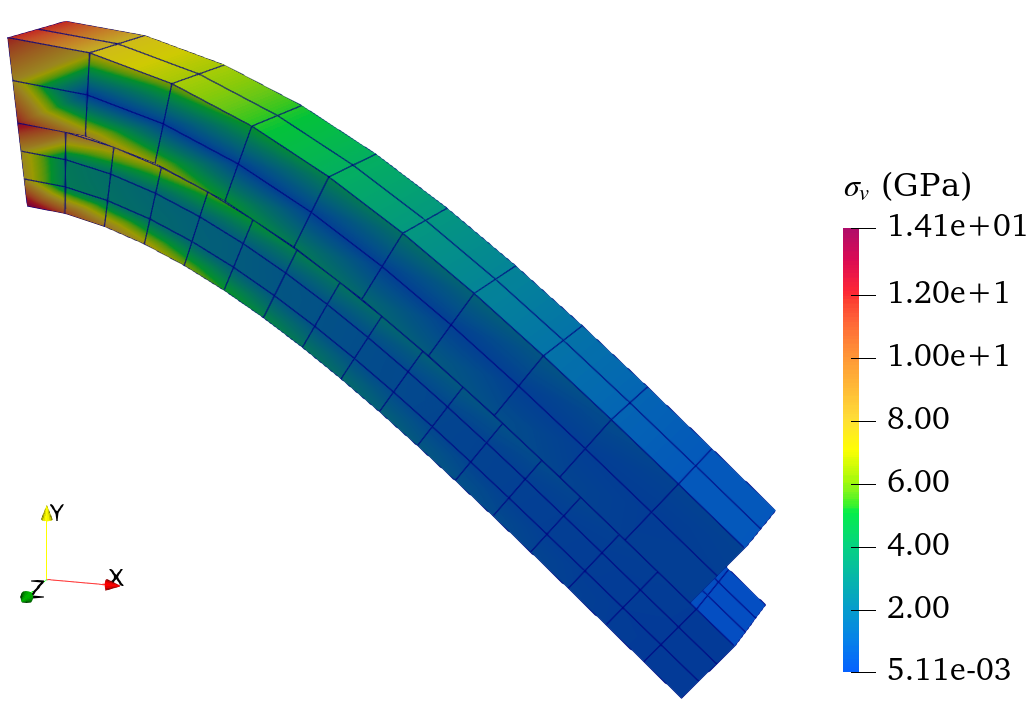}
        \caption{}
        \label{fig:two_beam_test_flat_interface_result_vonmises}
    \end{subfigure}
    \hfill
    \begin{subfigure}[b]{0.4\linewidth}
        \centering
        \includegraphics[width=\linewidth]{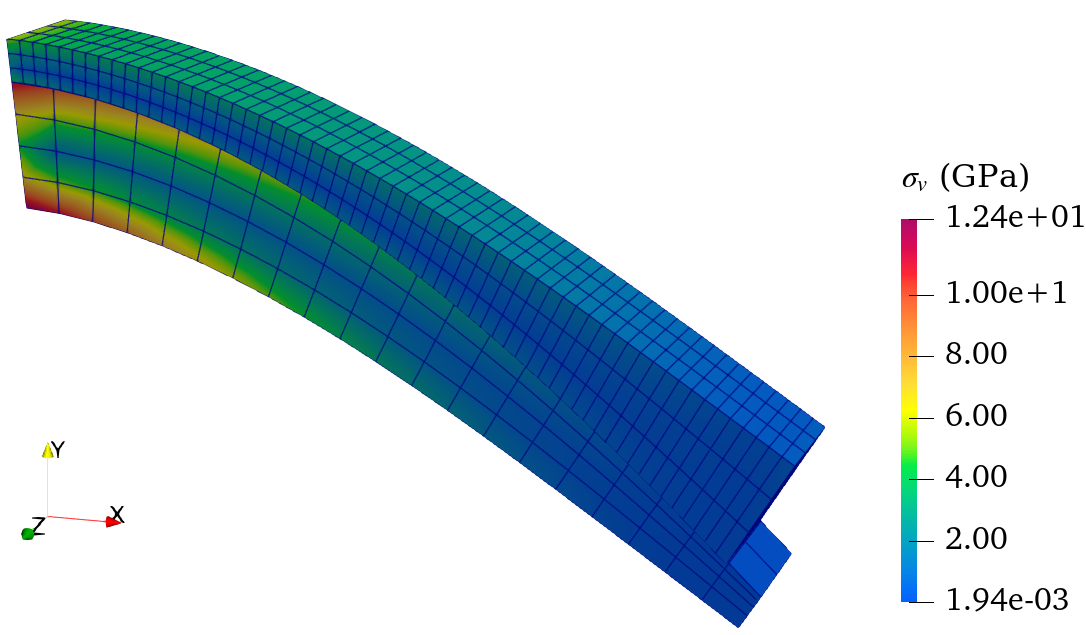}
        \caption{}
        \label{fig:two_beam_test_curved_interface_result_vonmises}
    \end{subfigure}
    \hfill    
    \caption{Two-beam bending test: (a),(b) show net displacements, and (c),(d) show von Mises stress for beams with flat and curved contact interface, respectively.}
    \label{fig:two_beam_bending_figs_deformed}
\end{figure}

In this test, the bending of two beams is carried out to check the locking tendency of the algorithm \cite{pusoMortarSegmenttosegmentContact2004}. For variation in contact conditions, two types of interfaces are considered between the non-matching meshes of beams - flat and curved, as shown in Figs. \ref{fig:two_beam_test_flat_interface_Underformed_mesh},\ref{fig:two_beam_test_curved_interface_Underformed_mesh}. The outer dimensions of the two-beam system in both cases are $100\times10\times20$ mm$^3$, with the two beams in the first case having identical sizes. The same linear elastic material is used for all beams having a modulus of elasticity, E$=100$ GPa and zero Poisson's ratio. The left ends of the two beams in both cases are fixed, and loading is applied to the top surface of the upper beam. A uniform pressure $p=0.1$ GPa is applied on the top surface, followed by a vertical line loading of $q=0.1$ kN/mm, with both acting through the vertical nodal forces applied on the nodes of the top surface and calculated in the undeformed configuration. 
As can be seen in Figs. \ref{fig:two_beam_test_flat_interface_result_displacement_y},\ref{fig:two_beam_test_curved_interface_result_displacement_y},\ref{fig:two_beam_test_flat_interface_result_vonmises},\ref{fig:two_beam_test_curved_interface_result_vonmises}, both beams undergo continuous bending for both types of interfaces, thereby demonstrating a locking-free behaviour. The penalty method allows interpenetration between the beams in both cases, enabling a continuous deformation along the interface through constraints on contact surfaces.

\subsection{\textit{Hertzian contact}}

Developed by Heinrich Hertz in 1882 \cite{hertzUeberBeruhrungFester1882}, the Hertzian contact theory has served as a gold standard in computational contact mechanics as it provides the analytical solution for nonconformal smooth surfaces in contact \cite{johnsonOneHundredYears1982}. The nonconformal surfaces have a point or line as the initial contact that turns into a finite area under deformation. This theory is based on the small deformation assumption for elastic solids with homogeneous and isotropic material.

\begin{figure}[!tbhp]
    \centering
    \begin{subfigure}{0.29\linewidth}
        \centering
        \includegraphics[width=\linewidth]{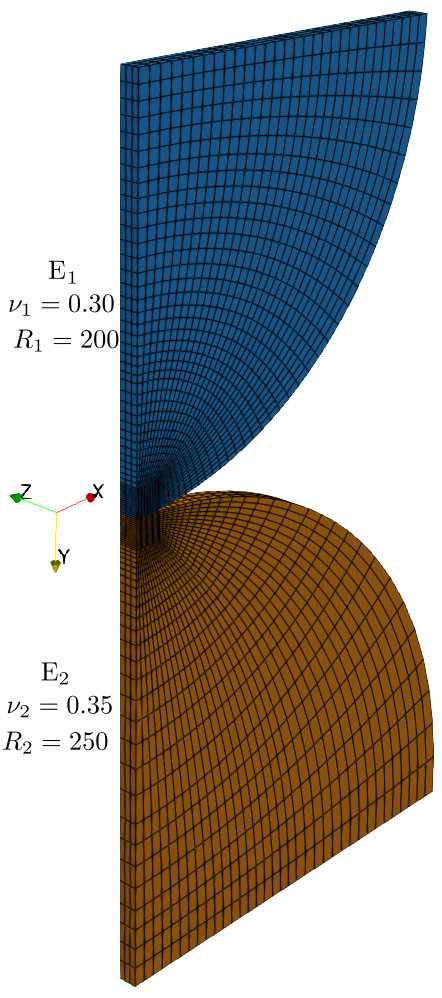}
        \caption{Undeformed Mesh 15:19}
        \label{fig:hertz_contact_full_undeformed_mesh_15to19}
    \end{subfigure}
    \begin{subfigure}{0.7\linewidth}
        \centering
        \begin{subfigure}{0.3\linewidth}
            \centering
            \includegraphics[width=\linewidth]{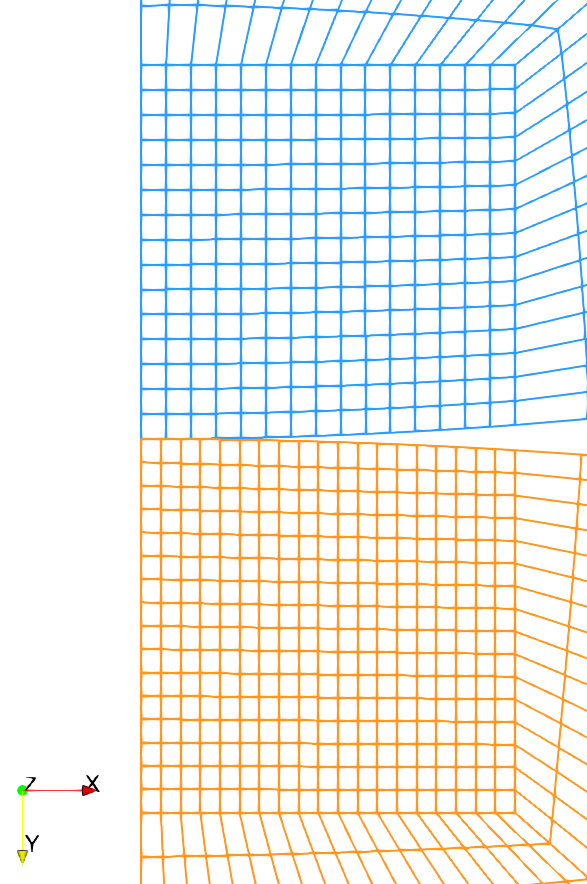}
            \caption{Mesh ratio 15:19}
            \label{fig:hertz_contact_undeformed_mesh_nearcontactzone_15to19}
        \end{subfigure}
        \begin{subfigure}{0.3\linewidth}
            \centering
            \includegraphics[width=\linewidth]{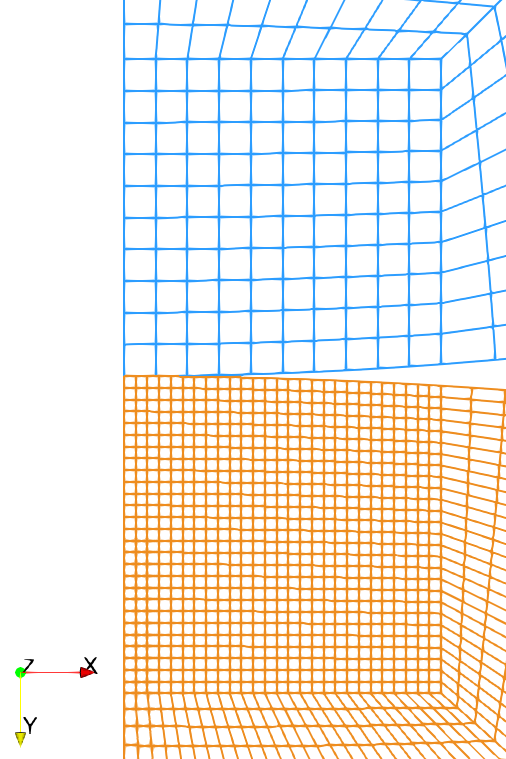}
            \caption{Mesh ratio 10:27}
            \label{fig:hertz_contact_undeformed_mesh_nearcontactzone_10to27}
        \end{subfigure}
        \begin{subfigure}{0.70\linewidth}
            \centering
            \includegraphics[width=\linewidth]{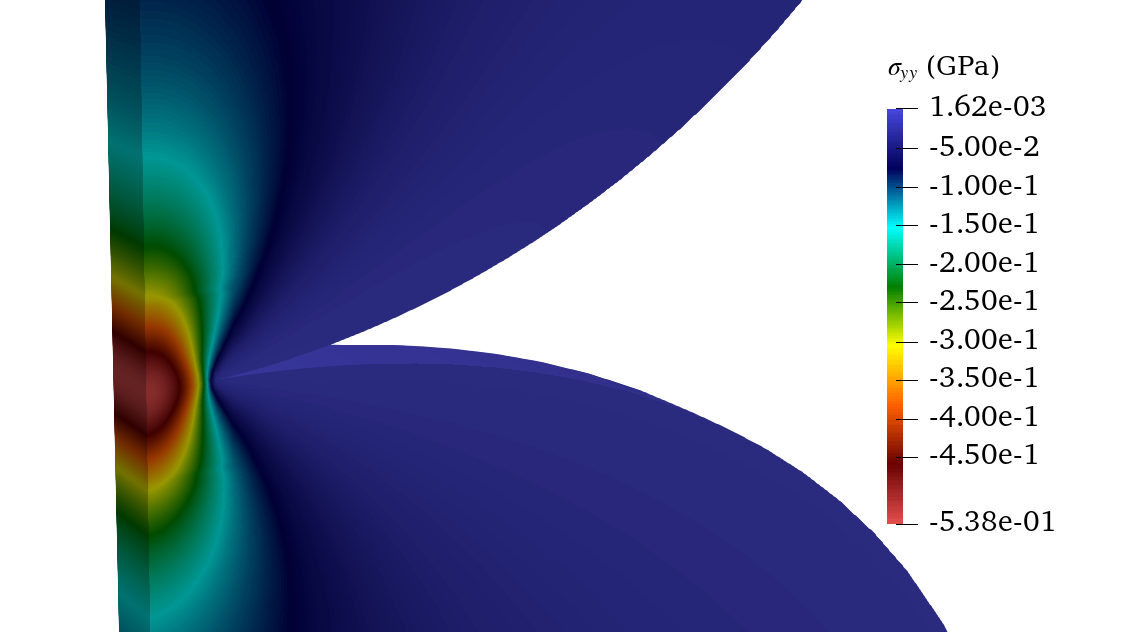}
            \caption{Stress distribution $\sigma_{yy}$ for mesh 15:19, E$_2$:E$_1=100$, $f_s=10$}
            \label{fig:hertz_contact_deformed_mesh_15to19_stress_distribution}
        \end{subfigure}
    \end{subfigure}
    \begin{subfigure}[b]{0.60\linewidth}
        \centering
        \includegraphics[width=\linewidth]{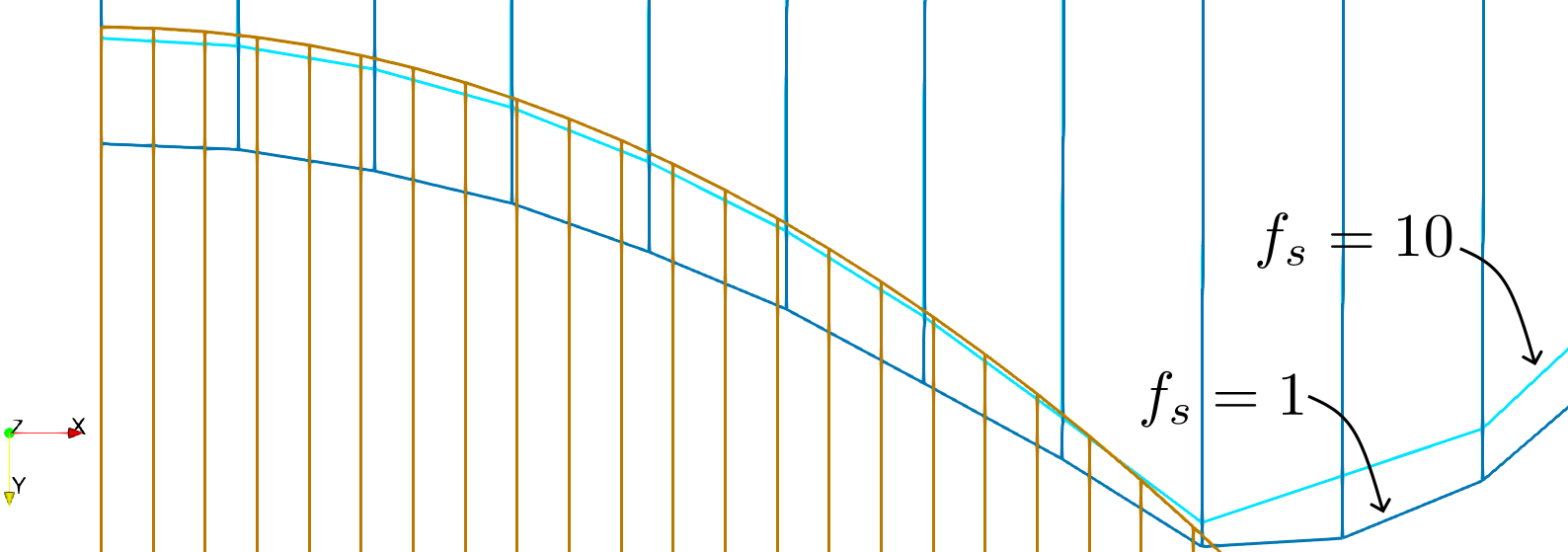}
        \caption{Two deformed meshes (ratio 10:27) for E$_2$:E$_1=100$ corresponding to two scaling factors ($f_s=1,10$). The two lower meshes in dark orange nearly overlap as they behave relatively rigidly. Both meshes are vertically scaled up to observe the interpenetration.}
        \label{fig:2024_07_17_hertzonact_341341_nonconformal_10to27_twocyl_minBulkfs10_E2toE1ratio_100_VS_fs01_marked}
    \end{subfigure}
    \hfill
    \begin{subfigure}[b]{0.25\linewidth}
        \centering
        \includegraphics[width=\linewidth]{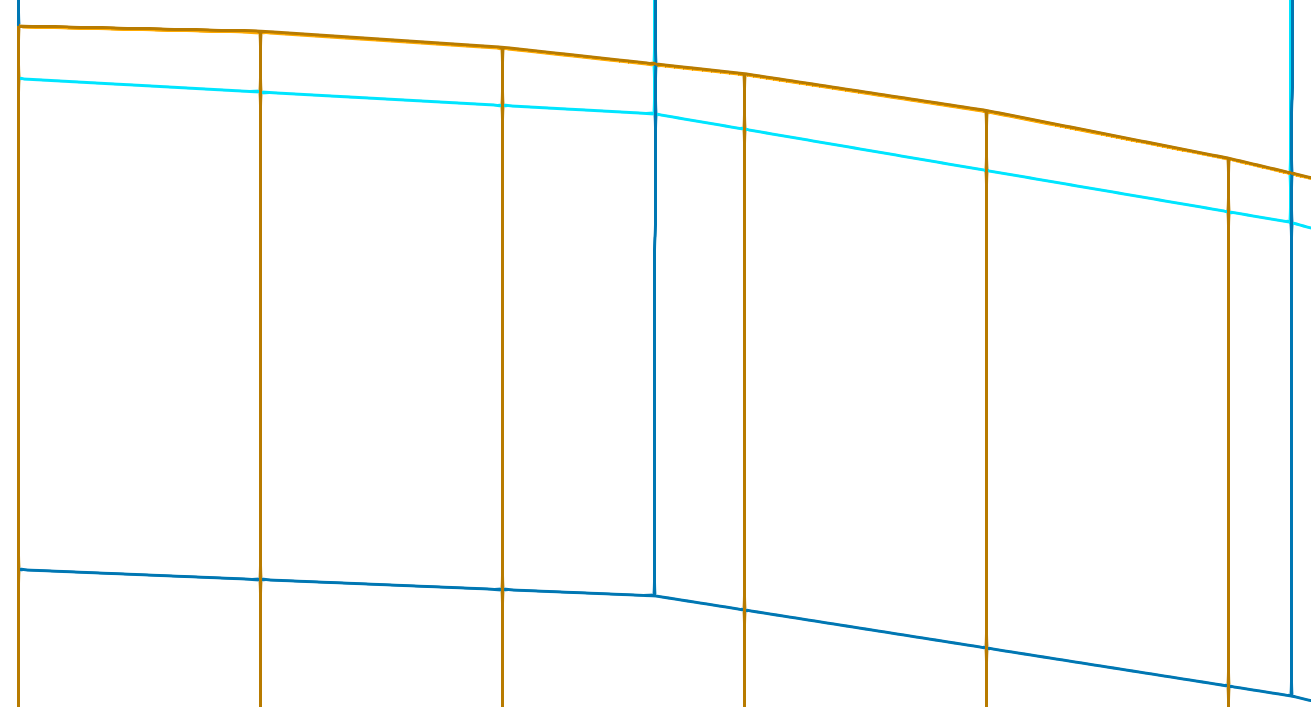}
        \caption{Zoomed in view near the first point of contact}        
        \label{fig:2024_07_17_hertzonact_341341_nonconformal_10to27_twocyl_minBulkfs10_E2toE1ratio_100_VS_fs01_Zoomed}
    \end{subfigure}
    
    \caption{Hertz contact: (a) geometry of two cylinder problem, (b) and (c) two types of meshes used, (d) stress distribution near contact region, and (e), (f) comparison of interpenetration gap variation for different penalties along the contact interface.}
    \label{fig:mesh_hertz_contact}
\end{figure}

To analyse the performance of the proposed algorithm, a canonical problem is considered wherein two cylindrical bodies having radii, R$_1=200$ mm and R$_2=250$ mm, are loaded against each other for comparison against Hertz's analytical solution. The symmetry of the problem is leveraged, and only a quarter of the slices of the two cylindrical bodies are used in the analysis, see Fig.~\ref{fig:hertz_contact_full_undeformed_mesh_15to19}. While the original problem considers a line loading atop the cylindrical surface, a uniform pressure is applied on top of the upper cylindrical mesh in this work \cite{lorenzisIsogeometricContactReview2014}\cite{carvalhoEfficientAlgorithmRigid2022}. The three lateral surfaces of the upper and lower cylinders and the bottom surface of the lower cylinder are all constrained against motion in a direction normal to their surface, thus maintaining the plane strain condition in the original problem.

Two different non-matching meshes and three different combinations of materials are studied to thoroughly examine the contact algorithm. The two meshes are discretised with different numbers of elements in the same geometrical region to have a ratio of 15:19 and 10:27 of elements at the interface, see Figs. \ref{fig:hertz_contact_undeformed_mesh_nearcontactzone_15to19},\ref{fig:hertz_contact_undeformed_mesh_nearcontactzone_10to27}. For the upper and lower cylinders, Young's modulus of Elasticity is E$_1=10$ GPa, E$_2\in\{10,100,1000\}$ GPa, and Poisson's ratio is $\nu_1=0.30$, $\nu_2=0.35$, respectively, and the contact definition in most cases uses a penalty scaling factor of $f_s=10$.

\begin{figure}[tbhp]
    \centering
    \begin{subfigure}[b]{0.49\linewidth}
        \centering
        \includegraphics[ width=1.0\linewidth]{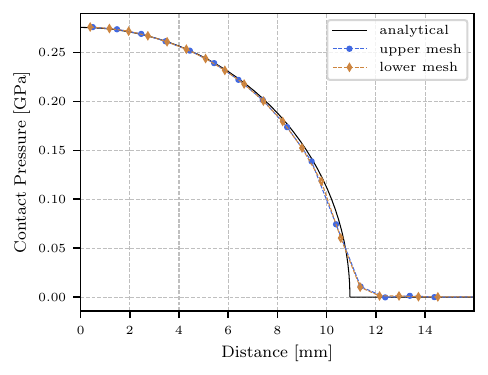}
        \caption{Mesh 15:19, E$_2$:E$_1=1$, $f_s=10$}
        \label{fig:hertzonact_341341_nonconformal_15to19_twocyl_minBulkfs10_E2toE1ratio_1}
    \end{subfigure}
    \begin{subfigure}[b]{0.49\linewidth}
        \centering
        \includegraphics[ width=1.0\linewidth]{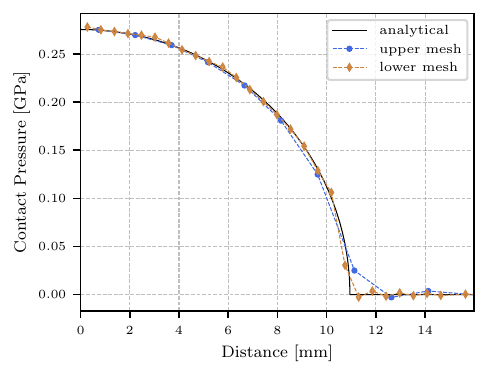}
        \caption{Mesh 10:27, E$_2$:E$_1=1$, $f_s=10$}
        \label{fig:hertzonact_341341_nonconformal_10to27_twocyl_minBulkfs10_E2toE1ratio_1}
    \end{subfigure}

    \begin{subfigure}[b]{0.49\linewidth}
        \centering
        \includegraphics[width=1.0\linewidth]{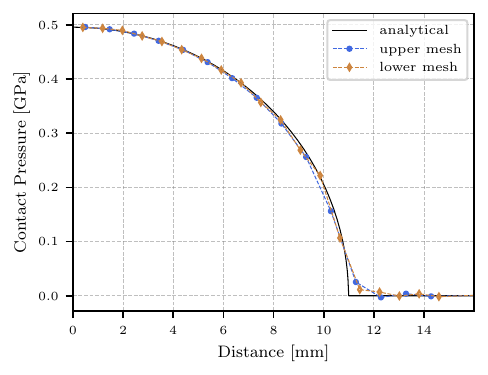}
        \caption{Mesh 15:19, E$_2$:E$_1=10$, $f_s=10$}
        \label{fig:hertzonact_341341_nonconformal_15to19_twocyl_minBulkfs10_E2toE1ratio_10}
    \end{subfigure}
    \begin{subfigure}[b]{0.49\linewidth}
        \centering
        \includegraphics[width=1.0\linewidth]{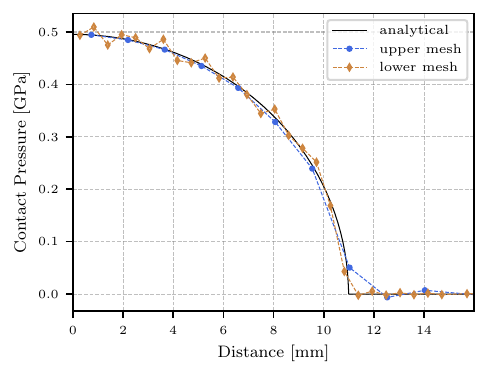}
        \caption{Mesh 10:27, E$_2$:E$_1=10$, $f_s=10$}
        \label{fig:hertzonact_341341_nonconformal_10to27_twocyl_minBulkfs10_E2toE1ratio_10}
    \end{subfigure}

    \begin{subfigure}[b]{0.49\linewidth}
        \centering
        \includegraphics[width=1.0\linewidth]{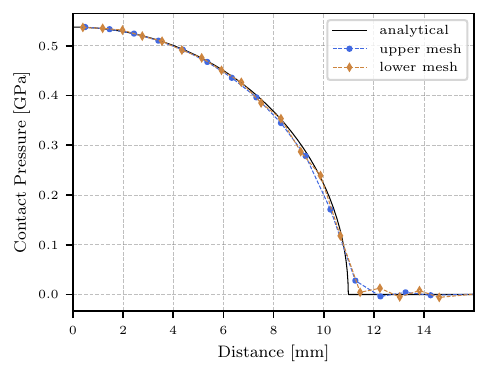}
        \caption{Mesh 15:19, E$_2$:E$_1=100$, $f_s=10$}
        \label{fig:hertzonact_341341_nonconformal_15to19_twocyl_minBulkfs10_E2toE1ratio_100}
    \end{subfigure}
    \begin{subfigure}[b]{0.49\linewidth}
        \centering
        \includegraphics[width=1.0\linewidth]{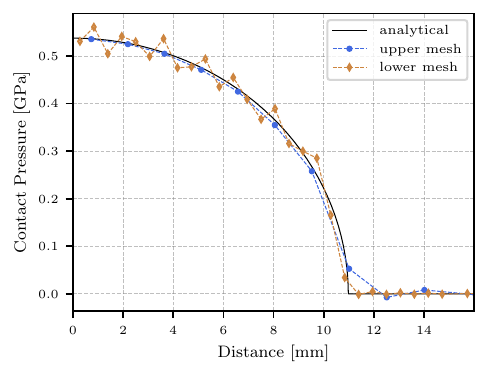}
        \caption{Mesh 10:27, E$_2$:E$_1=100$, $f_s=10$}
        \label{fig:hertzonact_341341_nonconformal_10to27_twocyl_minBulkfs10_E2toE1ratio_100}
    \end{subfigure}

    \caption{Hertz contact: contact pressure variation on both cylinders for mesh 15:19 (a,c,e), and mesh 10:27 (b,d,f), all with the same penalty scaling factor $f_s=10$.}
    \label{fig:hertz_contact_plots_part1}
\end{figure}

\begin{figure}[tbhp]
    \centering

    \begin{subfigure}[b]{0.49\linewidth}
        \centering
        \includegraphics[width=1.0\linewidth]{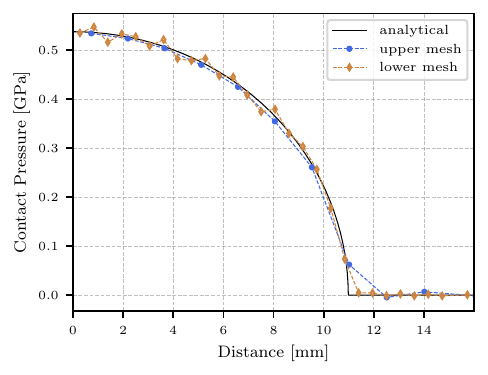}
        \caption{Mesh 10:27, E$_2$:E$_1=100$, $f_s=5$}
        \label{fig:hertzonact_341341_nonconformal_10to27_twocyl_minBulkfs05_E2toE1ratio_100}
    \end{subfigure}
    \begin{subfigure}[b]{0.49\linewidth}
        \centering
        \includegraphics[width=1.0\linewidth]{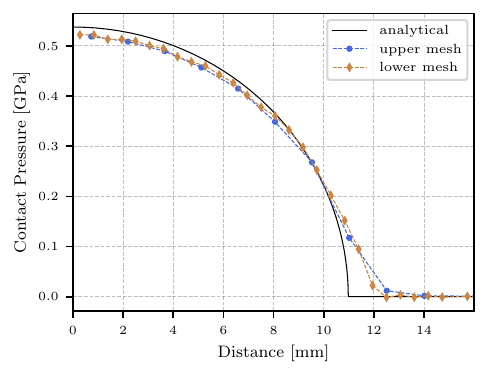}
        \caption{Mesh 10:27, E$_2$:E$_1=100$, $f_s=1$}
        \label{fig:hertzonact_341341_nonconformal_10to27_twocyl_minBulkfs01_E2toE1ratio_100}
    \end{subfigure}
        
    \caption{Hertz contact: contact pressure variation on both cylinders for mesh 10:27 with lowered penalty scaling factors of (a) 5, and (b) 1.}
    \label{fig:hertz_contact_plots_part2}
\end{figure}

The variation of the contact pressure for both upper and lower cylinders for two types of meshes discretised with under-integrated first-order elements with three different material combinations are shown in Figs. \ref{fig:hertz_contact_plots_part1},\ref{fig:hertz_contact_plots_part2} and stress state in the two cylinders are depicted for one case in Fig.~\ref{fig:hertz_contact_deformed_mesh_15to19_stress_distribution}. For the same material stiffness (E$_2$:E$_1=1$), albeit different Poisson's ratios, the variation of contact pressure in both upper and lower cylinders is relatively very smooth, Figs. \ref{fig:hertzonact_341341_nonconformal_15to19_twocyl_minBulkfs10_E2toE1ratio_1},\ref{fig:hertzonact_341341_nonconformal_10to27_twocyl_minBulkfs10_E2toE1ratio_1}. Both types of meshes show edge effect near the periphery of the contact zone, which is not only an outcome of the discrete nature of contacting segments but also depends on their sizes. When the lower cylinder is composed of material with increasingly higher stiffness (E$_2$:E$_1=10,100$), the discrepancy in the contact pressure in the case of mesh with larger element size ratio becomes evident and is more pronounced with higher stiffness ratio, see Figs\ref{fig:hertzonact_341341_nonconformal_15to19_twocyl_minBulkfs10_E2toE1ratio_10}, \ref{fig:hertzonact_341341_nonconformal_10to27_twocyl_minBulkfs10_E2toE1ratio_10}, \ref{fig:hertzonact_341341_nonconformal_15to19_twocyl_minBulkfs10_E2toE1ratio_100}, \ref{fig:hertzonact_341341_nonconformal_10to27_twocyl_minBulkfs10_E2toE1ratio_100}.

It can be observed that the oscillations only occur in lower cylinders (fine mesh) and result from the non-monotonically changing interpenetration gap between the discretised surfaces that also has a non-smooth rate of change on moving along the contact interface. These discretised surfaces have an under-representation of the originally curved geometry, with the coarser meshes having larger discrepancies. Moving along the discretised boundaries, the outward normal also suffers from a sharp change in direction at the segment corners. Theoretically, contact between originally curved geometries will have a smooth contact interface upon deformation, so no oscillations in contact pressure; however, the contact load induced deformed shapes of differently discretised surfaces do not adhere to each other, leading to a non-smooth gap variation. 

The oscillations increase with an increase in either the mesh ratio or the stiffness ratio. 
The higher oscillations with differences in the sizes of contact segments are due to the pronounced effect of the deformation of finite segments, not exhibiting the same smoothness as the deformation of the originally circular geometry. For the mesh ratio 10:27, more segments on the lower mesh are in contact per upper cylinder segment, compared to the mesh ratio of 15:19. So, the already present discrepancy with the original geometry gets worsened with the application of the contact constraints on the finite segments, causing higher oscillations in the fine mesh due to the non-smooth rate of change of gap that is also changing non-monotonically. 
When the lower cylinder has higher stiffness, it behaves more rigidly and requires the upper cylinder to have a larger curvature on deformation to adapt to the lower cylinder. The inability of the coarse finite segments to provide smoothness on deformation becomes more pronounced here, causing a rise in oscillations in the fine mesh.

The upper (coarse) mesh has larger segments than the fine mesh. The lower (fine) mesh has a better curvature upon deformation, which the upper mesh follows, albeit coarsely. The net effect of contact traction due to changes in interpenetration gaps over each coarse segment while moving along the interface is monotonically decreasing, so no oscillations are observed for the lower mesh. In contrast, the upper mesh cannot provide the required smooth curvature of deformation to the fine mesh, resulting in oscillations in the fine mesh. These oscillations get attenuated by using softer penalty scaling factors, see Figs. \ref{fig:hertzonact_341341_nonconformal_10to27_twocyl_minBulkfs05_E2toE1ratio_100},\ref{fig:hertzonact_341341_nonconformal_10to27_twocyl_minBulkfs01_E2toE1ratio_100}. Softer penalties allow higher interpenetration until the required counterbalancing contact forces are generated. However, the unit penalty, $f_s=1$, which can provide a better smoothness in the contact pressure, comes at the cost of an over-prediction of the contact zone and an under-prediction of the maximum contact pressure. A comparison of the interpenetration gap for two penalties ($f_s=1,10$) for the mesh 10:27 with stiffness ratio E$_2$:E$_1=100$ is shown in Figs. \ref{fig:2024_07_17_hertzonact_341341_nonconformal_10to27_twocyl_minBulkfs10_E2toE1ratio_100_VS_fs01_marked},\ref{fig:2024_07_17_hertzonact_341341_nonconformal_10to27_twocyl_minBulkfs10_E2toE1ratio_100_VS_fs01_Zoomed}. While a higher interpenetration implies a higher violation of the impenetrability constraint of contact, the rate of change of interpenetration going along the contact interface is smaller for
$f_s=1$ compared to $f_s=10$, leading to a smoother variation in contact traction for softer penalties. 

In all cases, the numerical solutions follow the analytical solutions, and disparities result from the piecewise finite element discretisation. So, the solution accuracy improves if the two contacting curved meshes have similar element sizes and/or stiffness. The proposed methodology reveals the importance of allowing and controlling the interpenetration between meshes for a desirable smoothness in the solution. This is in contrast to the traditional methods that seek conformity between contacting surfaces by disallowing interpenetration through the restriction of the motion of nodes. The distributed traction, calculated by penalisation of true 3D interpenetration, provides improved accuracy in contact pressure.

\subsection{\textit{Indentation test with a rigid flat punch}}
Indentation of an elastic half-space by a flat rigid punch is a standard problem in contact mechanics that also has an analytical solution \cite{johnsonContactMechanics1985}\cite{popovHandbookContactMechanics2019}. The analytical solution assumes that the elastic half-space has homogeneous and isotropic material, and the deformation is small, so the small-strain theory is applicable. In 2D, this problem exhibits a line loading over the half-space with a plane strain assumption. The contact pressure increases radially along the contact region and approaches a theoretically infinite value at the edges of the punch, where a stress singularity is seen. 

For setting up the numerical problem in 3D, the half symmetry of the original problem is used. With this, only half of the punch and the block underneath, representing the elastic half-space, are employed for the numerical test. Young's modulus of the block and Poisson's ratio are $210$GPa and zero, respectively. Similar to the Hertzian contact problem in the previous section, the non-free surfaces of both the punch and the block are restricted to in-plane motion only. To adhere to the rigid punch assumption in the analytical solution, all the nodes in the punch are prescribed a uniform downward displacement of 0.001mm. The half-width of the punch is $ a=5$ mm, and the vertical load $P$ is recovered from the contact forces applied on the interacting surfaces.

\begin{figure}[tbp!]
    \begin{subfigure}[b]{0.45\linewidth}
    
        \begin{subfigure}[b]{0.8\linewidth}
            \centering
            \includegraphics[width=1.0\linewidth]{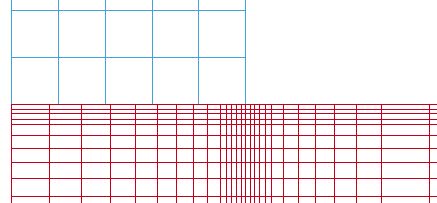}
            \caption{Coarse mesh}
            \label{fig:punch_coarse_mesh}
        \end{subfigure}
        
        \begin{subfigure}[b]{0.8\linewidth}
            \centering
            \includegraphics[width=1.0\linewidth]{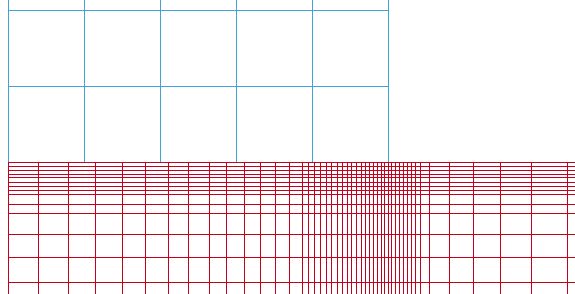}
            \caption{Refined mesh}
            \label{fig:punch_refined_mesh}
        \end{subfigure}        
        
    \end{subfigure}
    \begin{subfigure}[b]{0.48\linewidth}
        \centering
        \includegraphics[width=1.0\linewidth]{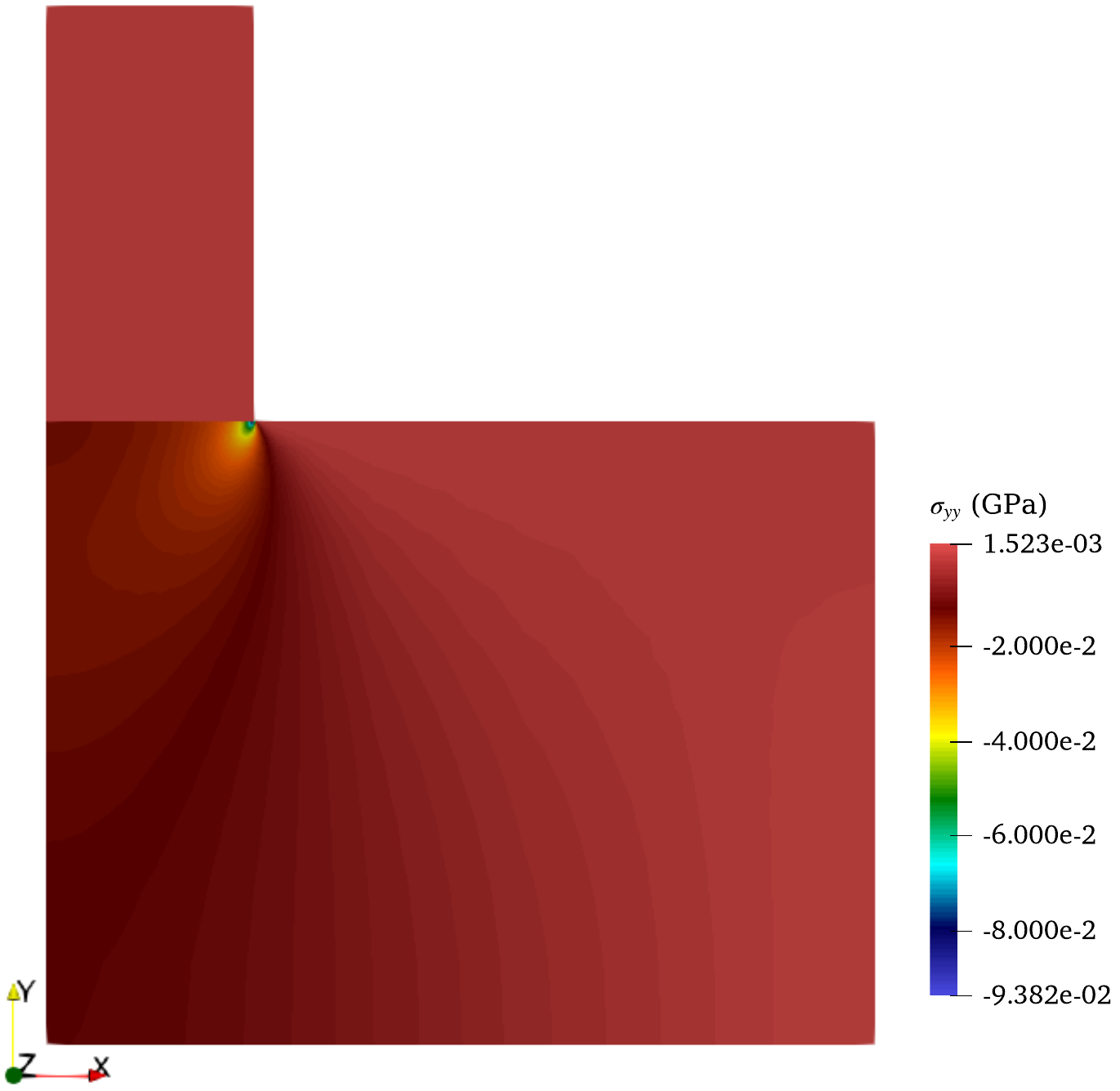}
        \caption{Distribution of stress $\sigma_{yy}$ }
    \end{subfigure}
    \caption{(a) and (b) show the meshes used for the punch test, and (c) shows the stress distribution in the block for the refined mesh with $f_s=40$.}
    \label{fig:punch_stress_distribution}
\end{figure}

Two nonconformal discretisation schemes are studied here - (a) coarse mesh shown in Fig.~\ref{fig:punch_coarse_mesh}, and (b) refined mesh shown in Fig.~\ref{fig:punch_refined_mesh}. Only the elements in the block have been subdivided for the case of the refined mesh, as the punch behaves in a rigid manner with a prescribed displacement applied to all of its nodes. For contact definition, three penalty scaling factors are studied here, i.e. $f_s=1, 10$ and $ 40$. 

\begin{figure}[htbp!]
    \begin{subfigure}[b]{0.48\linewidth}
        \centering
        \includegraphics[width=1.0\linewidth]{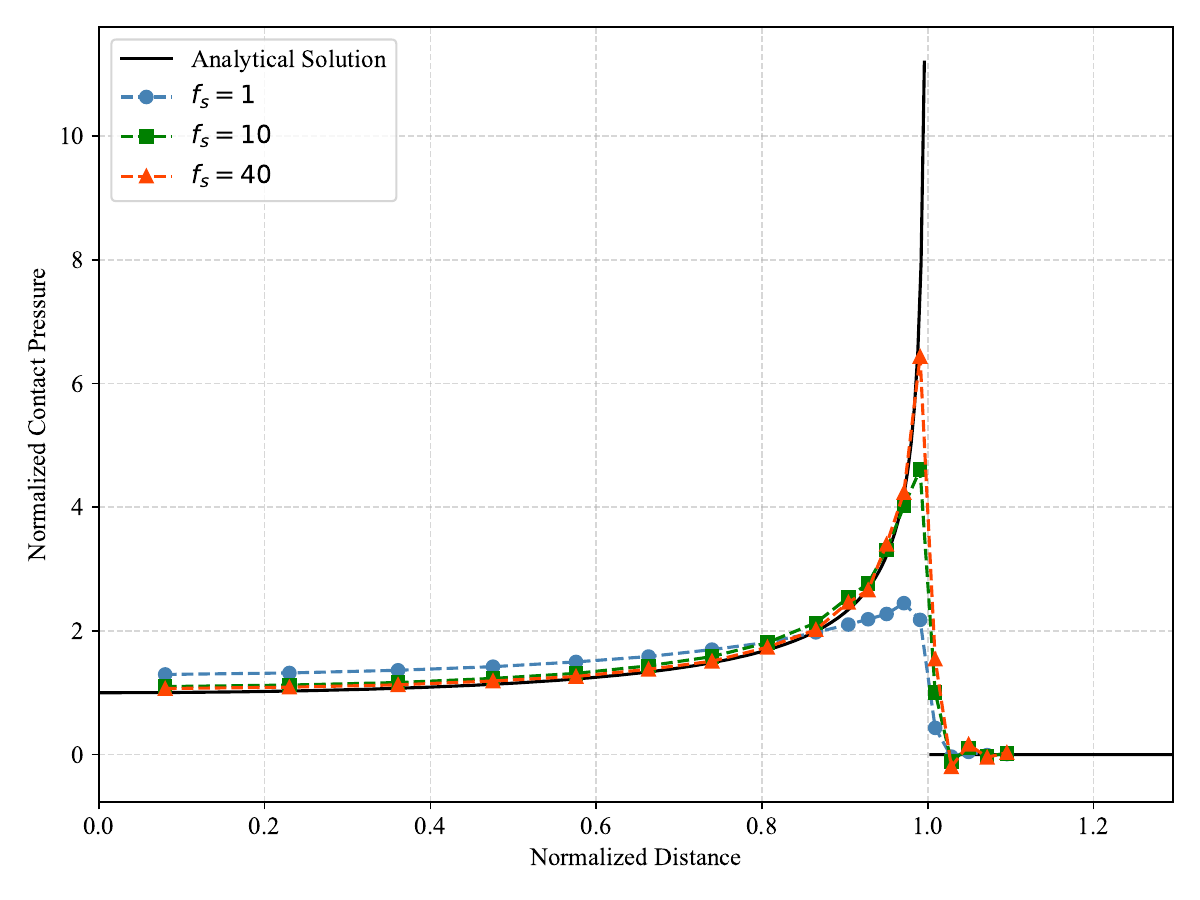}
        \caption{}
        \label{fig:punch_coarse_plot}
    \end{subfigure}
    \begin{subfigure}[b]{0.48\linewidth}
        \centering
        \includegraphics[width=1.0\linewidth]{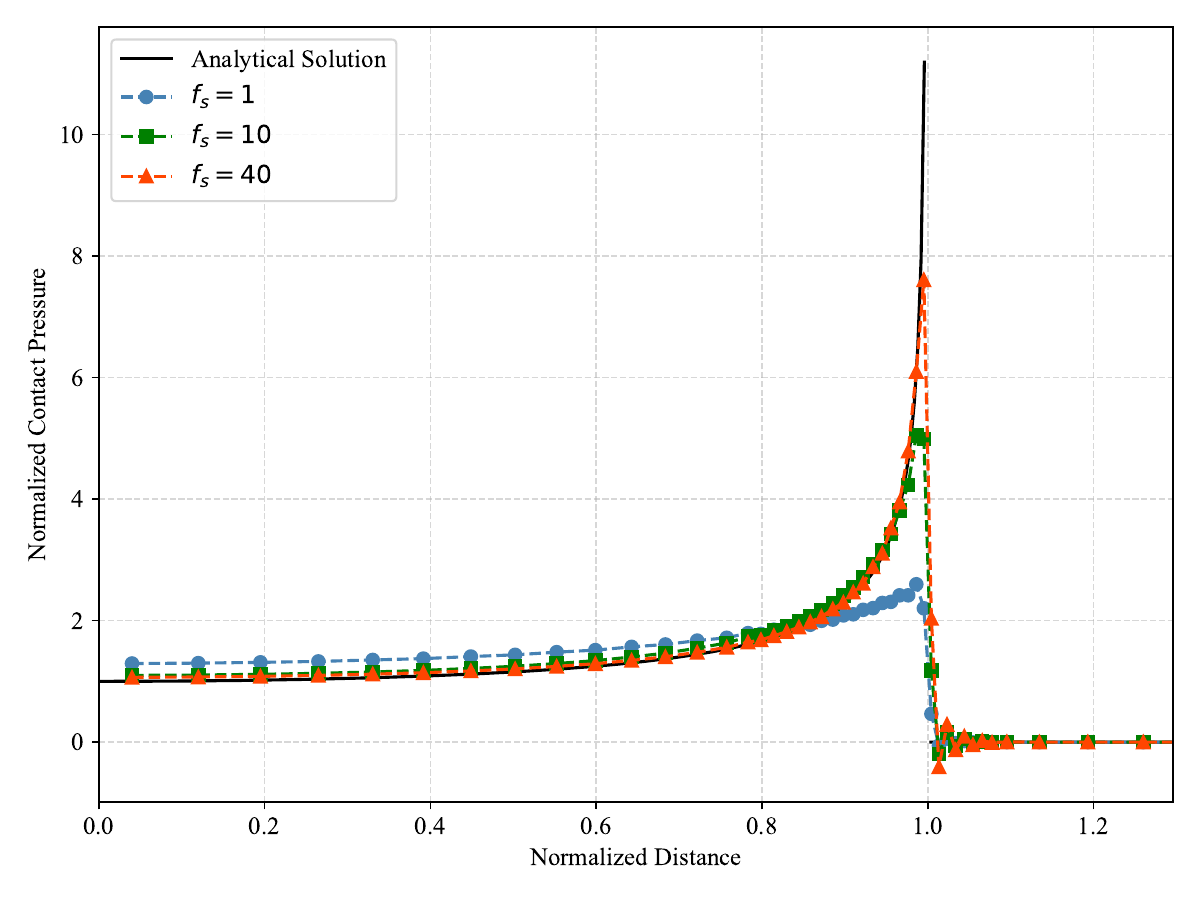}
        \caption{}
        \label{fig:punch_refined_plot}
    \end{subfigure}
    \caption{Plots for the coarse mesh (a) and the refined mesh (b), showing the variation of the normalised contact pressure with the normalised distance. }
    \label{fig:punch_both_plots}
\end{figure}

To compare the numerical solution in both discretisation schemes, the contact pressure is normalised with respect to the analytical contact pressure under the punch's axis in each case. The radial distance along the contact region is normalised with respect to the half-width of the punch. Fig.~\ref{fig:punch_stress_distribution} shows the stress distribution in the case of the refined mesh for the block. The coarse and refined mesh plots are shown in Fig.~\ref{fig:punch_both_plots}. In contrast to the Hertzian contact, using a lower penalty scaling factor in this test results in over-prediction of the contact pressure near the vertical axis of symmetry and under-prediction along the edges of the contact region. With the increase in the penalty scaling factor, the numerical solution converges towards the analytical solution and the numerical contact pressure near the edge becomes higher. When the refined mesh is used for this test, the increased number of elements results in better capturing of the contact pressure variation, and the normalised plots show an improvement over the coarse mesh for each penalty scaling factor. Thus, the proposed algorithm can be concluded to pass the indentation test.

\subsection{\textit{Self-contact test}}
    In some practical scenarios, the deformation of solid parts may result in contact between surfaces of the same original part, a phenomenon referred to as self-contact or auto-contact. This subsection investigates the developed methodology's capability to address such self-contact problems by assessing its performance for auto-detection and constraint application. The self-contact problem presented here entails self-contact in a boxed structure with cavities and cracks that deform under loading and also experience sliding.

\begin{figure}[!btp]
        \centering
        \begin{subfigure}{0.3\linewidth}
            \centering
            \includegraphics[width=\linewidth]{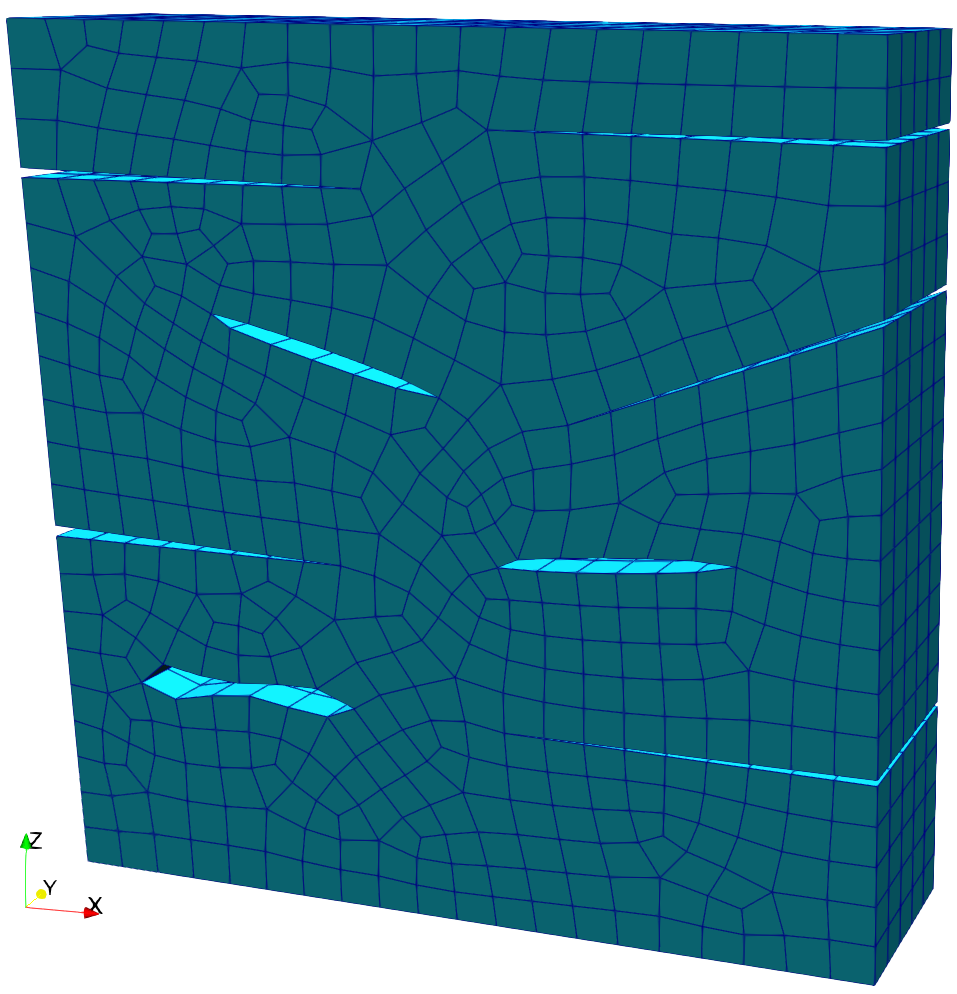}
            \caption{}
            \label{fig:self_contact_undeformed_box_with_cracks}
        \end{subfigure}
        \begin{subfigure}{0.4\linewidth}
            \centering
            \includegraphics[width=\linewidth]{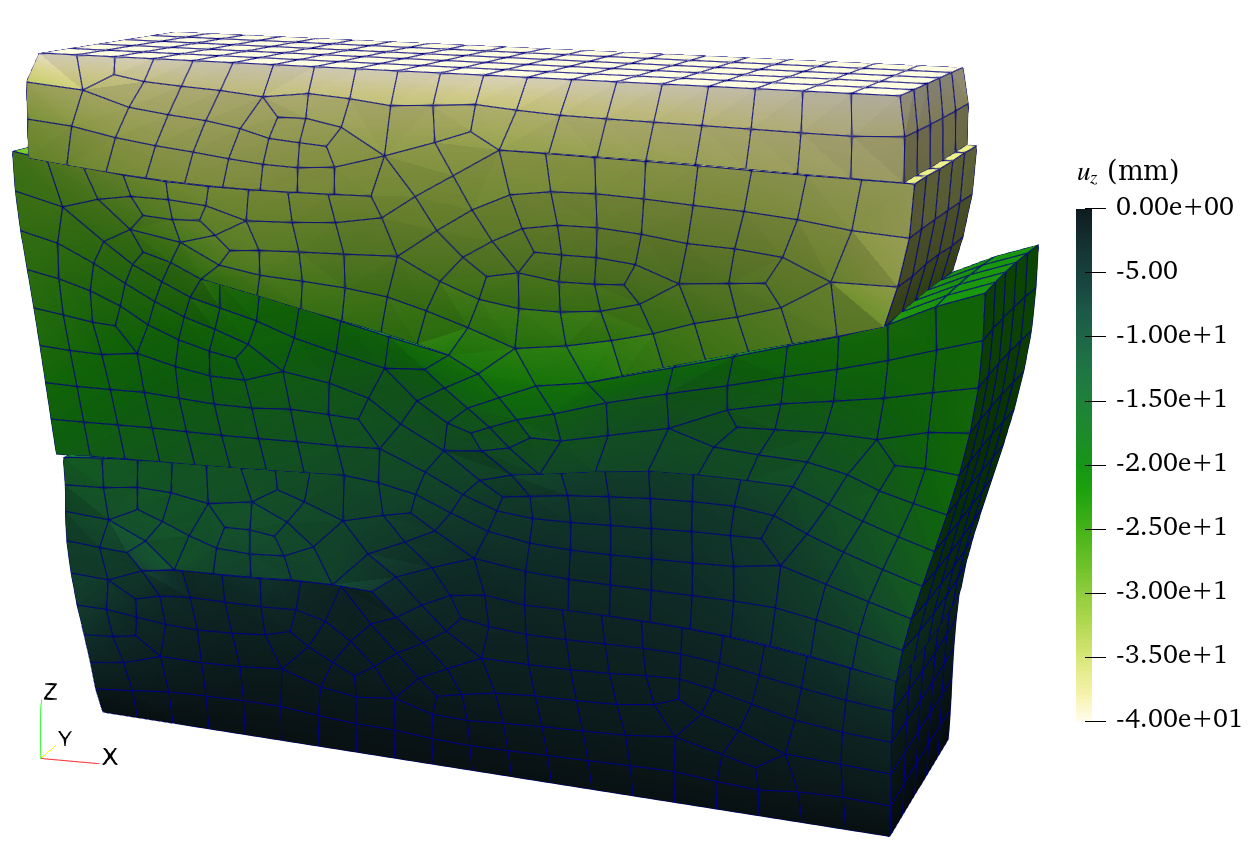}
            \caption{}
            \label{fig:self_contact_deformed_box_with_cracks}
        \end{subfigure}
        \caption{Self-contact in a block with presence of cracks: (a) undeformed geometry, (b) deformed geometry under downward displacement of the top surface.}
        \label{fig:self_contact_box_with_cracks}
    \end{figure}

    The discretised geometry of the box having size $200\times200\times50$ mm$^3$ is shown in Fig.~\ref{fig:self_contact_undeformed_box_with_cracks}. It is composed of isotropic elasto-plastic material with linear hardening having Young's modulus, E$=10$ GPa, Poisson's ratio set to 0, initial yield stress $\sigma_{y_0}=1.0$ GPa, and hardening modulus of H $=0.1$ GPa. The top surface of the box is uniformly pressed downwards by 40 mm while the bottom surface is held fixed. With deformation, cavities and cracks undergo interaction between their surfaces with auto-detection of contact, transferring load that results in deformation of the complete structure, Fig.~\ref{fig:self_contact_deformed_box_with_cracks}. Some domain regions also deform by sliding over the cracks due to the combined effect of material deformation and contact traction generated over contacting surfaces. Even for self-contact problems, contact search provides potential contact pairs based on the intersection of bounding boxes around all facets of the body, and contact detection finds oppositely facing facets with geometrical interpenetration. It should be noted that there is no redundancy problem at any contact edge \cite{pusoDualPassMortar2020} in the proposed method, as it applies contact constraints based on the continuous interpenetration between segments. The net distribution of nodal forces on the two opposite segments, sharing any edge, due to the contact traction might lead to net nodal forces acting on the edge nodes, which is an artefact of the mathematical nature of the elements and is inherently preserved in the proposed formulation to ensure the mathematical consistency.
    
    The self-contact tests presented above demonstrate the inherent fair treatment of surfaces while symmetrically evaluating contact tractions in just a single pass in each solution step. Also, the solution procedure here is similar to the contact between facets of different parts, with just the difference of contact search yielding potential contact facet pairs from the same parts.

\subsection{\textit{Impact between two bars}}
For evaluating the algorithm's performance in dynamic environments, two cases of symmetric collision between bars are considered - elastic bars with a flat interface and elasto-plastic bars with a curved interface. 

\subsubsection{Elastic symmetric collision}\label{sec:elastic_symmetric_collision}

\begin{figure}[!tbhp]
    \begin{subfigure}[b]{0.48\linewidth}
        \centering
        \includegraphics[width=1.0\linewidth]{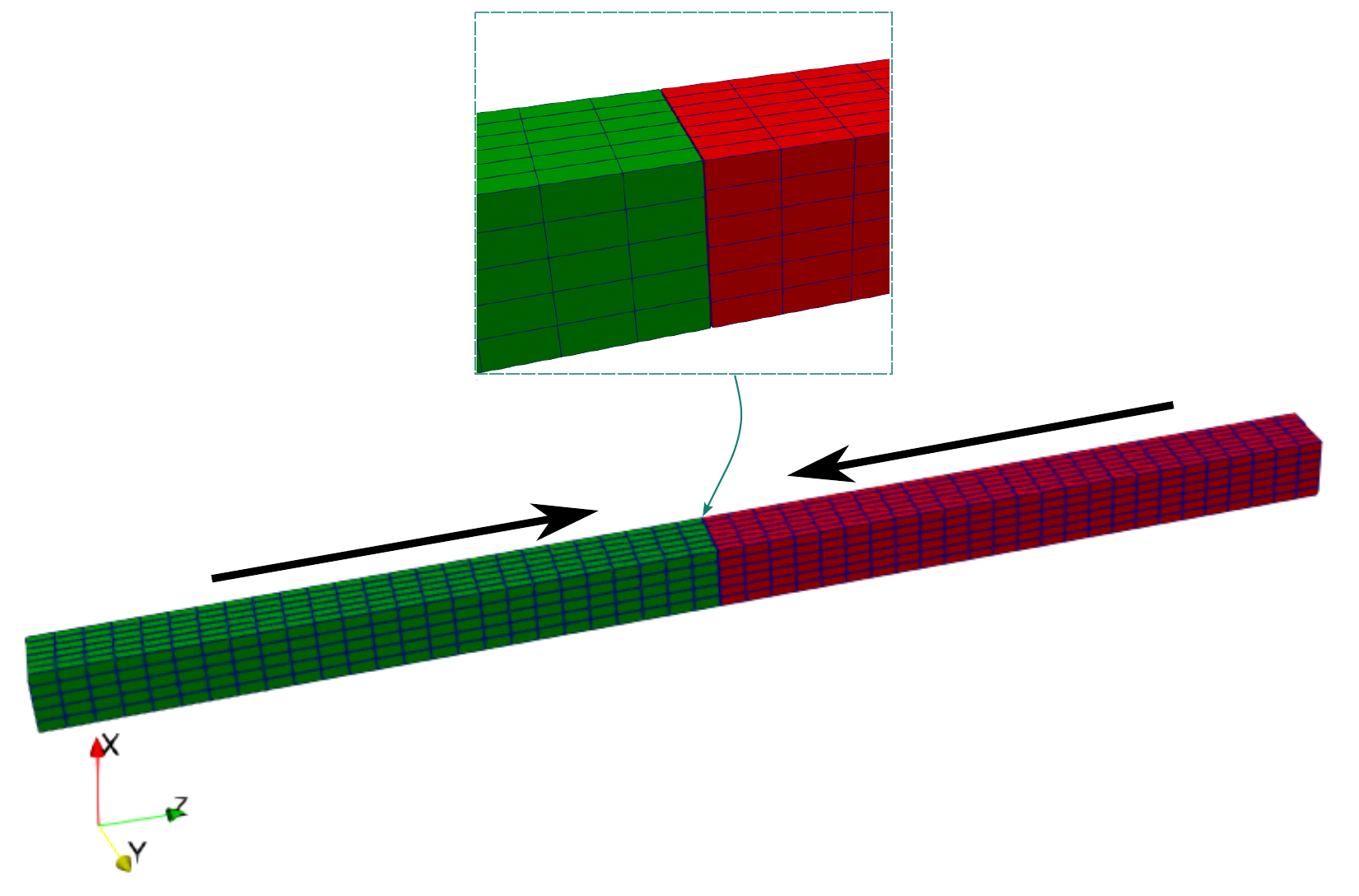}
        \caption{}
        \label{fig:bar_to_bar_impact_a}
    \end{subfigure}
    \begin{subfigure}[b]{0.48\linewidth}
        \centering
        \includegraphics[width=1.0\linewidth]{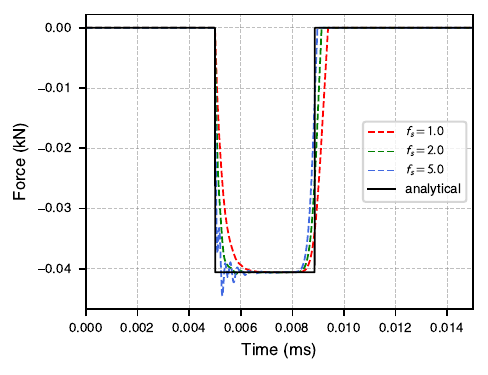}
        \caption{}
    \end{subfigure}
    
    \begin{subfigure}[b]{0.46\linewidth}
        \centering
        \includegraphics[width=1.0\linewidth]{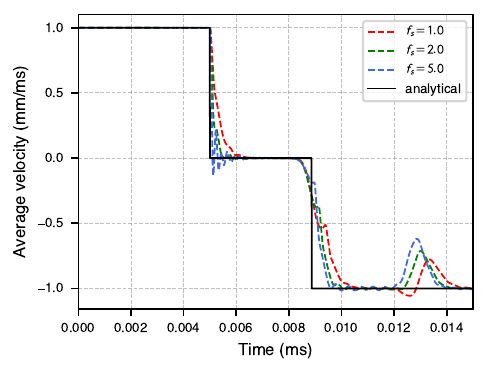}
        \caption{}
    \end{subfigure}
    \begin{subfigure}[b]{0.50\linewidth}
        \centering
        \includegraphics[width=1.0\linewidth]{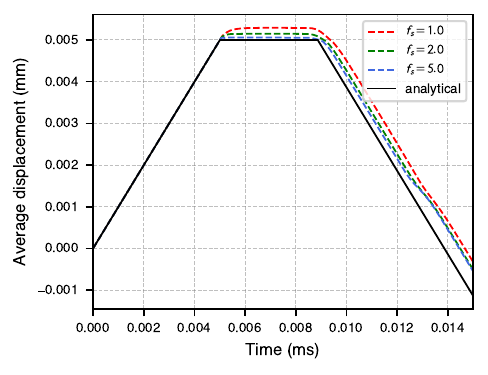}
        \caption{}
    \end{subfigure}    
    \caption{Impact between two bars : (a) discretisation in both bars, (b) variation of force on contacting face of the left bar, (c) variation of the average velocity of contacting face of the left bar, and (d) variation of average displacement of contacting face of the left bar.}
    \label{fig:bar_to_bar_impact}
\end{figure}

Two bars with equal and opposite velocities along the same direction are considered for collision \cite{carpenterLagrangeConstraintsTransient1991}\cite{kwonFullyNonlinearThreedimensional2023}. The dimensions of both bars are $1\times1\times10  \text{mm}^3$, and both are discretised to have non-matching meshes at the contact interface during the collision, as shown in Fig.~\ref{fig:bar_to_bar_impact_a}. Initially, both bars have a gap of 0.01 mm between them and an equal speed of 1.0 mm/ms along their axis. The elastic material in both bars has Young's modulus of 210 GPa, Poisson's ratio $\nu=0$, and density $7.85\times10^{-6}$ kg/mm$^3$. For contact interaction, three different penalty scaling factors are considered. Artificial bulk viscosity is used to smooth out oscillations caused by high-velocity gradients \cite{danielsonCurvedNodetofaceContact2022}\cite{danielsonReliableSecondorderHexahedral2011}\cite{danielsonSecondorderFiniteElements2016}\cite{browningHigherorderFiniteElements2020}\cite{browningSecondorderPyramidElement2023}, which arise from higher penalties. The linear and quadratic factors are taken as 0.6 and 1.5, respectively. While a linear factor of 0.06, commonly used in practice, is sufficient for cases with $f_s=1,2$, significant oscillations were observed for a higher penalty of $f_s=5$. These are effectively smoothed by increasing the linear factor to 0.6. Plots for the variation of contact force, average velocity and average displacement for nodes on the contacting face of the left bar are shown in Fig.~\ref{fig:bar_to_bar_impact}. While the contact force attains a constant analytical value for the duration of contact, the numerical solution has a gradual change in the contact force due to the penalisation of gradual interpenetration in the two bars. Similarly, a gradual change is observed in the average velocity of the contacting face of the left bar during contact, and some oscillations persist even after the collision. The overshoot in the average displacement of the contacting face of the left bar above the analytical solution during contact indicates penetration between the two bars, resulting in a lag in displacement in the further numerical solution. In all results, using a higher penalty scaling factor brings the numerical solution closer to the analytical solution, but it suffers from increased oscillations, as has also been reported in \cite{zangContactAlgorithm3D2011}. Higher penalties also decrease the size of stable time step increments and require increased artificial bulk viscosity. The overall deviation of the numerical solution compared to the analytical solution is an artefact of the penalty method, as can also be seen in \cite{carpenterLagrangeConstraintsTransient1991}. However, the use of non-matching meshes in the presented solution proves the validity of the proposed method in general contact problems.

\subsubsection{Inelastic collision between two bars with curved interface}
\begin{figure}[!tbhp]
    \centering
    \begin{subfigure}{0.55\linewidth}
        \centering
        \includegraphics[width=\linewidth]{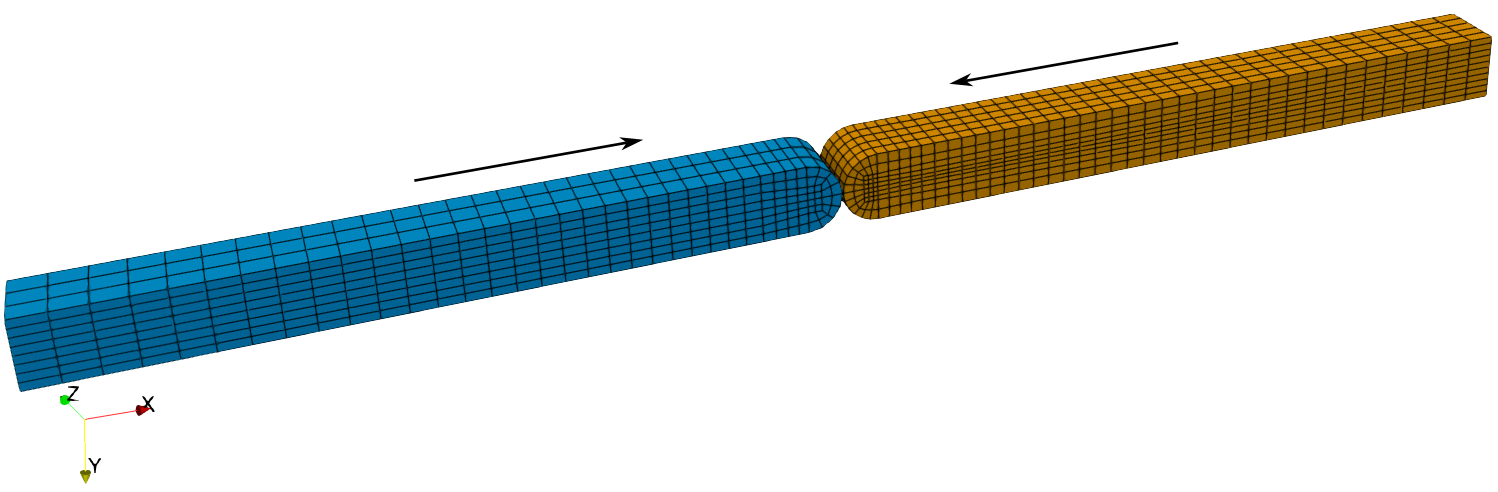}
        \caption{}
        \label{fig:two_bar_inelastic_collision_full_undeformed_dissimilar_mesh}
    \end{subfigure}
    \begin{subfigure}{0.20\linewidth}
        \centering
        \begin{subfigure}{\linewidth}
            \centering
            \includegraphics[width=\linewidth]{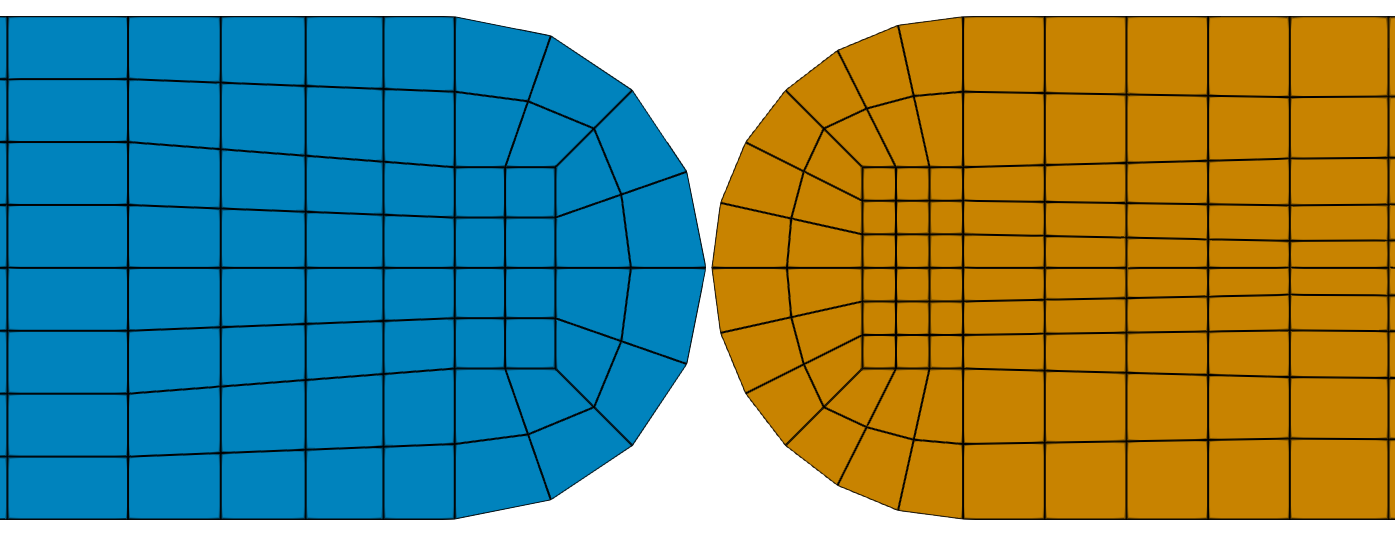}
            \caption{}
            \label{fig:two_bar_inelastic_collision_nearcontact_undeformed_dissimilar_mesh}
        \end{subfigure}
        \begin{subfigure}{\linewidth}
            \centering
            \includegraphics[width=\linewidth]{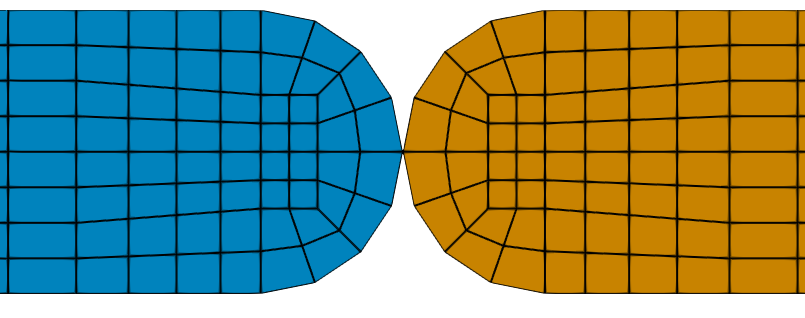}
            \caption{}
            \label{fig:two_bar_inelastic_collision_nearcontact_undeformed_similar_mesh}
        \end{subfigure}
    \end{subfigure}    
    \caption{Inelastic collision: (a) geometry of the two bars moving towards each other, (b) dissimilar discretisation and (c) same discretisation of curved surfaces.}
    \label{fig:inelastic_collision_undeformed_meshes}
\end{figure}

\begin{figure}[!tbhp]
    \centering
    \begin{subfigure}{0.49\linewidth}
        \centering
        \includegraphics[width=\linewidth]{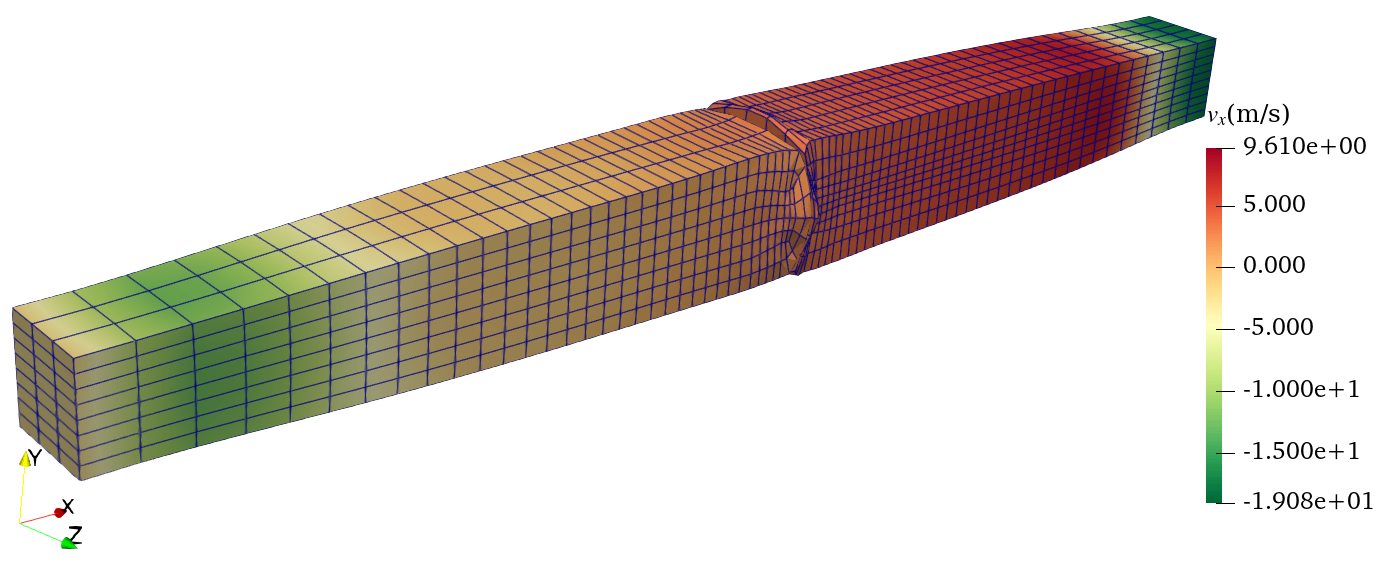}
        \caption{}
        \label{fig:two_bar_inelastic_collision_full_deformed_dissimilar_mesh_t_0_0139}
    \end{subfigure}
    \begin{subfigure}{0.49\linewidth}
        \centering
        \includegraphics[width=\linewidth]{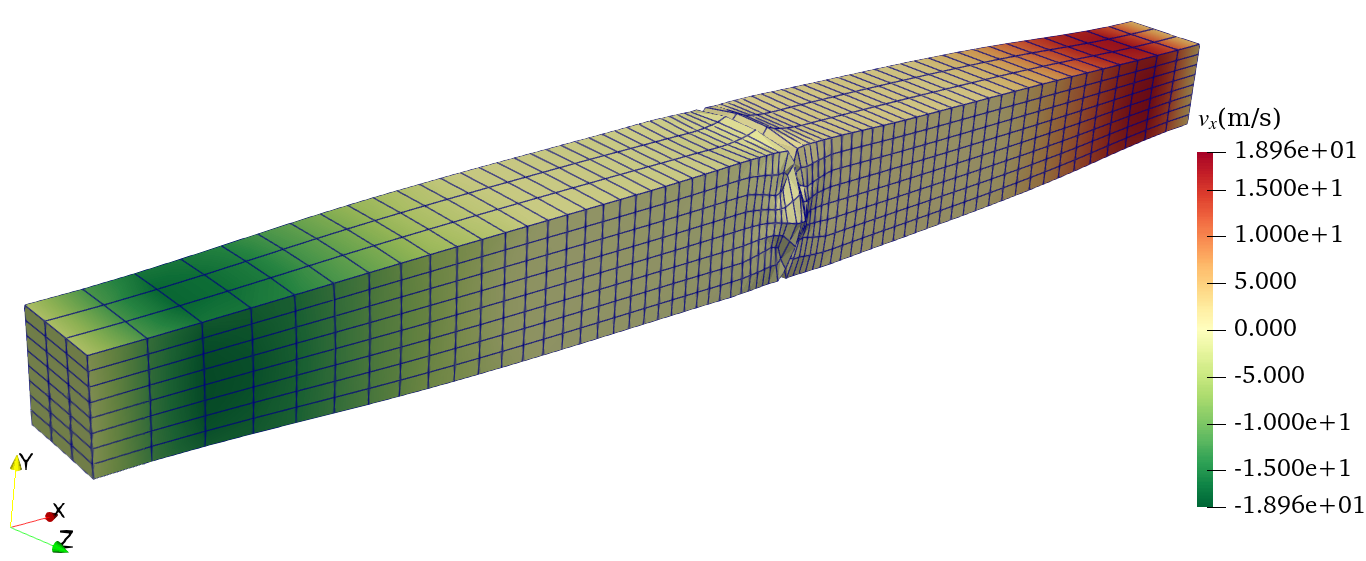}
        \caption{}
        \label{fig:two_bar_inelastic_collision_full_deformed_similar_mesh_t_0_0139}
    \end{subfigure}    
    \begin{subfigure}{0.70\linewidth}
        \centering
        \includegraphics[width=\linewidth]{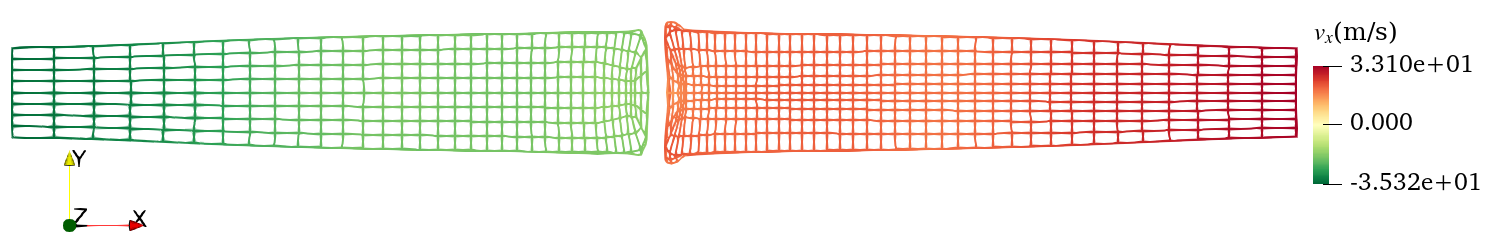}
        \caption{}
        \label{fig:two_bar_inelastic_collision_full_deformed_dissimilar_mesh_t_final_wireframe}
    \end{subfigure}    
    \begin{subfigure}{0.70\linewidth}
        \centering
        \includegraphics[width=\linewidth]{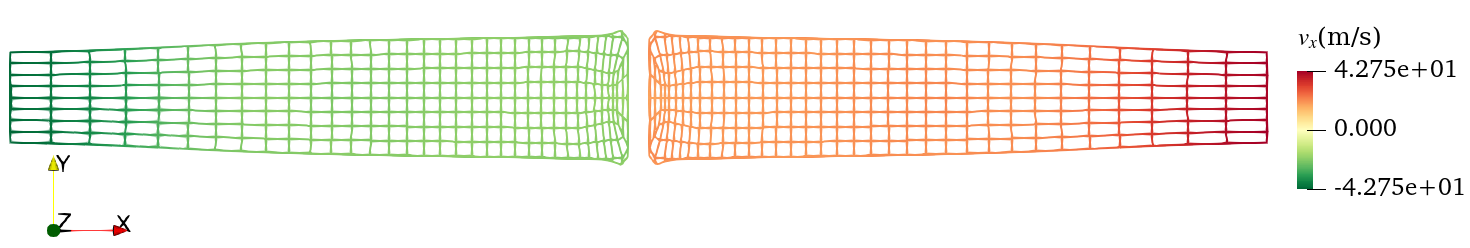}
        \caption{}
        \label{fig:two_bar_inelastic_collision_full_deformed_similar_mesh_t_final_wireframe}
    \end{subfigure}
    \caption{Inelastic collision: bars with (a) dissimilar and (b) same discretisation undergoing deformation at around t=0.0139 ms. The same bars upon rebound are shown for (c) dissimilar and (d) the same discretisation.}
    \label{fig:inelastic_collision_deformed_meshes}
    
\end{figure}

This test, inspired by the symmetric Taylor impact test, considers the high-speed symmetric collision between two elasto-plastic bars (linear-hardening) having a curved interface, see Fig.~\ref{fig:two_bar_inelastic_collision_full_undeformed_dissimilar_mesh}. Both bars have Young's modulus E $=69$ GPa, Poisson's ratio $\nu=0.33$, density $\rho=2.71\times10^{-6}$ kg/mm$^3$, initial yield stress $\sigma_{y_0}=0.276$ GPa, hardening modulus H $ =0.2$ GPa, and are moving co-axially towards each other with a speed of $v=330$ m/s. The maximum dimensions of the two bars are the same, $1\times1\times10.5$ mm$^3$, with the circular geometry having a diameter of 1 mm. 
The deformation of both bars with identical properties and symmetrical initial conditions should be symmetrical. However, to study the finite element deformation, two cases are considered here: (a) dissimilar discretisation, Fig.~\ref{fig:two_bar_inelastic_collision_nearcontact_undeformed_dissimilar_mesh}, and (b) same discretisation of meshes, Fig.~\ref{fig:two_bar_inelastic_collision_nearcontact_undeformed_similar_mesh}. 
In both cases, the two bars undergo plastic deformation upon collision, with lateral bulging and a gradual transition from initial edge contact to full surface-to-surface contact. Deformed states at around $t=0.0139$ ms are shown in Figs. \ref{fig:two_bar_inelastic_collision_full_deformed_dissimilar_mesh_t_0_0139}, \ref{fig:two_bar_inelastic_collision_full_deformed_similar_mesh_t_0_0139}. The symmetric mesh results in symmetric velocity and deformation fields, as expected. In contrast, the dissimilar mesh exhibits asymmetric deformation.
Deformation seen in the two meshes near the contact region is asymmetric or the same, depending on whether the two bars have dissimilar or the same discretisation, respectively, see Figs. \ref{fig:two_bar_inelastic_collision_full_deformed_dissimilar_mesh_t_final_wireframe}, \ref{fig:two_bar_inelastic_collision_full_deformed_similar_mesh_t_final_wireframe}.

The asymmetric deformation arises primarily from differences in the distribution of nodes. In the dissimilar discretisation case, the finely meshed bar has a higher non-axial nodal mass than the opposite, coarser bar. Additionally, first-order elements cannot accurately represent curvature, so the contacting discretised segments on both bars deviate asymmetrically from the original circular geometry. Since this is a high-speed impact problem, inertia effects are significant, leading to asymmetric plastic deformation and the formation of a concave-convex interface. In this case, the undeformed finer mesh also better captures the original curvature with its smaller segments, resulting in a slightly higher total mass compared to the opposite bar.

\begin{figure}[!tbhp]
    \centering
    \begin{subfigure}{0.40\linewidth}
        \includegraphics[width=\linewidth]{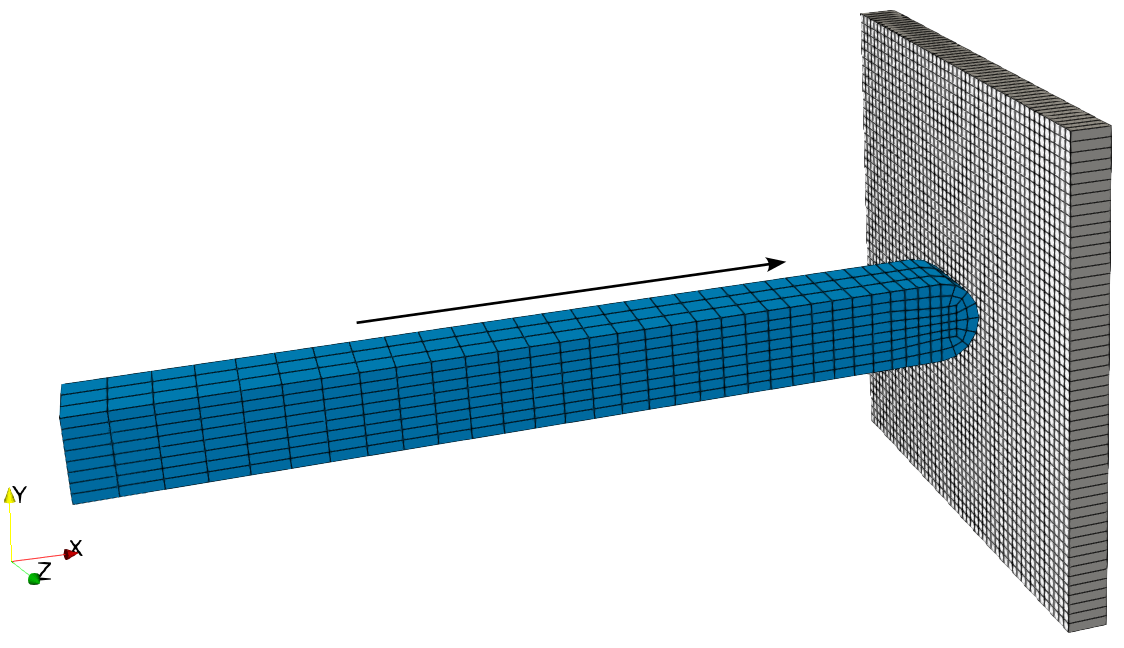}
        \caption{}
        \label{fig:taylor_impact_nonmatching_mesh_problem}
    \end{subfigure}
    \hspace{0.12\linewidth}
    \begin{subfigure}{0.20\linewidth}
        \includegraphics[width=\linewidth]{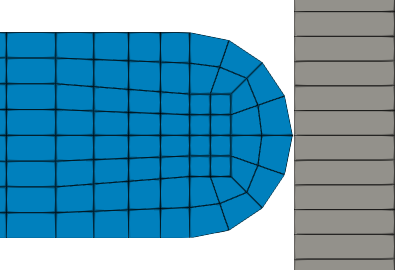}
        \caption{}
        \label{fig:taylor_impact_mesh_near_contact_view}
    \end{subfigure}

    \begin{subfigure}{0.30\linewidth}
        \includegraphics[width=\linewidth]{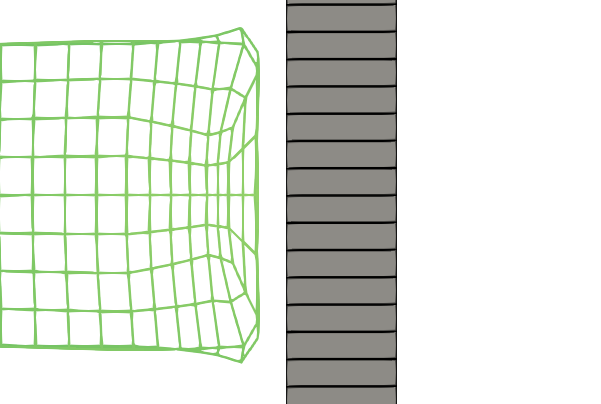}
        \caption{}
        \label{fig:taylor_impact_near_contact_deformed_wireframe_t_final}
    \end{subfigure}
    \hspace{0.03\linewidth}
    \begin{subfigure}{0.30\linewidth}
        \includegraphics[width=\linewidth]{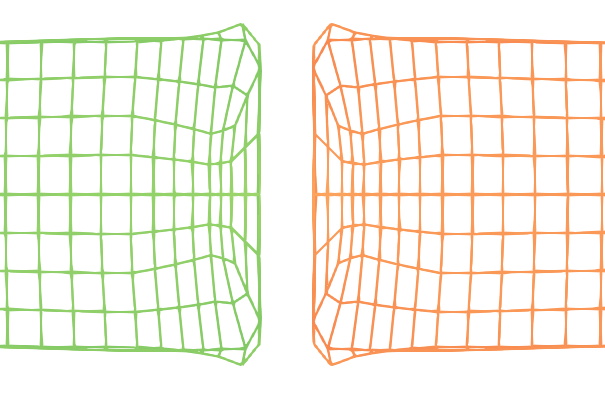}
        \caption{}
        \label{fig:two_bar_inelastic_collision_nearcontact_deformed_similar_mesh_t_final_wireframe}
    \end{subfigure}
    \hspace{0.05\linewidth}
    \begin{subfigure}{0.30\linewidth}
        \includegraphics[width=\linewidth]{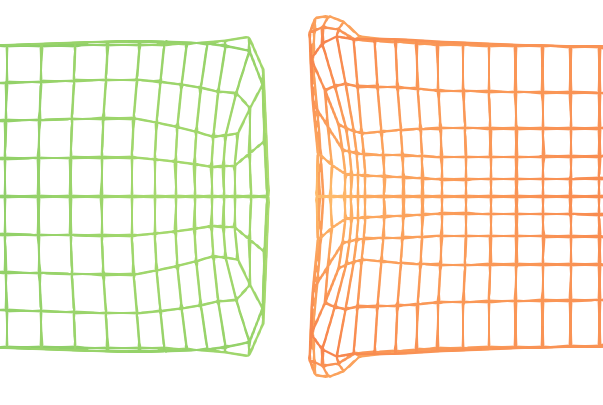}
        \caption{}
        \label{fig:two_bar_inelastic_collision_nearcontact_deformed_dissimilar_mesh_t_final_wireframe}
    \end{subfigure}    
    \caption{Single rod impact with rigid wall: (a) mesh used with differences in discretisation on rod and the wall, (b) showing its magnified view near contact region, (c) deformation upon rebound, compared with the deformation upon rebound for two rods with (d) same discretisation, (e) dissimilar discretisation.}
\end{figure}

To ascertain that the formation of curved interface in the case of dissimilar discretisation is not due to any inaccuracy in the contact algorithm, an additional test is performed where the coarsely meshed rod is impacted over a rigid wall, see Fig.~\ref{fig:taylor_impact_nonmatching_mesh_problem}. The rod deforms on collision and its configuration changes from Fig.~\ref{fig:taylor_impact_mesh_near_contact_view} to that shown in Fig.~\ref{fig:taylor_impact_near_contact_deformed_wireframe_t_final}, attaining a flat interface. This flat interface matches with the collision between similarly discretised meshes, Fig.~\ref{fig:two_bar_inelastic_collision_nearcontact_deformed_similar_mesh_t_final_wireframe} and is in contrast to the curved interface seen in the collision between dissimilarly discretised bars, Fig.~\ref{fig:two_bar_inelastic_collision_nearcontact_deformed_dissimilar_mesh_t_final_wireframe}.

For meaningful results in symmetric Taylor impact tests, highly refined meshes are typically used, unlike the coarser meshes used here for both the same and dissimilar discretisations. The coarse meshes in this work are used intentionally to study finite element behaviour. The case with the same discretisation produces symmetric deformation, demonstrating the high accuracy of the contact method in large deformation and high strain rate problems. In contrast, the asymmetric response in the case with dissimilar discretisation highlights the importance of using similar discretisation to avoid artificial asymmetry in large deformation contact problems with curved surfaces.

\subsection{\textit{Oblique collision of two deformable cylinders}}
\begin{figure}[htbp]
    \centering
    \begin{subfigure}[b]{0.3341179807146909\linewidth}
        \centering
        \includegraphics[width=\linewidth]{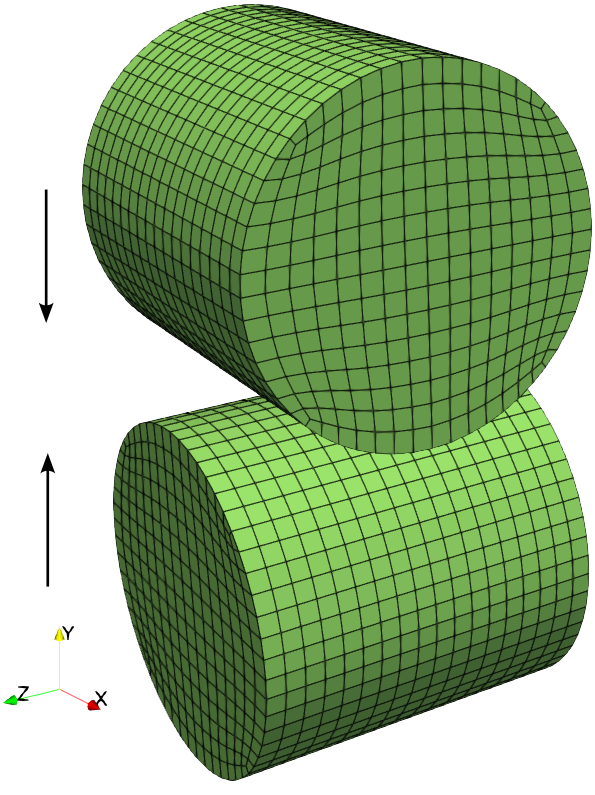}
        \caption{}
        \label{fig:two_deformable_cylinders_oblique_impact_mesh}
    \end{subfigure}
    \begin{subfigure}[b]{0.27852524106636417\linewidth}
        \centering
        \includegraphics[width=\linewidth]{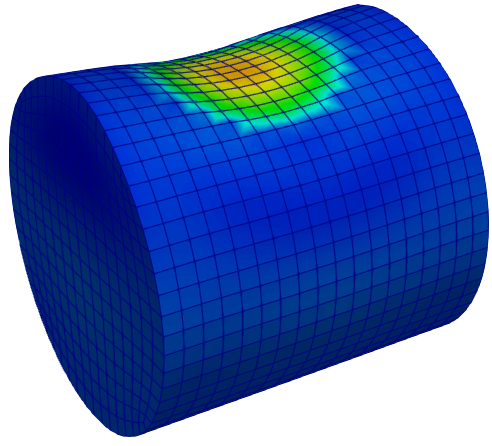}
        \caption{}
        \label{fig:two_deformable_cylinders_oblique_impact_cylinder1_stressyy_t_point055_legendastpt115}
    \end{subfigure}
    \begin{subfigure}[b]{0.37735677821894503\linewidth}
        \centering
        \includegraphics[width=\linewidth]{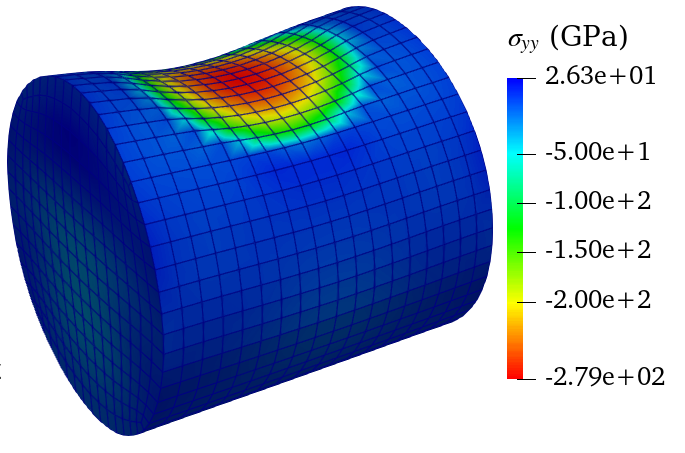}
        \caption{}
        \label{fig:two_deformable_cylinders_oblique_impact_cylinder1_stressyy_t_point115}
    \end{subfigure}
   
    \caption{Oblique collision of two deformable cylinders: (a) mesh, stress wave propagation at time (b) $t=0.055$ ms, and (c) $t=0.115$ ms.}
    \label{fig:oblique_collision_deformable_cylinders_meshes}
\end{figure}

For multibody benchmarking, a cylinder-to-cylinder oblique impact problem is tested to show deformable-to-deformable contact \cite{ottoExplicitDynamicsImpact2020}. Here, two cylinders, each having a radius of 1 mm and a length of 2 mm, are taken in a crossed configuration as shown in Fig.~\ref{fig:two_deformable_cylinders_oblique_impact_mesh}. Both are composed of the same material with Young's modulus E$=10^3$ GPa, Poisson's ratio $nu=0.2$, density $\rho=1$ kg/mm$^3$ and moving towards each other with a speed of $2.5$ m/s. With a scaling factor of $f_s=100$ in contact definition, and linear and quadratic factors of bulk viscosity being 0.6 and 1.5, respectively, the subsequently deformed configurations and wave propagation in the lower cylinder are shown in Figs.~\ref{fig:two_deformable_cylinders_oblique_impact_cylinder1_stressyy_t_point055_legendastpt115}, \ref{fig:two_deformable_cylinders_oblique_impact_cylinder1_stressyy_t_point115}.

\begin{figure}[htbp]
    \centering
    \begin{subfigure}[b]{0.3\linewidth}
        \centering
        \includegraphics[width=\linewidth]{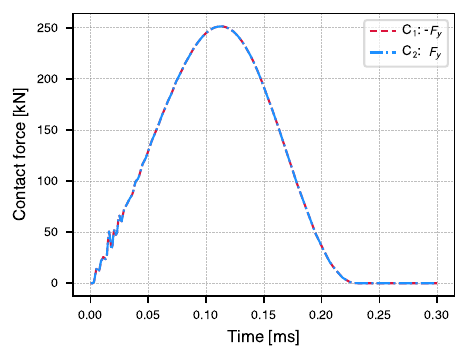}
        \caption{}
        \label{fig:two_cylinder_impact_contact_force_Fy}
    \end{subfigure}
    \begin{subfigure}[b]{0.3\linewidth}
        \centering
        \includegraphics[width=\linewidth]{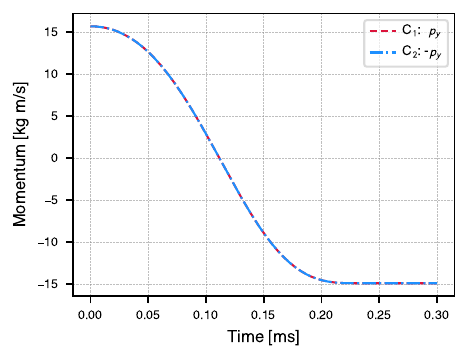}
        \caption{}
        \label{fig:two_cylinder_impact_momentum_y}
    \end{subfigure}    
    \begin{subfigure}[b]{0.3\linewidth}
        \centering
        \includegraphics[width=\linewidth]{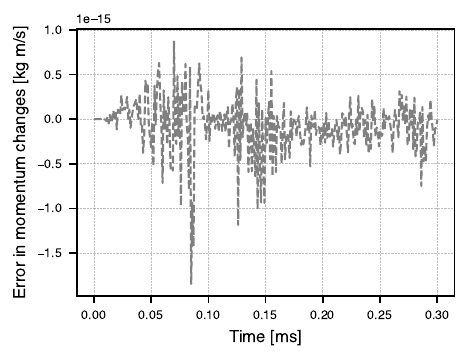}
        \caption{}
        \label{fig:two_cylinder_impact_momentum_z_error}
    \end{subfigure}    
    \caption{Oblique collision of two deformable cylinders: variation of (a) net force $F_y$, (b) momentum, and (c) differences in momentum of the two bodies with time.}
    \label{fig:oblique_collision_deformable_cylinders_plots}
\end{figure}

The variation of the contact force $F_y$ and the momentum of both cylinders are shown in Figs. \ref{fig:two_cylinder_impact_contact_force_Fy}, \ref{fig:two_cylinder_impact_momentum_y}, respectively. As can be seen with the plots of both quantities, the contact force and the momentum of both cylinders are symmetrically exchanged, and the error in the momentum variation ($p_y^1 + p_y^2$) is negligible, see Fig.~\ref{fig:two_cylinder_impact_momentum_z_error}. While not shown, momentum gained in the x and y directions is also negligible. 

\section{Conclusion}\label{sec_conclusion}
This work presents a robust contact processing methodology for evaluating contact tractions using a midplane-based approach for segment-to-segment contact pairs, enforcing contact constraints symmetrically using the penalty method, ensuring an unbiased treatment of the surfaces coming into contact. Consequently, it obviates the need for a second pass, generally applied to deal with the issue of biasing, while maintaining the force-moment equilibrium. While the computational cost of dual-pass scales as $2n$ with $n$ number of interacting contact pairs, the cost of the single-pass approach scales as $n$, providing an advantage for large-scale simulations. The development of a mathematical framework describing this approach in both continuum and discrete settings of the contact problems has been presented in this work, with a special focus on the algorithmic implementation of the methodology. An extensive exposition of the geometrical configurations of the facets coming into contact has been showcased, which allows for the continuous enforcement of contact constraints with high accuracy. The formulation has been rigorously tested through multiple examples for contact, including flat surfaces, curved surfaces, and sharp corners in contact for non-matching meshes. 
It successfully passed the contact patch test with an accuracy level matching the finite element accuracy. The two-beam test demonstrated its locking-free behaviour. The extensive study of Hertzian contact with different materials and nonconformal meshes highlighted the method's capability to represent contact between curved surfaces. Also, it emphasised the importance of allowing and controlling interpenetration for smoothness in contact traction between discretised curved surfaces. The method showed convergence towards analytical solutions in the Hertzian contact and indentation problem with increasing penalty factor. The proposed method's versatility also extends to problems of self-contact, where its inherent unbiased nature helps to provide accurate solutions while working in a single pass. Lastly, dynamic problems of coaxial collision between two elastic and inelastic rods with different meshes and oblique collision of two cylinders demonstrated its performance in dynamic contact problems. The inelastic collision also highlighted the importance of having similar discretisations of meshes undergoing large deformation problems, particularly those where mass distribution also affects the resulting solution. Numerical solutions to contact problems obtained through this method can be improved by choosing a suitably high penalty scaling factor and mesh refinement in the contact region. 

Penalty methods are known to be sensitive to the choice of penalty parameter. However, the proposed methodology provides high-accuracy solutions within the framework of pure penalty-based constraints. Unlike conventional penalty methods working to directly apply concentrated forces at nodes, this work applies distributed forces over the contacting segments by penalisation of true interpenetration by a constant penalty dependent upon material stiffness. This close consistency with the continuum description leads to higher accuracy in solutions. 
Based on the numerical tests studied in this work, a scaling factor of unity is sufficient for most general problems. For improved accuracy, it can be increased to 10 in static problems and to between 2 and 5 in dynamic problems with appropriate bulk viscosity. For materials with high inertia and low stiffness, a higher penalty is required. Overall, the choice of penalty depends on the contact problem and material properties such as stiffness and inertia. While only a linear penalty is considered here, future studies could explore nonlinear penalties that adaptively adjust with increasing interpenetration, potentially obviating the scaling factor.

For future work, this contact interaction methodology is being studied with frictional effects to formulate a tangential traction evaluation that preserves the unbiased nature of the contact. Additionally, the methodology is being extended to contact between discretised domains composed of elements with higher-order interpolation, which can better describe complex surfaces.

\section{Acknowledgement}
The authors extend their gratitude to the Engineering and Physical Sciences Research Council (EPSRC) and Rolls-Royce for providing financial support for this research through the Strategic Partnership in Computational Science for Advanced Simulation and Modelling of Engineering Systems - ASiMoV (EPSRC Reference EP/S005072/1).

\appendix

\section{Explicit time integration scheme}
    The original problem of governing equations is considered in a discretised setting with domains $\Omega_h^i \approx \Omega^i$. In the finite element framework, these discretised domains are composed of multiple finite elements -
    \begin{align}
        \Omega_h^i = \bigcup_{j=1}^{m} \Omega_{h_j}^i
    \end{align}
    where $m$ represents the total number of finite elements. In this work, the numerical solution of the weak form of the dynamic equilibrium over discrete domains utilises an explicit time integration scheme as presented in \cite{zienkiewiczFiniteElementMethod2014}\cite{batheKJBatheFinite2014}\cite{belytschkoNonlinearFiniteElements2014}. The virtual work equation is applied over constituting finite elements subdomains that represent the original geometry of the solids. For the solid $\Omega_h^i$, it can be written as -
    \begin{equation}
        \sum_{j=1}^{m} \left( \int_{\Omega_{h_j}^i}\rho\dmat{N}^{T}\dmat{N}\dmat{\overset{..}{u}}dV+\int_{\Omega_{h_j}^i}\dmat{B}^{T}\boldsymbol{\sigma}dV-\int_{\Omega_{h_j}^i}\dmat{N}^{T}\tens{b}dV-\int_{\gamma_{h_j}}\dmat{N}^{T}\tens{t}d\gamma \right)=0
    \end{equation}
    where,  $\textbf{N}$ is the matrix containing shape functions, $\textbf{B}$ is the strain-displacement interpolation matrix, $\dmat{\overset{..}{u}}$ is the acceleration vector, and $\tens{t}$ is the traction vector that also includes the contact traction for $\Omega_h^i$. The summation is implied over all the $m$ elements in their respective domain $\Omega_{h_j}^i$. Considering this equation over the entire domains of both $\Omega_h^1$, $\Omega_h^2$, and converting it into a set of coupled equations renders a semi-discrete problem in the form of a set of ordinary differential equations -
    \begin{equation}
        \dmat{M}\dmat{\ddot{u}} + \dmat{f}_{\text{int}} = \dmat{f}_{\text{ext}}
        \label{eq: dynamic_equilibrium_matrix_eqn}
    \end{equation}
    where, $\textbf{M}$ is the mass matrix, $\dmat{f}_{\text{int}}$ is the vector of internal nodal forces, and $\dmat{f}_{\text{ext}}$ is the vector of external nodal forces. To decouple the set of equations, a suitable mass lumping scheme can be utilised for practical computational efforts \cite{duczekMassLumpingTechniques2019}\cite{browningHigherorderFiniteElements2020}. This work considers the rate of deformation together with the Jaumann stress rate to update the stress state \cite{dunneIntroductionComputationalPlasticity2005}\cite{belytschkoNonlinearFiniteElements2014} and evaluate internal forces.   
    Using the central difference time integration method, the time-based evolution of the kinematical quantities can be written as 
    \begin{equation}
        \textbf{u}^{n+1} = \textbf{u}^n + \Delta t \dmat{\dot{u}}^{n+\frac{1}{2}}, \ \  \\
        \dmat{\dot{u}}_{n+\frac{1}{2}}= \dmat{\dot{u}}_{n-\frac{1}{2}}+\dmat{\ddot{u}}_n\Delta t , \ \  \\
        \dmat{\ddot{u}}_n = \frac{   \dmat{u}_{n+1} - 2\dmat{u}_n + \dmat{u}_{n-1}     }{(\Delta t)^2}
        \\
    \end{equation}

    In eq.~\ref{eq: dynamic_equilibrium_matrix_eqn}, the external force vector $\dmat{f}_{\text{ext}}$ contains the forces due to the Neumann boundary conditions specifying the surface tractions applied on the boundary, as well as the effect of Dirichlet boundary conditions prescribing the kinematic quantities. Also, the presence of any active contact constraint results in the addition of some entries in  $\dmat{f}_{ext}$. A core objective of this work is to detect those nodes getting constrained in physical contact and calculate the forces acting on them.

\section{Elastic collision of a bar with rigid wall}

\begin{figure}[tbhp]
    \begin{subfigure}[b]{0.48\linewidth}
        \centering
        \includegraphics[width=1.0\linewidth]{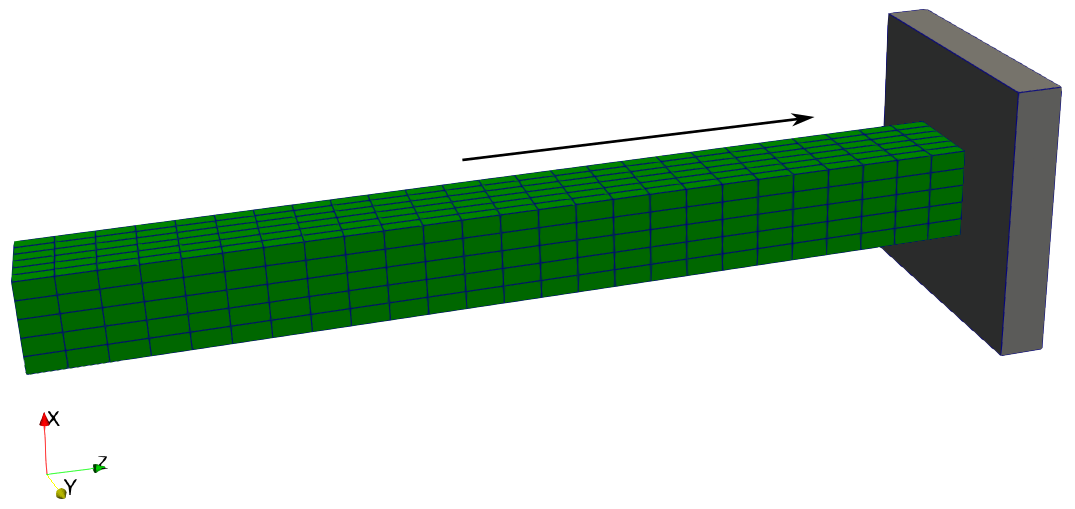}
        \caption{}
        \label{fig:bar_to_wall_impact_a}
    \end{subfigure}
    \begin{subfigure}[b]{0.48\linewidth}
        \centering
        \includegraphics[width=1.0\linewidth]{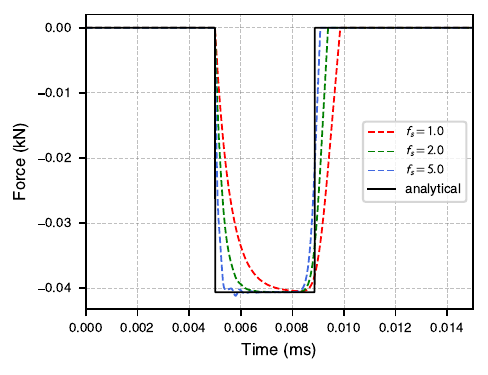}
        \caption{}
    \end{subfigure}
    
    \begin{subfigure}[b]{0.46\linewidth}
        \centering
        \includegraphics[width=1.0\linewidth]{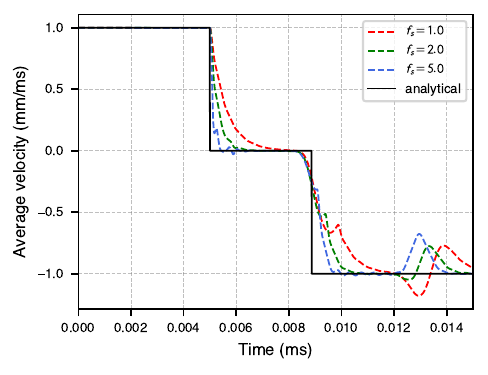}
        \caption{}
    \end{subfigure}
    \begin{subfigure}[b]{0.50\linewidth}
        \centering
        \includegraphics[width=1.0\linewidth]{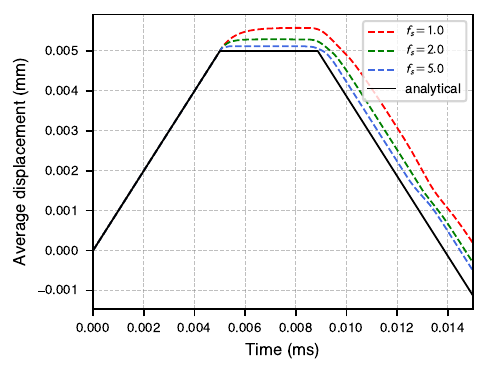}
        \caption{}
    \end{subfigure}    
    \caption{Elastic impact of a bar with rigid wall : (a) mesh, (b) variation of force on contacting face of the bar, (c) variation of the average velocity of contacting face of the bar, and (d) variation of average displacement of contacting face of the bar, shown with the analytical solution of the two-bar collision problem.}
    \label{fig:bar_to_wall_impact}
\end{figure}

The elastic collision problem given in the section \ref{sec_conclusion}, is compared against collision of a single bar against a rigid wall. The wall is placed such that the contact occurs at a distance midway between the previous case of a two-bar collision. With the same speed initial velocity and material properties, one bar impacts over the rigid wall, and the resulting plots of contact force, average velocity and average displacement of contact interface are shown in Fig.~\ref{fig:bar_to_wall_impact}. Here, to minimise oscillations, artificial bulk viscosity is used, with the linear and quadratic factors being 0.6 and 1.5, respectively. The lags observed in the quantities in all three plots, in comparison to two-bar collision, are due to the fact that the contact force rises much quicker on interpenetration with the second bar moving in the opposite direction. All quantities also have lower oscillations in this case compared to the two-bar collision.

 \bibliographystyle{elsarticle-num} 
 \bibliography{references}
 
\end{document}